\documentclass[12pt]{article}
\usepackage{graphicx}
\usepackage{amssymb}
\usepackage{theorem}
\usepackage{bbm}

\theorembodyfont{\upshape}
\newtheorem{theorem}{Theorem}[section]
\newtheorem{lemma}[theorem]{Lemma}
\newtheorem{corollary}[theorem]{Corollary}
\newtheorem{definition}{Definition}[section]

\let\boldface\bfseries
\renewcommand{\bfseries}{\boldface\boldmath}

\newcommand{\SLE}[1][]{\ensuremath{\mbox{SLE}_{#1}}}

\newcommand{\C}[1][]{\ensuremath{\mathbbm{C}^{#1}}}
\newcommand{\D}{\ensuremath{\mathbbm{D}}}
\renewcommand{\H}{\ensuremath{\mathbbm{H}}}
\newcommand{\Hbar}{\ensuremath{\overline{\mathbbm{H}}}}
\newcommand{\N}[1][]{\ensuremath{\mathbbm{N}^{#1}}}
\newcommand{\R}[1][]{\ensuremath{\mathbbm{R}^{#1}}}
\newcommand{\Z}[1][]{\ensuremath{\mathbbm{Z}^{#1}}}

\newcommand{\Prob}{\mbox{\bfseries P}}
\newcommand{\Exp}{\mbox{\bfseries E}}
\newcommand{\fld}[1]{{\mathcal{#1}}}
\newcommand{\fil}[1]{\hbox{\boldmath{$\mathcal{#1}$}}}

\newcommand{\defn}[1]{{\bfseries #1}}

\newcommand{\Hypergeom}{\null_2F_1}
\newcommand{\close}[1]{\overline{#1}}

\renewcommand{\epsilon}{\varepsilon}
\renewcommand{\theta}{\vartheta}
\renewcommand{\phi}{\varphi}

\renewcommand{\Re}{\mathop{\rm Re\,}\nolimits}
\renewcommand{\Im}{\mathop{\rm Im\,}\nolimits}
\newcommand{\im}{{\rm i}}

\newcommand{\goesto}{\mathrel{\rightarrow}}
\newcommand{\downto}{\mathrel{\downarrow}}
\newcommand{\upto}{\mathrel{\uparrow}}

\newcommand{\ahalf}{{\ensuremath{\frac{1}{2}}}}
\newcommand{\e}[1]{{\rm e}^{#1}}
\newcommand{\dif}[1][]{{\rm d}^{#1}}
\newcommand{\half}[1]{\frac{#1}{2}}
\newcommand{\inv}[2][1]{\frac{#1}{#2}}
\newcommand{\deriv}[2][]{\frac{{\rm d}^{#1}}{{\rm d}{#2}^{#1}}}
\newcommand{\pderiv}[2][]{\frac{\partial^{#1}}{\partial{#2}^{#1}}}

\newcommand{\fracpderiv}[3][]{\frac{\partial^{#1}{#3}}{\partial{#2}^{#1}}}

\newcommand{\abar}{\bar{a}}
\newcommand{\zbar}{\bar{z}}

\newcommand{\ftilde}{\tilde{f}}

\newcommand{\gammahat}{\hat{\gamma}}
\newcommand{\Ztilde}{\tilde{Z}}
\newcommand{\xitilde}{\tilde{\xi}}
\newcommand{\ghat}{\hat{g}}
\newcommand{\Khat}{\hat{K}}

\newcommand{\floor}[1]{\left\lfloor{#1}\right\rfloor}

\newcommand{\roival}[1]{\left[{#1}\right)}
\newcommand{\loival}[1]{\left({#1}\right]}

\newenvironment{proof}{\paragraph{Proof.}}{\hfill$\Box$\vskip1\baselineskip}

\title{A Guide to Stochastic L\"owner Evolution and its Applications}
\author{Wouter Kager and Bernard Nienhuis\\
 \null\\
 Institute for Theoretical Physics\\
 University of Amsterdam\\
 Valckenierstraat 65\\
 1018 XE Amsterdam, the Netherlands\\
 e-mail: kager@science.uva.nl and nienhuis@science.uva.nl}

\begin{document}

\maketitle


\begin{abstract}

This article is meant to serve as a guide to recent developments in the study
of the scaling limit of critical models. These new developments were made
possible through the definition of the Stochastic L\"owner Evolution ({\SLE})
by Oded Schramm. This article opens with a discussion of L\"owner's method,
explaining how this method can be used to describe families of random curves.
Then we define {\SLE} and discuss some of its properties. We also explain how
the connection can be made between {\SLE} and the discrete models whose
scaling limits it describes, or is believed to describe. Finally, we have
included a discussion of results that were obtained from {\SLE} computations.
Some explicit proofs are presented as typical examples of such computations.
To understand {\SLE} sufficient knowledge of conformal mapping theory and
stochastic calculus is required. This material is covered in the appendices.


\end{abstract}


\newpage

\section{Introduction}
\label{sec:introduction}

The Stochastic L\"owner Evolution appears as a new branch to an already
varied palette of techniques available for the study of continuous phase
transitions in two dimensions. Phase transitions are among the most striking
phenomena in physics. A small change in an environmental parameter, such as
the temperature or the external magnetic field, can induce huge changes in
the macroscopic properties of a system. Typical examples are liquid-gas
transitions and spontaneous magnetization in ferromagnets. Many more examples
are observed in nature in the most diverse systems, and for a long time
physicist have been searching for explanations of these phenomena (for a
nice introduction explaining the notions involved in the physical
interpretation of these phenomena, see for example~\cite{pfeuty:1977}).

To characterize phase transitions, one introduces an order parameter, a
quantity which vanishes on one side of the phase transition and is non-zero
on the other side. For magnets one uses the magnetization, while for
liquid-gas transitions the density difference between the two phases defines
the order parameter. At the phase transition, the change in the order
parameter can be either discontinuous or continuous. In the former case the
transition is called first-order, in the latter case it is called continuous
or second-order.

It is found experimentally that near continuous phase transitions many
observable quantities have a power-law dependence on their parameters with
non-integer powers, called critical exponents. Thus, the order parameter for
example typically behaves like~$(T_c-T)^\beta$ just below the transition
temperature~$T_c$, and observables such as the specific heat or the
susceptibility diverge as~$|T-T_c|^{-\alpha}$ near~$T_c$. Moreover, the
critical exponents appear to be universal in the sense that there are classes
of different systems, that show critical behaviour with the exact same values
of the critical exponents. This phenomenon is known as universality (some
examples are given in the standard reference~\cite{ma:1976} on the theory of
critical phenomena).

Universality allows one to draw parallels between different systems and
different types of phase transitions. Theoretically, it leads to the
conclusion that the behaviour near a critical point can be described by
just a few relevant parameters, and that many microscopic details of the
system become irrelevant near the critical point. It turns out that the
critical exponents are largely determined by just the dimension of the
system, and the dimension and symmetries of the order parameter. This
justifies the use of simple model systems, in which all the details of
the interactions have been neglected, to investigate critical behaviour.
Examples of such models are the Ising model, the $q$-state Potts models and
O($n$) models. The concept of universality is particularly useful in those
cases where one of these simple models can be solved exactly, because such
solutions determine the universal properties of a whole class of systems,
including those far too difficult to solve exactly.

The behaviour of a system near a transition point is governed by the
fluctuations in the system. When one approaches the critical point from the
disordered phase, these local fluctuations tend to be correlated over larger
and larger distances as one gets closer to the transition point. The
typical length-scale~$\xi$ of these correlations, the correlation length,
diverges as one approaches the critical point. This led physicists to
introduce the idea of length-rescaling~\cite{kadanoff:1966} as a tool for
studying critical phenomena, and ultimately led to the formulation of the
renormalization group approach~\cite{wilson:1974}. The idea of this approach
is that if we look at the system at successively larger scales, the
correlation length will be successively reduced. Away from the critical point
the correlation length is finite, and rescaling drives us further away from
the transition, but at the critical point the correlation length is infinite,
and the system is invariant under rescaling.

The hypothesis of scale-invariance led to the development of several
techniques for the computation of critical exponents and other observables
of critical behaviour, such as correlation functions. One of these
techniques is the successful Coulomb Gas method~\cite{nienhuis:1984}, which
produces exact results provided certain qualitative assumptions are valid.

Another development came from the idea that we need not restrict ourselves
to studying scale-invariance for a system as a whole, but that we might
consider scaling properties locally. More precisely, the system
can be rescaled with a factor that depends on the position, and we may
wonder if the system is invariant under such transformations. This approach
led people to believe that in the continuum limit, many model systems are
not just scale-invariant, but are in fact conformally
invariant~\cite{cardy:1987}. The natural realm for studying conformally
invariant behaviour is that of two-dimensional systems, since in two
dimensions the group of conformal transformations is so much richer than in
higher dimensions. Over the years, the assumption of conformal invariance
has indeed been successful in explaining critical behaviour in
two-dimensional systems. The assumption is supported by the agreement of the
results with the results from exactly solvable models~\cite{baxter:1982}.

Many questions remain, especially from the mathematical point of view. The
physicist intuitively believes that there exists a continuum or scaling limit
of his discrete models when the lattice spacing goes to zero. But when
exactly does this limit exist? What does the limit model look like? Is it
indeed conformally invariant? Such questions have puzzled both mathematicians
and physicists for a long time, and answers to these questions have seemed
quite far away.

A big step forward was made when Oded Schramm~\cite{schramm:2000} combined
an old idea of Karl L\"owner~\cite{loewner:1923} (who later changed his name
to Charles Loewner) from univalent-function theory with stochastic calculus.
This led to the definition of the one-parameter family of Stochastic L\"owner
Evolutions, \SLE[\kappa] (see~\cite{werner:2002} for a mathematical review).
Schramm proved that if the loop-erased random walk has a scaling limit, and
if this limit is conformally invariant, then it must be described by \SLE[2].
He made similar conjectures relating critical percolation to \SLE[6] and
uniform spanning trees to \SLE[8].

Schramm's conjectures for loop-erased random walks and uniform spanning
trees were later proved by Lawler, Schramm and Werner~\cite{lsw:pre0112234}
using \SLE{} techniques. Independently, and using different methods,
Smirnov~\cite{smirnov:2001b} proved the existence and conformal invariance
of the scaling limit of critical site percolation on the triangular lattice,
thus establishing the connection with \SLE[6]. It is believed that many other
models in two dimensions, such as the self-avoiding walk, the $q$-state Potts
models and the O($n$)~models also have a conformally invariant scaling limit
that is described by an \SLE[\kappa] for some characteristic value of~$\kappa$.

The goal of the present article is to explain Schramm's idea to an audience
of both physicists and mathematicians. We place emphasis on how the connection
between the discrete models and \SLE{} is made, and we have included several
typical {\SLE} computations to explain how results can be derived from \SLE{}.
The article is organized as follows. Section~\ref{sec:preliminaries}
contains a few preliminaries that are required to follow the main line of
thought. In section~\ref{sec:loewnerevolutions} we introduce L\"owner
evolutions, and define {\SLE}. The section includes a discussion of the
L\"owner equation in a deterministic setting as an aid to the reader in
understanding the relation between {\SLE} and random paths.

Section~\ref{sec:SLEproperties} then gives an overview of the main properties
of {\SLE} and the random {\SLE} paths. The connection between {\SLE} and
discrete models is discussed in section~\ref{sec:discretemodels}. We give
explicit descriptions of those models that are known rigorously to converge
to {\SLE}, and we also consider the conjectured connections between {\SLE}
and self-avoiding walks, Potts models and O($n$) models. Examples of results
obtained from {\SLE} are given in section~\ref{sec:results}. We have included
several worked-out proofs in the text, explaining how things can be calculated
from {\SLE}. The article ends with a short discussion.

To make the article self-contained, the appendices deal with the background
material that is needed to fully understand all the details of the main text.
In addition, we have intended these appendices to make the mathematical
literature on~{\SLE} more accessible to interested readers, who may not have
all the required background knowledge. For this reason, the appendices cover
more material than is strictly required for the present article.
Appendix~\ref{sec:conformalmappingtheory} deals with conformal mapping
theory. We present some general results of the theory, and focus on topics
that are specific for~{\SLE}. Appendix~\ref{sec:stochasticprocesses} is about
stochastic processes, and includes an introduction to the measure-theoretic
background of probability theory.


\section{Preliminaries}
\label{sec:preliminaries}

In this short section we give a quick overview of notations and some basic
results concerning conformal mapping theory that are used throughout this
article. For a more comprehensive treatment of this material including
illustrations we refer to appendix~\ref{sec:conformalmappingtheory}.

First, some notation. We shall write~$\C$ for the complex plane, and~$\R$
for the set of real numbers. The open upper half-plane $\{z:\Im z>0\}$ is
denoted by~$\H$, and the open unit disk $\{z:|z|<1\}$ by~$\D$. We shall
only consider domains whose boundary is a continuous curve, and this
implies that the conformal maps we work with have well-defined limit values
on the boundary.

Now suppose that~$D$ is a simply connected domain with continuous boundary,
and that $z_1$, $z_2$, $z_3$ and~$z_4$ are distinct points on~$\partial D$,
ordered in the counter-clockwise direction. Then we can map~$D$ onto a
rectangle $(0,L)\times(0,\im\pi)$ in such a way that the arc $[z_1,z_2]$ of
$\partial D$ maps onto~$[0,\im\pi]$, and $[z_3,z_4]$ maps onto~$[L,L+\im\pi]$.
The length~$L>0$ of this rectangle is determined uniquely, and is called
the \defn{$\pi$-extremal distance} between $[z_1,z_2]$ and~$[z_3,z_4]$
in~$D$.

A compact subset~$K$ of~$\Hbar$ such that $\H\setminus K$ is simply connected
and $K=\close{K\cap\H}$ is called a \defn{hull} (it is basically a compact set
bordering on the real line). For any hull~$K$ there exists a unique conformal
map, denoted by $g_K$, which sends $\H\setminus K$ onto $\H$ and satisfies the
normalization
\begin{equation}
 \lim_{z\goesto\infty}\big(g_K(z)-z\big)=0.
\end{equation}
This map has an expansion for $z\goesto\infty$ of the form
\begin{equation}
 g_K(z) = z + \frac{a_1}{z} + \ldots + \frac{a_n}{z^n} + \ldots
\end{equation}
where all expansion coefficients are real. The coefficient $a_1=a_1(K)$ is
called the \defn{capacity} of the hull~$K$.

The capacity of a nonempty hull~$K$ is a positive real number, and satisfies
a \defn{scaling rule} and a \defn{summation rule}. The scaling rule says that
if~$r>0$ then $a_1(rK) = r^2 a_1(K)$. The summation rule says that if
$J\subset K$ are two hulls and $L$ is the closure of $g_J(K\setminus J)$, then
$g_K = g_L\circ g_J$ and $a_1(K) = a_1(J) + a_1(L)$. The capacity of a hull
is bounded from above by the square of the radius of the smallest half-disk
that contains the hull and has its centre on the real line.



\section{L\"owner evolutions}
\label{sec:loewnerevolutions}

This section is devoted to the L\"owner equation and its relation to paths
in the upper half-plane. In this section, we will first discuss the L\"owner
equation in a deterministic setting. We will show how one can describe a
given continuous path by a family of conformal maps, and we will prove that
these maps satisfy L\"owner's differential equation. Then we will prove that
conversely, the L\"owner equation generates a family of conformal maps, that
may of may not describe a continuous curve. Finally, we move on to the
definition of the stochastic L\"owner evolution. This section is based on
ideas from Lawler, Schramm and Werner~\cite{lsw:I}, 
Lawler~\cite{lawler:url2001}, and Rohde and Schramm~\cite{rohde:2001}.

\subsection{Describing a path by the L\"owner equation}
\label{ssec:pathtoloewner}

Suppose that~$\gamma(t)$ (where~$t\geq0$) is a continuous path in~$\Hbar$
which starts from $\gamma(0)\in\R$. We allow the path to hit itself or the
real line, but if it does, we require the path to reflect off into open
space immediately. In other words, the path is not allowed to enter a region
which has been disconnected from infinity by $\gamma[0,t]\cup\R$. To be
specific, let us denote by~$H_t$ for~$t\geq0$ the unbounded connected
component of~$\H\setminus\gamma[0,t]$, and let~$K_t$ be the closure of
$\H\setminus H_t$. Then we require that for all $0\leq s<t$, $K_s$ is a
proper subset of~$K_t$. See figure~\ref{fig:Path} for a picture of a path
satisfying these conditions.

We further impose the conditions that for all~$t\geq0$ the set~$K_t$ is
bounded, so that~$\{K_t:t\geq0\}$ is a family of growing hulls, and that
the capacity of these hulls eventually goes to infinity,
i.e.\ $\lim_{t\goesto\infty}a_1(K_t)=\infty$. The latter condition implies
that the path eventually has to escape to infinity, but there do exist paths
to infinity whose capacities remain finite (a formula for the capacity is
given at the end of appendix~\ref{ssec:hullsandcapacity}). Now let us state
the purpose of this subsection.

For every~$t\geq0$ we set~$g_t:=g_{K_t}$, and we further define the
real-valued function~$U_t:=g_t(\gamma(t))$ (this is the point to which the
tip of the path is mapped). The purpose of this subsection is to prove that
the maps~$g_t$ satisfy a simple differential equation, which is driven
by~$U_t$. Ideas for the proof were taken from~\cite{lsw:I}. For a different,
probabilistic approach, see~\cite{lawler:url2001}. The first thing that we
show, is that we can choose the time parameterization of~$\gamma$ such that
the capacity grows linearly in time. Clearly, this fact is a direct
consequence of the following theorem.

\begin{theorem}
 \label{the:reparameterization}
 Both~$a_1(K_t)$ and~$U_t$ are continuous in~$t$.
\end{theorem}

\begin{proof}
 The proof relies heavily on properties of $\pi$-extremal distance, and we
 refer to the chapter on extremal length, sections 4.1--4.5 and 4.11--4.13,
 in Ahlfors~\cite{ahlfors:1973} for the details. We shall prove
 left-continuity first.

 Without loss of generality we may assume that~$\gamma(0)=0$. Fix~$t>0$,
 let~$R$ be a large number, say at least several times the radius of~$K_t$,
 and let~$C_R$ be the upper half of the circle with radius~$2R$ centred
 at the origin. Fix~$\epsilon>0$. Then by continuity of~$\gamma(t)$, there
 exists a $\delta>0$ such that $|\gamma(t)-\gamma(u)|<\epsilon/2$ for all
 $u\in(t-\delta,t)$. Now let~$C_\epsilon$ be the circle with radius~$\epsilon$
 and centre~$\gamma(t)$, and let~$S$ be the arc of this circle in the
 domain~$H_t$. Then this set~$S$ disconnects $K_t\setminus K_{t-\delta}$ from
 infinity in~$H_{t-\delta}$, see figure~\ref{fig:Path}. Observe that the set
 $K_t\setminus K_{t-\delta}$ may be just a piece of~$\gamma$, but that it can
 also be much larger, as in the figure.

\begin{figure}
  \centering\includegraphics[scale=1.17]{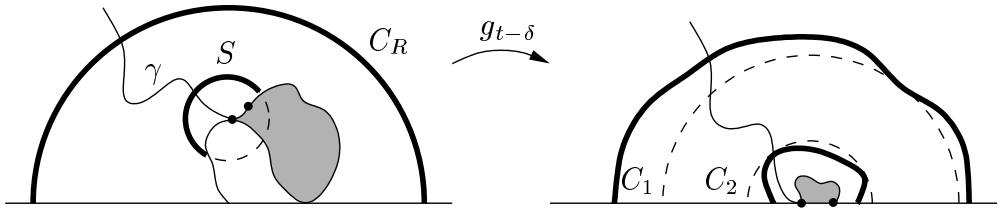}
  \caption{A path~$\gamma$. The two points represent $\gamma(t)$
   and~$\gamma(t-\delta)$, and the shaded area is the set $K_t\setminus
   K_{t-\delta}$. For clarity, the arc~$C_R$ is drawn much smaller than
   it is in the proof.}
  \label{fig:Path}
\end{figure}

 For convenience let us denote by~$\Omega$ the part of the
 domain~$H_{t-\delta}$ that lies below~$C_R$. Let~$\mathcal{L}$ be the
 $\pi$-extremal distance between~$S$ and~$C_R$ in~$\Omega$. By the properties
 of $\pi$-extremal distance, because the circle with radius~$R$ and centre
 at~$\gamma(t)$ lies below~$C_R$, $\mathcal{L}$ must be at least
 $\log(R/\epsilon)/2$. Note that since $\pi$-extremal distance is invariant
 under conformal maps, $\mathcal{L}$ is also the $\pi$-extremal distance
 between $g_{t-\delta}(C_R)$ and $g_{t-\delta}(S)$ in~$g_{t-\delta}(\Omega)$.
 This allows us to find an upper bound on~$\mathcal{L}$.

 To get this upper bound, we draw two concentric semi-circles $C_1$ and~$C_2$,
 the first hitting $g_{t-\delta}(C_R)$ on the inside, and the second hitting
 $g_{t-\delta}(S)$ on the outside as in figure~\ref{fig:Path} (this is always
 possible if~$R$ was chosen large enough). Note that by the hydrodynamic
 normalization of the map $g_{t-\delta}$, we have an upper bound on the
 radius of~$C_1$, which depends only on~$R$ (this follows for example from
 theorem~\ref{the:hullarea}). As is explained in Ahlfors, this means that the
 $\pi$-extremal distance~$\mathcal{L}$ satisfies an inequality of the form
 $\mathcal{L}\leq\log(C(R)/r)$, where~$C(R)$ depends only on our choice
 of~$R$, and~$r$ is the radius of the inner half-circle~$C_2$.
 But~$\mathcal{L}$ was at least~$\log(R/\epsilon)/2$, implying that~$r$ can
 be made arbitrarily small by choosing~$\delta$ small enough. It follows
 that for every $\epsilon>0$ there exists a $\delta>0$ such that the set
 $K_{t,\delta}:=g_{t-\delta}(K_t\setminus K_{t-\delta})$ is contained
 in a half-disk of radius~$\epsilon$. But then by the summation rule of
 capacity $a_1(K_t)-a_1(K_{t-\delta})=a_1(K_{t,\delta})\leq\epsilon^2$,
 proving left-continuity of~$a_1(K_t)$.

 To prove left-continuity of~$U_t$, let $\delta$ and $\epsilon$ be as
 above, and denote by $g_{t,\delta}$ the normalized map $g_{K_{t,\delta}}$
 associated with the hull $K_{t,\delta}$. It is clearly sufficient to show
 that $g_{t,\delta}$ converges uniformly to the identity as $\delta\downto0$
 (remember that $U_t$ is defined as $g_t(\gamma(t))$ and refer to
 figure~\ref{fig:Path}). To prove this, we may assume without loss of
 generality that the set~$K_{t,\delta}$ is contained within the disk
 of radius~$\epsilon$ centred at the origin, since the claim remains valid
 under translations over the real line. But then theorem~\ref{the:hullarea}
 says that if $|z|\geq2\epsilon$, then
 \begin{equation}
  |g_{t,\delta}(z)-z| \leq \sum_{n=1}^\infty \frac{a_n(K_{t,\delta})}{|z|^n}
   \leq \epsilon \sum_{n=1}^\infty \frac{\epsilon^n}{(2\epsilon)^n}
   = \epsilon.
 \end{equation}
 This shows that the map~$g_{t,\delta}$ converges uniformly to the identity.
 Left-continuity of~$U_t$ follows. In the same way we can prove
 right-continuity of $a_1(K_t)$ and~$U_t$.
\end{proof}

\begin{theorem}
 \label{the:pathtoloewner}
 Let $\gamma(t)$ be parameterized such that $a_1(K_t)=2t$. Then for all
 $z\in\H$, as long as $z$ is not an element of the growing hull, $g_t(z)$
 satisfies the L\"owner differential equation
 \begin{equation}
  \pderiv{t} g_t(z) = \frac{2}{g_t(z)-U_t},\quad g_0(z)=z.
 \end{equation}
\end{theorem}

\begin{proof}
 Our proof is based on the proof of theorem~\ref{the:reparameterization}
 and the Poisson integral formula, which states that the map~$g_{t,\delta}$
 satisfies
 \begin{equation}
  \label{equ:Poisson}
  g_{t,\delta}(z)-z = \inv{\pi}\int_{-\infty}^\infty
   \frac{\Im g_{t,\delta}^{-1}(\xi)}{g_{t,\delta}(z)-\xi}\,\dif{\xi},
   \qquad z\in\H\setminus K_{t,\delta}
 \end{equation}
 while the capacity~$a_1(K_{t,\delta})$ is given by the integral
 \begin{equation}
  \label{equ:Poissoncapacity}
  a_1(K_{t,\delta}) = \inv{\pi}\int_{-\infty}^\infty
   \Im g_{t,\delta}^{-1}(\xi)\,\dif{\xi}.
 \end{equation}
 See appendix~\ref{ssec:hullsandcapacity} for more information.

 First consider the left-derivative of~$g_t(z)$. Using the same notations as
 in the proof of theorem~\ref{the:reparameterization} we can write $g_t=
 g_{t,\delta}\circ g_{t-\delta}$. We know that $g_{t,\delta}$ converges to the
 identity as $\delta\downto0$, and that the support of $\Im g_{t,\delta}^{-1}$
 shrinks to the point~$U_t$. Moreover, using the summation rule of capacity
 and our choice of time parameterization, equation~(\ref{equ:Poissoncapacity})
 gives $\int\Im g_{t,\delta}^{-1}(\xi)\,\dif{\xi}=2\pi\delta$. Hence from
 equation~(\ref{equ:Poisson}) we get
 \begin{equation}
  \lim_{\delta\downto0}\frac{g_t(z)-g_{t-\delta}(z)}{\delta}
   = \lim_{\delta\downto0}\inv{\pi\delta} \int
   	\frac{\Im g_{t,\delta}^{-1}(\xi)}
 		 {g_{t,\delta}\big(g_{t-\delta}(z)\big)-\xi}
 	\,\dif{\xi}
   = \frac{2}{g_t(z)-U_t}.
 \end{equation}
 In the same way one obtains the right-derivative.
\end{proof}

\subsection{The solution of the L\"owner equation}
\label{ssec:loewnertopath}

In the previous subsection, we started from a continuous path~$\gamma$ in the
upper half-plane. We proved that the corresponding conformal maps satisfy the
L\"owner equation, driven by a suitably defined continuous function~$U_t$.
In this subsection, we will try to go the other way around. Starting from
a driving function~$U_t$, we will prove that the L\"owner equation generates
a (continuous) family of conformal maps~$g_t$ onto~$\H$. The proof follows
Lawler~\cite{lawler:url2001}.

So suppose that we have a continuous real-valued function~$U_t$. Consider
for some point~$z\in\Hbar\setminus\{0\}$ the L\"owner differential equation
\begin{equation}
 \label{equ:dle}
 \pderiv{t}g_t(z) = \frac{2}{g_t(z)-U_t}, \qquad g_0(z)=z.
\end{equation}
This equation gives us some immediate information on the behaviour of
$g_t(z)$. For instance, taking the imaginary part we obtain
\begin{equation}
 \label{equ:Imdle}
 \pderiv{t}\Im g_t(z) =
  \frac{-2\,\Im g_t(z)}{(\Re g_t(z) - U_t)^2 + (\Im g_t(z))^2}.
\end{equation}
This shows that for fixed~$z\in\H$, $\partial_t\Im[g_t(z)] < 0$, and hence
that $g_t(z)$ moves towards the real axis. Further, points on the real axis
will stay on the real axis.

For a given point~$z\in\Hbar\setminus\{0\}$, the solution of the L\"owner
equation is well-defined as long as~$g_t(z)-U_t$ stays away from zero. This
suggests that we define a time~$\tau(z)$ as the first time~$\tau$ such that
$\lim_{t\uparrow\tau}(g_t(z)-U_t)=0$, setting $\tau(z)=\infty$ if this never
happens. Note that as long as $g_t(z)-U_t$ is bounded away from zero,
equation~(\ref{equ:Imdle}) shows that the time derivative of~$\Im[g_t(z)]$
is bounded in absolute value by some constant times~$\Im[g_t(z)]$. For
points $z\in\H$ this shows that in fact, $\tau(z)$ must be the first time
when~$g_t(z)$ hits the real axis. We set
\begin{equation}
 \label{equ:dlehull}
 H_t := \{z\in\H:\tau(z)>t\}, \qquad
 K_t := \{z\in\Hbar:\tau(z)\leq t\}.
\end{equation}
Then~$H_t$ is the set of points in the upper half-plane for which~$g_t(z)$
is still well-defined, and~$K_t$ is the closure of its complement, i.e.\ it
is the hull which is excluded from~$H_t$. Our goal is now to prove the
following theorem.

\begin{theorem}
\label{the:loewnertomaps}
 Let~$U_t$ be a continuous real-valued function, and for every $t\geq0$ let
 $g_t(z)$ be the solution of the L\"owner equation~(\ref{equ:dle}). Define
 the set~$H_t$ as in~(\ref{equ:dlehull}). Then~$g_t(z)$ is a conformal map
 of the domain~$H_t$ onto~$\H$ which satisfies
 \begin{equation}
  \label{equ:dlemapexpansion}
   g_t(z) = z + \frac{2t}{z} + O\left(z^{-2}\right),
   \qquad z\rightarrow\infty.
  \end{equation}
\end{theorem}

\begin{proof}
 It is easy to see from~(\ref{equ:dle}) that~$g_t$ is analytic on~$H_t$.
 We will prove (i) that the map~$g_t$ is conformal on the domain~$H_t$, (ii)
 that this map is of the form~(\ref{equ:dlemapexpansion}), and (iii) that
 $g_t(H_t)=\H$.
 
 To prove~(i), we have to verify that~$g_t$ has nonzero derivative on~$H_t$,
 and that it is injective. So consider equation~(\ref{equ:dle}) for
 times~$t<\tau(z)$. Then the differential equation behaves nicely, and we
 can differentiate with respect to~$z$ to obtain
 \begin{equation}
  \pderiv{t}\log g'_t(z) = -\frac{2}{(g_t(z)-U_t)^2}.
 \end{equation}
 This gives $|\partial_t\log g'_t(z)|\leq2/[\Im g_t(z)]^2$. But we know that
 $\Im[g_t(z)]$ is decreasing. Hence, if we fix~$t_0<\tau(z)$, then the change
 in $\log g'_t(z)$ is uniformly bounded for all times~$t<t_0$. It follows
 that~$\log g'_{t_0}(z)$ is well-defined and bounded and hence,
 that~$g'_t(z)$ is well-defined and nonzero for all~$t<\tau(z)$.
 
 Next, choose two different points~$z,w\in\H$ and
 let~$t<\min\{\tau(z),\tau(w)\}$. Then
 \begin{equation}
  \pderiv{t}\log[g_t(z)-g_t(w)] = -\frac{2}{(g_t(z)-U_t)(g_t(w)-U_t)}.
 \end{equation}
 It follows that~$g_t(z)\neq g_t(w)$ for all~$t<\min\{\tau(z),\tau(w)\}$,
 using a similar argument as above. We conclude that~$g_t(z)$ is conformal
 on the domain~$H_t$.

 For the proof of~(ii), we note that~(i) implies that the map~$g_t(z)$
 can be expanded around infinity. We can determine the form of the expansion
 by integrating the L\"owner differential equation from~$0$ to~$t$. This
 yields
 \begin{equation}
  g_t(z)-z = \int_0^t \frac{2\dif{s}}{g_s(z)-U_s}.
 \end{equation}
 Consider this equation in the limit~$z\rightarrow\infty$. Then it is easy
 to see that the expansion of~$g_t(z)$ has no terms of quadratic or higher
 power in~$z$, and no constant term. The form~(\ref{equ:dlemapexpansion})
 follows immediately.

\begin{figure}
  \centering\includegraphics[scale=1.2]{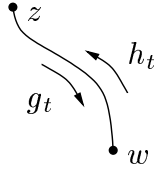}
  \caption{If the flow of a point~$z$ up to a time~$t_0$ is described
   by~$g_t(z)$, then~$h_t(w)$ as defined in the text describes the inverse
   flow.}
  \label{fig:Flow}
\end{figure}

 Finally, we prove~(iii), i.e.\ we will show that $g_t(H_t) = \H$. To see
 this, let~$w$ be any point in~$\H$, and let~$t_0$ be a fixed time. Define
 $h_t(w)$ for $0\leq t\leq t_0$ as the solution of the problem
 \begin{equation}
 \label{equ:inverseflow}
  \pderiv{t}h_t(w) = -\inv[2]{h_t(w)-U_{t_0-t}}, \qquad h_0(w)=w.
 \end{equation}
 The imaginary part of this equation says that~$\partial_t\Im[h_t(w)]>0$ and
 hence, that~$\Im[h_t(w)]$ is increasing in time. Since~$|\partial_t h_t(w)|
 \leq2/\Im[h_t(w)]$, it follows that $h_t(w)$ is well-defined for
 all~$0\leq t\leq t_0$.

 We defined $h_t(w)$ such that it describes the inverse of the flow of
 some point $z\in H_{t_0}$ under the L\"owner evolution~(\ref{equ:dle}) (see
 figure~\ref{fig:Flow}). To see that this is indeed the case, suppose that
 for some~$t$ between $0$ and~$t_0$, $h_{t_0-t}(w)=g_t(z)$ for some~$z$.
 Then it follows from the differential equation for~$h_t(w)$, that~$g_t(z)$
 satisfies equation~(\ref{equ:dle}). This observation holds for all
 times~$t$ between $0$ and~$t_0$. It follows that such a point~$z$ exists,
 and that it is in fact determined by $z=g_0(z)=h_{t_0}(w)$. In other words,
 for all~$w\in\H$ we have~$g_{t_0}(z)=w$ for some~$z\in H_{t_0}$. This
 completes the proof.
\end{proof}

We have just proved that a continuous function~$U_t$ leads, via the
L\"owner evolution equation~(\ref{equ:dle}), to a collection of conformal
maps~$\{g_t:t\geq0\}$. These conformal maps are defined on subsets of the
upper half-plane, namely the sets~$H_t=\H\setminus K_t$, with~$K_t$ a growing
hull. At this point we still don't know if the maps~$g_t(z)$ also
correspond to a path~$\gamma(t)$. But in the next subsection we shall
take~$U_t$ to be a scaled Brownian motion, and it is known~\cite{rohde:2001}
that in this case the L\"owner evolution does correspond to a path.

\subsection[Chordal SLE in the half-plane]{Chordal {\SLE} in the half-plane}
\label{ssec:slehalfplane}

In the previous subsection we showed that the L\"owner equation~(\ref{equ:dle})
driven by a continuous real-valued function generates a set of conformal
maps. Furthermore, these conformal maps may correspond to a path in the upper
half-plane, as is suggested by the conclusions of
section~\ref{ssec:pathtoloewner}. Chordal \SLE[\kappa] in the half-plane is
obtained by taking scaled Brownian motion as the driving process. We give a
precise definition in this subsection.

Let~$B_t$, $t\in\roival{0,\infty}$, be a standard Brownian motion on~$\R$,
starting from~$B_0=0$, and let $\kappa>0$ be a real parameter. For each
$z\in\Hbar\setminus\{0\}$, consider the L\"owner differential equation
\begin{equation}
 \label{equ:sle}
 \pderiv{t} g_t(z) = \frac{2}{g_t(z)-\sqrt{\kappa}B_t}, \quad g_0(z)=z.
\end{equation}
This has a solution as long as the denominator~$g_t(z)-\sqrt{\kappa}B_t$
stays away from zero.

For all~$z\in\Hbar$, just as in the previous subsection, we define~$\tau(z)$
to be the first time~$\tau$ such that~$\lim_{t\uparrow\tau}(g_t(z)-
\sqrt{\kappa}B_t)=0$, $\tau(z)=\infty$ if this never happens, and we set
\begin{equation}
 \label{equ:slehull}
 H_t := \{z\in\H : \tau(z)>t\}, \qquad
 K_t := \{z\in\Hbar : \tau(z) \leq t\}.
\end{equation}
That is, $H_t$ is the set of points in the upper half-plane for which~$g_t(z)$
is well-defined, and~$H_t=\H\setminus K_t$. The definition is such that~$K_t$
is a hull, while~$H_t$ is a simply-connected domain. We showed in the previous
subsection that for every~$t\geq0$, $g_t$ defines a conformal map of~$H_t$
onto the upper half-plane~$\H$, that satisfies the normalization
$\lim_{z\rightarrow\infty}(g_t(z)-z)=0$.

\begin{definition}[Stochastic L\"owner Evolution]
 The family of conformal maps $\{g_t:t\geq0\}$ defined through the
 stochastic L\"owner equation~(\ref{equ:sle}) is called
 \defn{chordal \SLE[\kappa]}. The sets~$K_t$~(\ref{equ:slehull}) are the
 \defn{hulls} of the process.
\end{definition}

The \SLE[\kappa] process defined through equation~(\ref{equ:sle}) is called
chordal, because its hulls are growing from a point on the boundary (the
origin) to another point on the boundary (infinity). We will keep using the
term chordal for processes going between two boundary points (and not only
for {\SLE} processes). Other kinds of processes might for instance grow
from a point on the boundary to a point in the interior of a domain. An
example of such a process is radial {\SLE}, see section~\ref{ssec:radialsle}.

It turns out that the hulls of chordal {\SLE} in fact are the hulls of a
continuous path~$\gamma(t)$, that is called the \defn{trace} of the {\SLE}
process. It is through this trace that the connection with discrete models
can be made. We shall discuss properties of the trace in
section~\ref{sec:SLEproperties}, and we will look at the connection with
discrete models in section~\ref{sec:discretemodels}. The precise definition
of the trace is as follows.

\begin{definition}[Trace]
 The {\bf trace}~$\gamma$ of \SLE[\kappa] is defined by
 \begin{equation}
  \label{equ:sletrace}
  \gamma(t) := \lim_{z\rightarrow0}g_t^{-1}(z+\sqrt{\kappa}B_t)
 \end{equation}
 where the limit is taken from within the upper half-plane.
\end{definition}

At this point we would like to make some remarks about the choice of time
parameterization. Chordal {\SLE} is defined such that the capacity of the
hull~$K_t$ satisfies $a_1(K_t)=2t$, and this may seem somewhat arbitrary.
But in practice, the choice of time parameterization does not matter for
our calculations. The point is, that in {\SLE} calculations we are usually
interested in expectation values of random variables at the first time when
some event happens, that is, at a stopping time. These expectation values
are clearly independent from the chosen time parameterization (even if we
make a random change of time). For examples of such calculations, see
sections \ref{ssec:phases} and~\ref{ssec:SLEcalculations}.

Still, it is interesting to examine how a time-change affects the L\"owner
equation. So, let~$c(t)$ be an increasing and differentiable function defining
a change of time. Then $\ghat_t:=g_{c(t)/2}$ is a collection of conformal
transformations parameterized such that~$a_1(\Khat_t):=a_1(K_{c(t)/2})=c(t)$.
This family of transformations satisfies the equation
\begin{equation}
 \pderiv{t}\ghat_t(z) = \frac{\deriv{t}c(t)}{\ghat_t(z)-\sqrt{k}B_{c(t)/2}},
 \qquad \ghat_0(z)=z.
\end{equation}
In particular, if we choose~$c(t)=2\alpha t$ for some constant~$\alpha>0$,
then the conformal maps~$\ghat_t$ satisfy
\begin{equation}
 \pderiv{t}\inv{\sqrt{\alpha}}\ghat_t(\sqrt{\alpha}z) =
 \frac{2}{\inv{\sqrt{\alpha}}\ghat_t(\sqrt{\alpha}z) -
          \sqrt{\frac{\kappa}{\alpha}} B_{\alpha t}},
 \qquad \inv{\sqrt{\alpha}}\ghat_0(\sqrt{\alpha}z)=z.
\end{equation}
But the scaling property of Brownian motion
(appendix~\ref{ssec:brownianmotion}) shows that the driving term of this
L\"owner equation is again a standard Brownian motion multiplied
by~$\sqrt{\kappa}$. This proves the following lemma.

\begin{lemma}[Scaling property of {\SLE[\kappa]}]
 \label{lem:slescaling}
 If~$g_t$ are the transformations of \SLE[\kappa] and~$\alpha$ is a positive
 constant, then the process $(t,z)\mapsto\ghat_t(z):=
 \alpha^{-1/2}g_{\alpha t}(\sqrt{\alpha}z)$ has the same distribution
 as the process~$(t,z)\mapsto g_t(z)$. Furthermore, the process~$t\mapsto
 \alpha^{-1/2}K_{\alpha t}$ has the same distribution as the process
 $t\mapsto K_t$.
\end{lemma}

This lemma is used frequently in {\SLE} calculations. Its significance will be
shown already in the following subsection, where we define the \SLE[\kappa]
process in an arbitrary simply connected domain. Meanwhile, the strong Markov
property of Brownian motion implies that chordal \SLE[\kappa] has another
basic property, which is referred to as stationarity. Indeed, for any
stopping time~$\tau$ the process $\sqrt{\kappa}(B_{t+\tau}-B_\tau)$ is itself
a standard Brownian motion multiplied by~$\sqrt{\kappa}$. So if we use this
process as a driving term in the L\"owner equation, we will obtain a
collection of conformal maps~$\ghat_t$ which is equal in distribution to the
normal \SLE[\kappa] process.

It is not difficult to see that the process~$\ghat_t(z)$ in question is
in fact the process defined by
\begin{equation}
 \label{equ:stationarityprocess}
 \ghat_t(z) := g_{t+\tau}\left( g^{-1}_\tau(z+\sqrt{\kappa}B_\tau) \right)
  -\sqrt{\kappa}B_\tau.
\end{equation}
Indeed, taking the derivative of~$\ghat_t(z)$ with respect to~$t$, we find
that this process satisfies the L\"owner equation
\begin{equation}
 \pderiv{t}\ghat_t(z) =
  \frac{2}{\ghat_t(z)-\sqrt{\kappa}\left( B_{t+\tau}-B_\tau \right)},
  \qquad \ghat_0(z) = z.
\end{equation}
This result establishes the following lemma.

\begin{lemma}[Stationarity of {\SLE[\kappa]}]
 \label{lem:slestationarity}
 Let~$g_t(z)$ be an \SLE[\kappa] process in~$\H$, and let~$\tau$ be a
 stopping time. Define $\ghat_t(z)$ by~(\ref{equ:stationarityprocess}).
 Then~$\ghat_t$ has the same distribution as~$g_t$, and it is
 independent from~$\{g_t : t\in[0,\tau]\}$.
\end{lemma}

Observe that the process~$\ghat_t$ of this lemma is just the original
\SLE[\kappa] process from the time~$\tau$ onwards, but shifted in such a
way that the new process starts again in the origin. The content of the
lemma is that this new process is the same in distribution as the standard
\SLE[\kappa] process, and independent from the history up to time~$\tau$.
So it is in this sense that the \SLE[\kappa] process is stationary.

\subsection[Chordal SLE in an arbitrary domain]{Chordal {\SLE} in an arbitrary domain}
\label{ssec:slegeneraldomain}

Suppose that~$D\subsetneq\C$ is a simply connected domain. Then the Riemann
mapping theorem says that there is a conformal map~$f:D\rightarrow\H$.
Now, let~$f_t$ be the solution of the L\"owner equation~(\ref{equ:sle})
with initial condition~$f_0(z)=f(z)$ for~$z\in D$. Then we will call the
process~$\{f_t : t\geq0\}$ the \SLE[\kappa] in~$D$ under the map~$f$.
The connection with the solution~$g_t$ of~(\ref{equ:sle}), with initial
condition~$g_0(z)=z$, is easily established. Obviously we
have~$f_t=g_t\circ f$, and if~$K_t$ are the hulls associated with~$g_t$,
then the hulls associated with~$f_t$ are~$f^{-1}(K_t)$.

Now suppose that we want to consider an \SLE[\kappa] trace that crosses
some domain~$D$ from a specified point to another specified point. To be
definite, let the starting point be~$a\in\partial D$, and let the ending
point be~$b\in\partial D$, $a\neq b$. Then we can find a conformal map
$f:D\rightarrow\H$ such that $f(a)=0$ and~$f(b)=\infty$. The \SLE[\kappa]
process from $a$ to~$b$ in~$D$ under the map~$f$ is then defined as we
discussed above, with starting point~$f(a)=0$.

The map~$f$, however, is not determined uniquely. But any other map~$\ftilde$
of $D$ onto~$\H$ that sends $a$ to~$0$ and $b$ to~$\infty$, must satisfy
$\ftilde(z)=\alpha f(z)$ for some~$\alpha>0$ by theorem~\ref{the:selfmapsofH}.
Lemma~\ref{lem:slescaling} then tells us that the trace of the \SLE[\kappa]
process in~$D$ under~$\ftilde$  is given simply by a linear time-change of
the \SLE[\kappa] process under~$f$. But we explained in the previous
subsection that a time-change does not affect our calculations, and may
therefore be ignored. Hence, in the sequel, we can simply speak of {\SLE}
processes in an arbitrary domain, without mentioning the conformal maps that
take these processes to the upper half-plane.

\subsection[Radial SLE]{Radial \SLE{}}
\label{ssec:radialsle}

So far we have looked only at chordal L\"owner evolution processes, which
grow from one point on the boundary of a domain to another point on the
boundary. One can also study L\"owner evolution processes which grow from
a boundary point to a point in the interior of the domain. We call such
processes \defn{radial} L\"owner evolutions. Radial \SLE[\kappa] in the
unit disk, for example, is defined as follows.

Let~$B_t$ again be Brownian motion, and~$\kappa>0$.
Set~$W_t:=\exp(\im\sqrt{\kappa}B_t)$, so that~$W_t$ is Brownian motion on the
unit circle starting from~$1$. Then radial \SLE[\kappa] is defined to be
the solution of the L\"owner equation
\begin{equation}
 \pderiv{t}g_t(z) = g_t(z)\frac{W_t+g_t(z)}{W_t-g_t(z)},\quad
 g_0(z)=z,\quad z\in\close{\D}.
\end{equation}
The solution again exists up to a time~$\tau(z)$ which is defined to be
the first time~$\tau$ such that $\lim_{t\upto\tau}(g_t(z)-W_t)=0$.

If we set
\begin{equation}
 H_t := \{z\in\D:\tau(z)>t\}, \qquad
 K_t := \{z\in\close{\D}:\tau(z)\leq t\},
\end{equation}
then $g_t$ is a conformal map of $\D\setminus K_t=H_t$ onto~$\D$. The maps
are in this case normalized by $g_t(0)=0$ and~$g_t'(0)>0$. In fact it is
easy to see from the L\"owner equation that $g_t'(0)=\exp(t)$, and this
specifies the time parameterization.

The trace of radial \SLE[\kappa] is defined by $\gamma(t):=\lim_{z\goesto W_t}
g_t^{-1}(z)$, where now the limit is to be taken from within the unit disk.
The trace goes from the starting point~$1$ on the boundary to the origin. By
conformal mappings, one can likewise define radial \SLE{} in an arbitrary
simply connected domain, growing from a given point on the boundary to a
given point in the interior.


\section[Properties of SLE]{Properties of {\SLE}}
\label{sec:SLEproperties}

In this section we describe some of the properties of {\SLE}. In particular,
we shall see that the family of conformal maps $\{g_t:t\geq0\}$ that is the
solution of the stochastic L\"owner equation~(\ref{equ:sle}) does describe a
continuous path. We will look at the properties of this path, and we shall
describe the connection with the hulls $\{K_t:t\geq0\}$ of the process. To
give the reader an impression of the kind of computations involved, we spell
out a few of the shorter proofs. All of this work was done originally by
Rohde and Schramm~\cite{rohde:2001}. We shall also see that {\SLE} has some
special properties in the cases $\kappa=6$ (locality) and $\kappa=8/3$
(restriction), as was shown in~\cite{lsw:I} and~\cite{lsw:2003}. We
end the section by giving the Hausdorff dimensions of the {\SLE} paths,
calculated by Beffara~\cite{beffara:pre0204208,beffara:pre0211322}.

\subsection{Continuity and transience}
\label{ssec:continuity}

In section~\ref{ssec:loewnertopath} we proved that the solution of the
L\"owner equation is a family of conformal maps onto the half-plane. We
then raised the question whether these conformal maps describe a continuous
path. Rohde and Schramm~\cite{rohde:2001} proved that for chordal
\SLE[\kappa] this is indeed the case, at least for all~$\kappa\neq8$.
The proof by Rohde and Schramm does not work for~$\kappa=8$. But later,
Lawler, Schramm and Werner~\cite{lsw:pre0112234} proved that \SLE[8] is the
scaling limit of the Peano curve winding around a uniform spanning tree (more
details follow in section~\ref{sec:discretemodels}). Thereby, they showed
indirectly that the trace is a continuous curve in the case~$\kappa=8$ as
well. More precisely, the following theorem holds.

\begin{theorem}[Continuity]
 For all $\kappa\geq0$ almost surely the limit
 \begin{equation}
  \gamma(t) := \lim_{z\goesto0} g_t^{-1}(z+\sqrt{\kappa}B_t)
 \end{equation}
 exist for every~$t\geq0$, where the limit is taken from within the upper
 half-plane. Moreover, almost surely $\gamma:\roival{0,\infty}\goesto\Hbar$
 is a continuous path and~$H_t$ is the unbounded connected component of
 $\H\setminus\gamma[0,t]$ for all~$t\geq0$.
\end{theorem}

In the same paper, Rohde and Schramm also showed that the trace of
\SLE[\kappa] is transient for all~$\kappa\geq0$, that is,
$\lim_{t\goesto\infty}|\gamma(t)|=\infty$ almost surely. This proves
that the {\SLE} process in the half-plane is indeed a chordal process
growing from~$0$ to infinity.

\subsection[Phases of SLE]{Phases of {\SLE}}
\label{ssec:phases}

\begin{figure}
 \centering\includegraphics[scale=1.2]{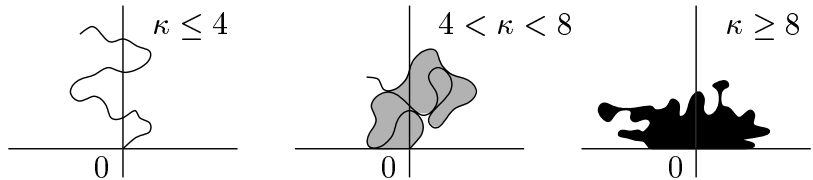}
 \caption{Simplified impression of {\SLE} in the three different phases.
  The trace of the {\SLE} process is shown in black. The union of the black
  path and the grey areas represents the hull.}
 \label{fig:Phases}
\end{figure}

The behaviour of the trace of \SLE[\kappa] depends naturally on the value
of the parameter~$\kappa$. It is the purpose of this subsection to point
out that we can discern three different phases in the behaviour of this
trace. The two phase transitions take place at the values $\kappa=4$
and~$\kappa=8$. A sketch of what the three different phases look like is
given in figure~\ref{fig:Phases}.

For~$\kappa\in[0,4]$ the \SLE[\kappa] trace~$\gamma$ is almost surely a
simple path, i.e.\ $\gamma(s)\neq\gamma(t)$ for all~$0\leq t<s$. Moreover,
the trace a.s.\ does not hit the real line but stays in the upper half-plane
after time~$0$. Clearly then, the hulls~$K_t$ of the process coincide with
the trace~$\gamma[0,t]$.

When $\kappa$ is larger than~$4$, the trace is no longer simple. In fact,
for all~$\kappa>4$ we have that every point~$z\in\Hbar\setminus\{0\}$
a.s.\ becomes part of the hull in finite time. This means that every point
is either on the trace, or is disconnected from infinity by the trace. But as
long as~$\kappa<8$, it can be shown that the former happens with probability
zero. Therefore, for $\kappa\in(4,8)$ we have a phase where the trace
is not dense but does eventually disconnect all points from infinity. In
other words, the trace now intersects both itself and the real line, and
the hulls~$K_t$ now consist of the union of the trace~$\gamma[0,t]$ and
all bounded components of $\Hbar\setminus\gamma[0,t]$.

Finally, when $\kappa\geq8$ the trace becomes dense in~$\H$. In fact, we are
then in a phase where $\gamma\roival{0,\infty}=\Hbar$ with probability~$1$,
and the hulls~$K_t$ coincide with the trace~$\gamma[0,t]$ again.

The proofs of the properties of \SLE[\kappa] for~$\kappa\in[0,4]$ are not
too difficult and illustrate nicely some of the techniques involved in
{\SLE} calculations. For these reasons, we reproduce these proofs from Rohde
and Schramm~\cite{rohde:2001} below. Details of the stochastic methods
involved can be found in appendix~\ref{sec:stochasticprocesses}. Readers
who are not so much interested in detailed proofs may skip directly to
section~\ref{ssec:localityandrestriction}.

\begin{lemma}
 \label{lem:firstphase}
 Let $\kappa\in[0,4]$ and let $\gamma$ be the trace of~\SLE[\kappa]. Then
 almost surely $\gamma\roival{0,\infty}\subset\H\cup\{0\}$.
\end{lemma}

\begin{proof}
 Let $0<a<b$ be real and~$x\in[a,b]$. Define the process~$Y_x(t)$ by
 $Y_x(t):=g_t(x)-\sqrt{\kappa}B_t$, and let~$F(x)$ be the probability that
 $Y_x(t)$ hits~$b$ before it hits~$a$. Let~$T$ denote the first time when
 $Y_x(t)$ hits either of these points, and let $t<T$. The stationarity
 property of {\SLE}, lemma~\ref{lem:slestationarity}, shows that the process
 $Y_{Y_x(t)}(t')$ has the same distribution as the process $Y_x(t+t')$ and is
 independent from $\{Y_x(s):0\leq s\leq t\}$(set $\tau\mapsto t$, $t\mapsto t'$
 and $z\mapsto g_t(x)-\sqrt{\kappa}B_t$ in the lemma, and use time homogeneity
 of Brownian motion). It follows that
 \begin{equation}
  \Exp[1_{\displaystyle \{Y_x(T)=b\}}\mid\mathcal{F}_t] =
   \Exp[1_{\displaystyle \{Y_{Y_x(t)}(T)=b\}}\mid\mathcal{F}_t] = F(Y_x(t)),
 \end{equation}
 where $\mathcal{F}_t$ is the $\sigma$-field generated by $\sqrt{\kappa}B_t$
 up to time~$t$. Thus, if $s<t$ then
 \begin{eqnarray}
  \Exp[F(Y_x(t))\mid\fld{F}_s]
   &=& \Exp\Big[\Exp[1_{\displaystyle \{Y_x(T)=b\}}
   		\mid\mathcal{F}_t] \mathrel{\Big|} \fld{F}_s\Big] \nonumber\\
   &=& \Exp[1_{\displaystyle \{Y_x(T)=b\}}\mid\mathcal{F}_s] = F(Y_x(s))
 \end{eqnarray}
 because $\fld{F}_s\subset\fld{F}_t$, which shows that $F(Y_x(t))$ is a
 martingale.

 It\^o's formula for~$F(Y_x(t))$ (theorem~\ref{the:Ito}) is easily derived
 from the differential equation for~$Y_x(t)$,
 \begin{equation}
  \dif{Y_x} = \inv[2]{Y_x}\,\dif{t}-\sqrt{\kappa}\,\dif{B_t}.
 \end{equation}
 Since the drift term in It\^o's formula for~$F(Y_x(t))$ must be zero at
 $t=0$, we find that $F(x)$ satisfies the differential equation
 \begin{equation}
  \half{\kappa}F''(x)+\inv[2]{x}F'(x)=0.
 \end{equation}
 The boundary conditions obviously are~$F(a)=0$ and~$F(b)=1$. The solution is
 given by
 \begin{equation}
  F(x)=\frac{f(x)-f(a)}{f(b)-f(a)}
 \end{equation}
 where
 \begin{equation}
  f(x) = \left\{
  	\begin{array}{ll}
		x^{(\kappa-4)/\kappa} & \mbox{\ if\ }\kappa\neq4, \\
		\log(x) & \mbox{\ if\ }\kappa=4.
  	\end{array} \right.
 \end{equation}
 One can easily verify that this solution satisfies $F(x)\goesto1$ when
 $a\downto0$ for $\kappa\leq4$ (but not for $\kappa>4$) and arbitrary~$b$.

 Hence, for~$\kappa\leq4$ the process~$Y_x(t)$ is going to reach~$\infty$
 before reaching~$0$. The differential equation for~$Y_x(t)$ shows that
 $Y_x(t)$ changes only slowly when~$Y_x(t)$ is large, and we conclude that
 almost surely $Y_x(t)$ does not escape to infinity in finite time. It is
 also clear that $Y_{x'}(t)\geq Y_x(t)$ if $x'>x$, because under the L\"owner
 evolution the order of points on the real line must be conserved. Therefore,
 almost surely for every $x>0$, $Y_x(t)$ is well-defined for all $t\geq0$,
 and $Y_x(t)\in(0,\infty)$. It follows that almost surely the trace
 $\gamma\roival{0,\infty}$ does not intersect~$(0,\infty)$. In the same way
 it can be proved that the trace does not intersect the negative real line.
\end{proof}

\begin{theorem}
 \label{the:firstphase}
 For all $\kappa\in[0,4]$, the trace~$\gamma$ of \SLE[\kappa] is almost surely
 a simple path.
\end{theorem}

\begin{proof}
\begin{figure}
  \centering\includegraphics[scale=1.2]{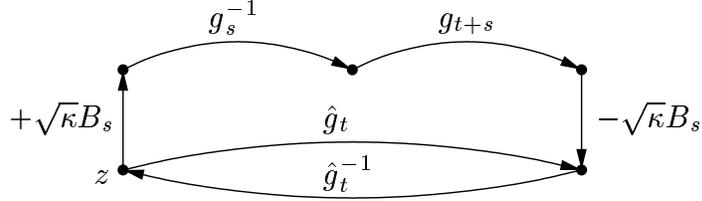}
  \caption{The definition of the map~$\ghat_t(z)$ depicted graphically.}
  \label{fig:Firstphase}
\end{figure}

 Let $t_2>t_1>0$. We need to prove that $\gamma[0,t_1]\cap\gamma
 \roival{t_2,\infty}=\emptyset$. To do so, note that there exists a rational
 $s\in(t_1,t_2)$ such that $\gamma(s)\not\in\R\cup K_{t_1}$, since the
 capacity is strictly increasing between $t_1$ and~$t_2$. In the following
 paragraphs we will prove that
 \begin{equation}
  \label{equ:firstphase}
  \gamma\roival{s,\infty} \cap (\R\cup K_s) = \{\gamma(s)\}.
 \end{equation}
 Suppose for now that this is true, and assume that there is a point~$z$ that
 is both in $\gamma[0,t_1]$ and in $\gamma\roival{t_2,\infty}$. Then clearly
 $z\in\gamma\roival{s,\infty}$ since~$s<t_2$, and $z\in\R\cup K_s$ since
 $s>t_1$. Hence $z=\gamma(s)$ by~(\ref{equ:firstphase}). But then it follows
 that $\gamma(s)=z\in\R\cup K_{t_1}$, a contradiction. This proves the theorem,
 so it only remains to establish~(\ref{equ:firstphase}).
 
 To prove~(\ref{equ:firstphase}), for fixed~$s$ as above consider the
 process~$\ghat_t(z)$ defined by
 \begin{equation}
  \ghat_t(z) := g_{t+s}\big(g_s^{-1}(z+\sqrt{\kappa}B_s)\big)
   - \sqrt{\kappa}B_s.
 \end{equation}
 By stationarity of {\SLE} (lemma~\ref{lem:slestationarity}), this process
 has the same distribution as~$g_t(z)$; we saw in the derivation of the
 stationarity property that its driving process is $\sqrt{\kappa}
 (B_{t+s}-B_s)$. Let us denote by~$\gammahat_s(t)$ the trace corresponding
 to the maps $\{\ghat_t:t\geq 0\}$. Then we have
 \begin{eqnarray}
  \gammahat_s(t) &:=& \ghat^{-1}_t\big(\sqrt{\kappa}(B_{t+s}-B_s)\big) =
   g_s\big(g^{-1}_{t+s}(\sqrt{\kappa}B_{t+s})\big) - \sqrt{\kappa}B_s
   \nonumber\\
   &=& g_s\big(\gamma(t+s)\big) - \sqrt{\kappa}B_s
 \end{eqnarray}
 as can be seen from figure~\ref{fig:Firstphase}. Applying the map~$g^{-1}_s$
 to this result gives
 \begin{equation}
  \gamma(t+s) = g^{-1}_s\big( \gammahat_s(t)+\sqrt{\kappa}B_s \big).
 \end{equation}
 Now, lemma~\ref{lem:firstphase} tells us that for every~$t\geq0$,
 $\gammahat_s(t)\in\H\cup\{0\}$. Hence, because~$g^{-1}_s$ maps~$\H$ onto
 $H_s=\H\setminus K_s$, (\ref{equ:firstphase}) follows. The proof is now
 complete.
\end{proof}

\subsection{Locality and restriction}
\label{ssec:localityandrestriction}

We discussed above the two special values of~$\kappa$ where {\SLE}
undergoes a phase transition. Two other special values of~$\kappa$ are
$\kappa=6$ and~$\kappa=8/3$. At these values, \SLE[\kappa] has some very
specific properties, that will be discussed in detail below.

\subsubsection[The locality property of SLE(6)]{The locality property of \SLE[6]}

\begin{figure}
 \centering\includegraphics[scale=1.2]{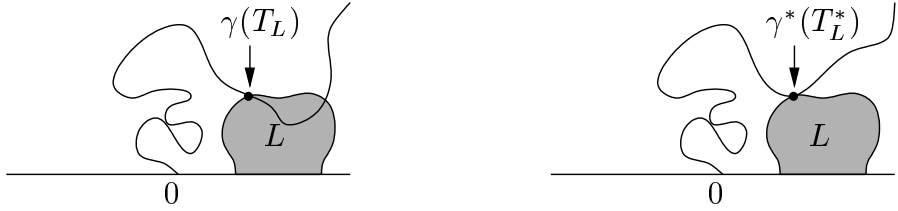}
 \caption{Comparison of two \SLE[\kappa] processes from $0$ to~$\infty$,
  in the domain~$\H$ (left) and in the domain~$\H\setminus L$ (right). If
  these processes have the same distribution up to the hitting time of the
  set~$L$, then we say that \SLE[\kappa] has the locality property.}
 \label{fig:Locality}
\end{figure}

Let us start by giving a precise definition of the locality property. Assume
for now that~$\kappa>0$ is fixed. Suppose that~$L$ is a hull in~$\H$ which is
bounded away from the origin. Let~$K_t$ be the hulls of a chordal \SLE[\kappa]
process in~$\H$, and let~$K^\ast_t$ be the hulls of a chordal \SLE[\kappa]
process in~$\H\setminus L$, both processes going from $0$ to~$\infty$. Denote
by~$T_L$ the first time at which $K_t$ intersects the set~$L$. Likewise, let
$T^\ast_L$ be the first time when~$K^\ast_t$ intersects $L$ (note that in
this case, $T^\ast_L$ is the hitting time of an arc on the boundary of
the domain). See figure~\ref{fig:Locality} for an illustration comparing the
traces of the two processes in their respective domains.

Chordal \SLE[\kappa] is said to satisfy the locality property if for all
hulls~$L$ bounded away from the origin, the distribution of the
hulls~$\{K_t:t<T_L\}$ is the same as the distribution of the hulls
$\{K^\ast_t:t<T^\ast_L\}$, modulo a time re-parameterization. Loosely
speaking, suppose that \SLE[\kappa] has the locality property, and that we
are only interested in the process up to the first time when it hits~$L$.
Then it doesn't matter whether we consider chordal \SLE[\kappa] from $0$
to~$\infty$ in the domain~$\H$, or chordal \SLE[\kappa] from $0$ to~$\infty$
in the smaller domain~$\H\setminus L$: up to a time-change, these processes
are the same. It was first  proved in~\cite{lsw:I} that chordal \SLE[\kappa]
has the locality property for~$\kappa=6$, and for no other values of~$\kappa$.
Later, a much simpler proof appeared in~\cite{lsw:2003}. A sketch of
the proof with a discussion of some consequences appears
in~\cite{lawler:url2001}.

So far, we defined the locality property for a chordal process in~$\H$, but
it is clear that by conformal invariance we can translate the property to
an arbitrary simply connected domain. It is also true that radial \SLE[6]
has the same property. We shall not go into this further, but we would like
to point out one particular consequence of the locality property of \SLE[6].

\begin{figure}
 \centering\includegraphics[scale=1.2]{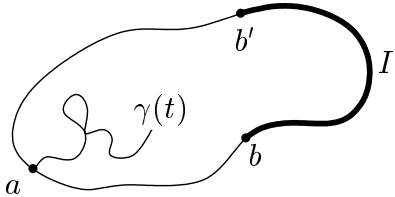}
 \caption{An \SLE[\kappa] process aimed towards an arc~$I$ on the boundary
  of a domain.}
 \label{fig:Splitting}
\end{figure}

Suppose that~$D$ is a simply connected domain with continuous boundary,
and let $a$, $b$ and~$b'$ be three distinct points on the boundary of~$D$.
Denote by~$I$ the arc of~$\partial D$ between $b$ and~$b'$ which does not
contain~$a$ (see figure~\ref{fig:Splitting} for an illustration). Let $K_t$
(respectively $K_t'$) be the hulls of a chordal \SLE[6] process from $a$
to~$b$ (respectively $b'$) in~$D$, and let~$T$ (respectively $T'$) be the
first time when the process hits~$I$. Then modulo a time-change, $\{K_t:
t<T\}$ and $\{K'_t:t<T'\}$ have the same distribution. As a consequence, if
you are interested in the behaviour of an \SLE[6] process up to the first
time when it hits an arc~$I$, then you may choose any point of~$I$ as the
endpoint for the {\SLE} process without affecting its behaviour.

\subsubsection[The restriction property of SLE(8/3)]{The restriction property of \SLE[8/3]}

To define the restriction property, assume that~$\kappa\leq4$ is fixed. Then
the trace~$\gamma$ of \SLE[\kappa] is a simple path. Now suppose, as in our
discussion of the locality property above, that~$L$ is a hull in the
half-plane which is bounded away from the origin. Let~$\Psi$ be the map
defined by $\Psi(z):=g_L(z)-g_L(0)$. Then $\Psi$ is the unique conformal
map of $\H\setminus L$ onto~$\H$ such that $\Psi(0)=0$, $\Psi(\infty)=\infty$
and $\Psi'(\infty)=1$. Now suppose that $\gamma$ never hits~$L$. Then we
let~$\gamma^\ast$ be the image of $\gamma$ under the map~$\Psi$, that is
$\gamma^\ast(t):=\Psi(\gamma(t))$.

We say that \SLE[\kappa] has the restriction property if for all hulls~$L$
that are bounded away from the origin, conditional on the event
$\{\gamma\roival{0,\infty}\cap L=\emptyset\}$, the distribution of
$\gamma^\ast\roival{0,\infty}$ is the same as the distribution of the trace
of a chordal \SLE[\kappa] process in~$\H$, modulo a time re-parameterization.
In words, suppose that \SLE[\kappa] has the restriction property. Then the
distribution of all paths that are restricted not to hit~$L$, and which are
generated by \SLE[\kappa] in the half-plane, is the same as the distribution
of all paths generated by \SLE[\kappa] in the domain~$\H\setminus L$.

{\SLE} has the restriction property for $\kappa=8/3$ and for no other values
of~$\kappa$. A proof is given in~\cite{lsw:2003} (a sketch of a proof
appears in~\cite{lawler:url2001}), and in the same article it was also shown
that
\begin{equation}
 \Prob\big[\gamma\roival{0,\infty}\cap L=\emptyset\big]
 = |\Psi'(0)|^{5/8}.
\end{equation}
Again, the restriction property can be translated into a similar property
for arbitrary domains, and radial \SLE[8/3] also satisfies the restriction
property. We refer to Lawler, Schramm and Werner~\cite{lsw:2003} and
Lawler~\cite{lawler:url2001} for more information.

\subsection{Hausdorff dimensions}
\label{ssec:dimensions}

Consider an \SLE[\kappa] process in the upper half-plane. If $\kappa\geq8$
the trace of the process is space-filling, and therefore the Hausdorff
dimension of the set~$\gamma\roival{0,\infty}$ is~$2$. But for
$\kappa\in(0,8)$ the Hausdorff dimension of~$\gamma\roival{0,\infty}$ is a
non-trivial number. Rohde and Schramm~\cite{rohde:2001} showed that its
value is bounded from above by $1+\kappa/8$, and the proof that for
$\kappa\neq4$ the Hausdorff dimension is in fact $1+\kappa/8$ was completed
by Beffara~\cite{beffara:pre0204208,beffara:pre0211322}. In the physics
literature the Hausdorff dimensions of the curves that are believed to
converge to {\SLE} were predicted by Duplantier and
Saleur~\cite{duplantier:2000,saleur:1987}.

In the case~$\kappa>4$ the hull of \SLE[\kappa] is not a simple path, and
it is natural to consider also the Hausdorff dimension of the boundary
of~$K_t$ for some fixed value of~$t>0$. Its value is conjectured to
be~$1+2/\kappa$, because (based on a duality relation derived by
Duplantier~\cite{duplantier:2000}) it is believed that the boundary of the
hull for~$\kappa>4$ is described by \SLE[16/\kappa]. The dimension of the
hull boundary is known rigorously only for $\kappa=6$ (where it is~$4/3$)
and for $\kappa=8$ (where it is~$5/4$). For $\kappa=6$ this follows from
the study of the ``conformal restriction measures'' in~\cite{lsw:2003},
for $\kappa=8$ this is a consequence of the strong relation between
loop-erased random walks and uniform spanning trees~\cite{lsw:pre0112234}
(section~\ref{ssec:LERWandUST}).


\section[SLE and discrete models]{{\SLE} and discrete models}
\label{sec:discretemodels}

In this section we take a look at the connection between {\SLE} and discrete
models. The connection is made by defining a path in these discrete models,
which in the scaling limit converges to the trace of a chordal or radial
{\SLE} process. In the first subsection, we describe how this works for the
exploration process of critical percolation, which is known to converge to
\SLE[6]. Then we describe the harmonic explorer and its convergence to
\SLE[4]. In section~\ref{ssec:LERWandUST} we consider the loop-erased random
walk and the Peano curve associated with the uniform spanning tree. These
paths converge to the traces of \SLE[2] and~\SLE[8] respectively.
Section~\ref{ssec:SAW} is about the conjectured connection between
self-avoiding walks and \SLE[8/3]. The final two subsections relate Potts
models and O($n$)-model to their {\SLE} counterparts.

\subsection{Critical percolation}
\label{ssec:percolation}

We define site percolation on the triangular lattice as follows. All vertices
of the lattice are independently coloured blue with probability~$p$ or yellow
with probability~$1-p$. An equivalent, and perhaps more convenient, viewpoint
is to say that we colour all hexagons of the dual lattice blue or yellow with
probabilities~$p$ and~$1-p$, respectively. It is well-known that
for~$p\leq1/2$, there is almost surely no infinite cluster of connected blue
hexagons, while for~$p>1/2$ there a.s.\ exist a unique infinite blue cluster.
This makes $p=1/2$ the critical point for site percolation on the triangular
lattice. For the remainder of this subsection we assume that we are at this
critical point.

\begin{figure}
 \centering\includegraphics[scale=1.2]{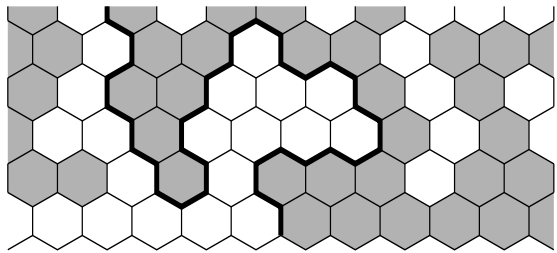}
 \caption{Part of the percolation exploration process in the half-plane.}
 \label{fig:Exploration}
\end{figure}

Let us for now restrict ourselves to the half-plane. Suppose that as our
boundary conditions, we colour all hexagons intersecting the negative real
line yellow, and all hexagons intersecting the positive real line blue. All
other hexagons in the half-plane are independently coloured blue or yellow
with equal probabilities. Then there exists a unique path over the edges of
the hexagons, starting in the origin, which separates the cluster of blue
hexagons attached to the positive real half-line from the cluster of yellow
hexagons attached to the negative real half-line. This path is called the
chordal exploration process from~$0$ to~$\infty$ in the half-plane. See
figure~\ref{fig:Exploration} for an illustration.

The exploration process can also be described as the unique path from the
origin such that at each step there is a blue hexagon on the right, and a
yellow hexagon on the left. This path can also be generated dynamically, as
follows. Initially, only the hexagons on the boundary receive a colour. Then
after each step, the exploration process meets a hexagon. If this hexagon
has not yet been coloured, we have to choose whether to make it blue or
yellow, and the exploration process can turn left or right with equal
probabilities. But if the hexagon has already been coloured blue or yellow,
the exploration path is forced to turn left or right, respectively.

Note that in this dynamic formulation it is clear that the trajectory of the
exploration process is determined completely by the colours of the hexagons
in the direct vicinity of the path. Further, it is clear that the tip of
the process can not become trapped, because it is forced to reflect off into
the open if it meets an already coloured hexagon. This suggests that in the
continuum limit, when we send the size of the hexagons to zero, the
exploration process may be described by a L\"owner evolution. The only
candidate is~\SLE[6], because only then we have the locality property.

Smirnov~\cite{smirnov:2001b} proved that in the continuum
limit, the exploration process is conformally invariant. Together with the
results on~\SLE[6] developed by Lawler, Schramm and Werner, this should
prove that the exploration process converges to the trace of~\SLE[6] in
the half-plane (although explicit proofs linking \SLE[6] to critical
percolation have not yet appeared). Thus, \SLE[6] may be used to calculate
properties of critical percolation that can be formulated in terms of the
behaviour of the exploration process. Some examples of how this can be done
are described in section~\ref{sec:results}.

So far, we have restricted percolation to a half-plane, but we can of
course consider other domains as well. For example, let~$D$ be a
simply connected domain with continuous boundary, and let $a$ and~$b$ be
two points on the boundary. In an approximation of the domain by hexagons,
colour all hexagons that intersect the arc of~$\partial D$ from $a$ to~$b$
in the counter-clockwise direction blue, and all remaining hexagons
intersecting~$\partial D$ yellow. Then there is a unique exploration process
in~$D$ which goes from $a$ to~$b$, and by Smirnov's result it converges in
the scaling limit to a chordal \SLE[6] trace in~$D$ going from $a$ to~$b$.
On that note we end our discussion of the connection between critical
percolation and \SLE[6].

\subsection{The harmonic explorer}
\label{ssec:harmonicexplorer}

\begin{figure}
 \centering\includegraphics[scale=1.2]{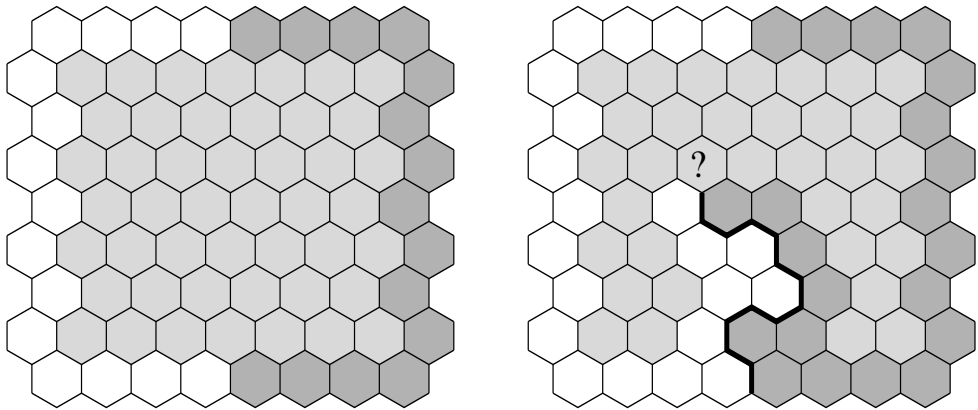}
 \caption{Left: the initial configuration for the harmonic explorer, with
  blue hexagons (dark faces), yellow hexagons (white faces) and uncoloured
  hexagons (light faces). Right: a part of the harmonic explorer process.
  The colour of the marked hexagon is determined as described in the text.}
 \label{fig:Explorer}
\end{figure}

The harmonic explorer is a random path similar to the exploration process
of critical percolation. It was defined recently by Schramm and Sheffield
as a discrete process that converges to \SLE[4]~\cite{schramm:2003}. To
define the harmonic explorer, consider an approximation of a bounded domain
with hexagons, as in figure~\ref{fig:Explorer}. As we did for critical
percolation, we partition the set of hexagons on the boundary of our domain
into two components, and colour the one component yellow and the other blue.
The hexagons in the interior are uncoloured initially.

The harmonic explorer is a path over the edges of the hexagons that starts
out on the boundary with a blue hexagon on its right and a yellow hexagon
on its left. It turns left when it meets a blue hexagon, and it turns right
when it meets a yellow hexagon. The only difference with the exploration
process of critical percolation is in the way the colour of an as yet
uncoloured hexagon is determined. For the harmonic explorer this is done
as follows.

Suppose that the harmonic explorer meets an uncoloured hexagon (see
figure~\ref{fig:Explorer}). Let~$f$ be the function, defined on the faces of
the hexagons, that takes the value~$1$ on the blue hexagons, the value~$0$
on the yellow hexagons, and is discrete harmonic on the uncoloured hexagons.
Then the probability that the hexagon whose colour we want to determine is
made blue, is given by the value of~$f$ on this hexagon. Proceeding in this
way, we obtain a path crossing the domain between the two points on the
boundary where the blue and yellow hexagons meet. In the scaling limit this
path converges to the trace of chordal \SLE[4].

\subsection{Loop-erased random walks and uniform spanning trees}
\label{ssec:LERWandUST}

In this subsection we consider loop-erased random walks (LERW's) and
uniform spanning trees (UST's). We shall define both models first, and we
will point out the close relation between the two. Then we will discuss
the connection with {\SLE} in the scaling limit. Schramm~\cite{schramm:2000}
already proved that the LERW converges to \SLE[2] under the assumption that
the scaling limit exists and is conformally invariant. In the same work, he
also conjectured the relation between UST's and \SLE[8]. The final proofs of
these connections were given by Lawler, Schramm and Werner
in~\cite{lsw:pre0112234}. Their proofs hold for general lattices, but for
simplicity, we shall restrict our description here to finite subgraphs
of the square grid~$\delta\Z[2]$ with mesh~$\delta>0$.

Suppose that~$G$ is a finite connected subgraph of~$\delta\Z[2]$, let~$u$
be a vertex of~$G$ and let~$V$ be a collection of vertices of~$G$ not
containing~$u$. Then the LERW from~$u$ to~$V$ in~$G$ is defined by taking 
a simple random walk in~$G$ from~$u$ to~$V$ and erasing all its loops in
chronological order. More precisely, if $\big(\omega(0),\ldots,\omega(T_V)
\big)$ are the vertices visited by a simple random walk starting from~$u$
and stopped at the first time~$T_V$ when it visits a vertex in~$V$, then
its loop-erasure $\big(\beta(0),\ldots,\beta(T)\big)$ is defined as follows.
We start by setting $\beta(0)=\omega(0)$. Then for $n\in\N$ we define
inductively: if $\beta(n)\in V$ then $T=n$ and we are done, and otherwise
we set $\beta(n+1)=\omega(1+\max\{m\leq T_V:\omega(m)=\beta(n)\})$. The path
$\big(\beta(0),\ldots,\beta(T)\big)$ is then a sample of the LERW in~$G$
from~$u$ to~$V$.

A spanning tree~$T$ in~$G$ is a subgraph of~$G$ such that every two vertices
of~$G$ are connected via a unique simple path in~$T$. A \emph{uniform}
spanning tree (UST) in~$G$ is a spanning tree chosen with the uniform
distribution from all spanning trees in~$G$. It is well-known that the
distribution of the unique simple path connecting two distinct vertices
$u$ and~$v$ of~$G$ in the UST is the same as that of the LERW from $u$
to~$\{v\}$ in~$G$.

In fact, the connection between LERW's and UST's is even stronger. For
suppose that we fix an ordering $(v_0,\ldots,v_n)$ of the vertices in~$G$.
Let $T_0=\{v_0\}$ and inductively define $T_{m+1}$ as the union of $T_m$
and a LERW from $v_{m+1}$ to~$T_m$, $T_{m+1}=T_m$ if $v_{m+1}\in T_m$.
Then~$T_n$ is a UST in~$G$, regardless of the chosen ordering of the
vertices of~$G$. This algorithm for generating UST's from LERW's is known
as Wilson's algorithm~\cite{wilson:1996}. See also~\cite{lsw:pre0112234,
schramm:2000} and references therein for more information.

\begin{figure}[t!]
 \centering\includegraphics[scale=1.2]{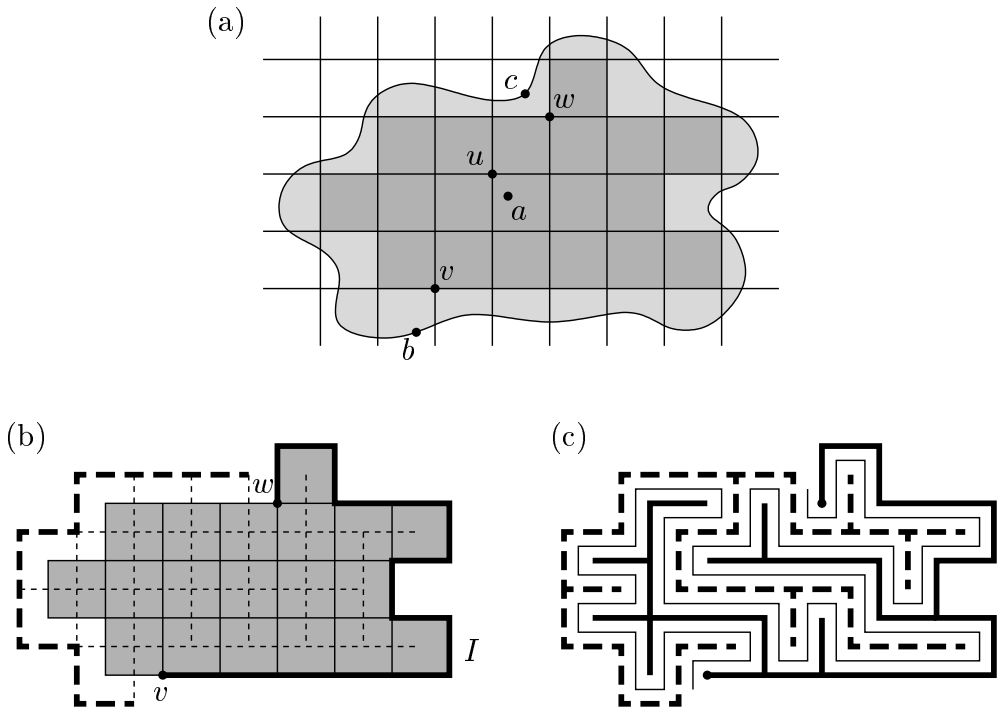}
 \caption{Part~(a) shows the discrete approximation (dark shaded region) of a
  domain. In part~(b), the discrete approximation is shown again, with the
  graph~$G$ in solid lines and the dual graph~$G^\dagger$ in dashed lines. The
  thick lines connect vertices that are identified. In part~(c) we see a
  spanning tree on~$G$ and its dual on~$G^\dagger$ (thick lines), and the
  Peano curve winding between them (thin line).}
 \label{fig:Peano}
\end{figure}

Let us now describe the scaling limit of the LERW and the connection with
{\SLE}. We shall work with a fixed, bounded, simply connected domain~$D$.
Fix the mesh~$\delta>0$, and let~$G$ be the subgraph of~$\delta\Z[2]$
consisting of all vertices and edges that are contained in~$D$. Then the
set of all points that are disconnected from~$\infty$ by~$G$ is a discrete
approximation~$D'$ of the domain~$D$, see figure~\ref{fig:Peano}, part~(a).
Suppose that~$a$ is a fixed interior point of~$D$ and let~$u$ be the vertex
of~$G$ which is closest to~$a$. Consider the LERW on~$\delta\Z[2]$ from~$u$
to the set of vertices that are not in~$G$. In the scaling limit, the
time-reversal of this LERW converges to the trace of a radial \SLE[2] process
in~$D$ from~$\partial D$ to~$a$. Here, the starting point of the \SLE[2]
process is defined by choosing the starting point of the Brownian motion
driving the L\"owner evolution on the unit disk uniformly on the unit
circle.

The fact that the LERW converges to an \SLE[2] process proves that the
LERW is conformally invariant in the scaling limit. Because of the close
connection between LERW's and UST's, this leads to the conclusion that
the UST has a conformally invariant scaling limit as well. Moreover, we
can define a path associated to the UST, that converges in the scaling
limit to the trace of \SLE[8]. This path is called the UST Peano curve,
and can be defined as we describe below (figure~\ref{fig:Peano} provides
an illustration).

Consider again the domains $D$, $D'$ and graph~$G$ as before. This time,
let $b$ and~$c$ be distinct points of~$\partial D$, and let $v$ and~$w$ be
distinct vertices of~$G$ on~$\partial D'$ closest to $b$ and~$c$,
respectively. We denote by~$I$ the counter-clockwise arc from~$v$ to~$w$
of~$\partial D'$, and identify all vertices of~$G$ that are on~$I$. Now
let~$G'$ be the graph consisting of all edges (and corresponding vertices)
of the lattice dual to~$\delta\Z[2]$, that intersect edges of~$G$ but
not~$I$. Then we define the dual graph~$G^\dagger$ of~$G$ as the union
of~$G'$ and those edges (and corresponding vertices) outside~$D'$ needed
to connect the vertices of~$G'$ outside~$D'$ via the shortest possible
path outside~$D'$, see figure~\ref{fig:Peano}, part~(b). On this dual
graph, we identify all vertices that lie outside~$D'$.

Now suppose that~$T$ is a UST in~$G$. Then there is a dual tree~$T^\dagger$
in~$G^\dagger$, consisting of all those edges that do not intersect edges
of the tree~$T$, see figure~\ref{fig:Peano}, part~(c). Observe that~$T^\dagger$
is a UST in~$G^\dagger$. The Peano curve is defined as the curve winding
between $T$ and~$T^\dagger$ on the square lattice with vertices at the points
$\frac{\delta}{2}\Z[2]+(\frac{\delta}{4},\frac{\delta}{4})$. Note that this
curve is space-filling, in that it visits all vertices of the lattice that
are disconnected from~$\infty$ by~$G\cup G^\dagger$. In the scaling limit,
the Peano curve defined as above converges to the trace of a chordal~\SLE[8]
process from~$b$ to~$c$ in~$D$.

\subsection{Self-avoiding walks}
\label{ssec:SAW}

A self-avoiding walk (SAW) of length~$n$ on the square lattice~$\delta\Z[2]$
with mesh~$\delta>0$ is a nearest-neighbour path $\omega=\big(\omega(0),
\omega(1),\ldots,\omega(n)\big)$ on the vertices of the lattice, such that
no vertex is visited more than once. In this subsection we shall restrict
ourselves to SAW's that start in the origin and stay in the upper half-plane
afterwards. The idea is to define a stochastic process, called the half-plane
infinite SAW, that in the scaling limit $\delta\downto0$ is believed to
converge to chordal~\SLE[8/3].

Following~\cite{lsw:pre0204277} we write $\Lambda_n^+$ for the set of all
SAW's~$\omega$ of length~$n$ that start in the origin, and stay above the
real line afterwards. For a given~$\omega$ in~$\Lambda_n^+$, let
$Q_k^+(\omega)$ be the fraction of walks $\omega'$ in~$\Lambda_{n+k}^+$
whose beginning is~$\omega$, i.e.\ such that $\omega'(i)=\omega(i)$ for
$0\leq i\leq n$. Define $Q^+(\omega)$ as the limit of $Q_k^+(\omega)$ as
$k\goesto\infty$. Then $Q^+(\omega)$ is roughly the fraction of very long
SAW's in the upper half-plane whose beginning is~$\omega$. It was shown by
Lawler, Schramm and Werner that the limit~$Q^+(\omega)$
exists~\cite{lsw:pre0204277}.

Now we can define the \emph{half-plane infinite self-avoiding walk} as the
stochastic process~$X_i$ such that for all $\omega=\big(0,\omega(1),\ldots,
\omega(n)\big)\in\Lambda^+_n$,
\begin{equation}
 \Prob[X_0=0, X_1=\omega(1), \ldots, X_n=\omega(n)] = Q^+(\omega).
\end{equation}
We believe that the scaling limit of this process as the mesh~$\delta$ tends
to~$0$ exists and is conformally invariant. By the restriction property the
scaling limit has to be \SLE[8/3], as pointed out in~\cite{lsw:pre0204277}.
At this moment it is unknown how the existence, let alone the conformal
invariance, of the scaling limit could be proved. However, there is very
strong numerical evidence for the conformal invariance of the scaling limit
of self-avoiding walks \cite{kennedy:2001,kennedy:2002}, confirming the
{\SLE} predictions of its restriction property.

Lawler, Schramm and Werner~\cite{lsw:pre0204277} also explain how one can
define a natural measure on SAW's with arbitrary starting points, leading
to conjectures relating SAW's to chordal and radial \SLE[8/3] in bounded
simply-connected domains. The article further discusses similar conjectures
for self-avoiding polygons, and predictions for the critical exponents of
SAW's that can be obtained from~{\SLE}. We shall not go into these topics
here.

\subsection{The Potts model}
\label{ssec:PottsModel}

So far in this section we discussed relations between {\SLE} at specific
values of~$\kappa$ to certain statistical lattice models. The results of
{\SLE} however suggest a further connection to continuous families of models,
of which we will discuss the two most obvious examples in this and the
following subsection. This subsection deals with the $q$-state Potts model.
Below we will show a standard treatment~\cite{baxter:1976}, which relates the
partition sum of the Potts model to an ensemble of multiple paths on the
lattice. In the scaling limit these paths will be the candidates for the
{\SLE} processes. The second example, allowing a similar treatment, is the
O($n$) model. This model will be discussed in the following subsection.

The Potts model has on each site of a lattice a variable $s_j$ which can
take values in $\{1,2,\ldots,q\}$. Of these variables only nearest neighbours
interact such that the energy is~$-1$ if both variables are in the same
state and~$0$ otherwise. The canonical partition sum is
\begin{equation}
Z = \sum_{\{s\}} \exp\left(\beta \sum_{\langle j,k \rangle}
	\delta_{s_j,s_k} \right).
\end{equation}
The summation in the exponent is over all nearest-neighbour pairs of
sites, and the external summation over all configurations of the~$s_j$.
The model is known to be disordered at high temperatures, and ordered at
low temperatures. One of the signatures of order is that the probability
that two distant $s$-variables are in the same state does not decay to
zero with increasing distance. We are interested in the behaviour at the
transition.

\begin{figure}
 \centering\includegraphics[scale=1.2]{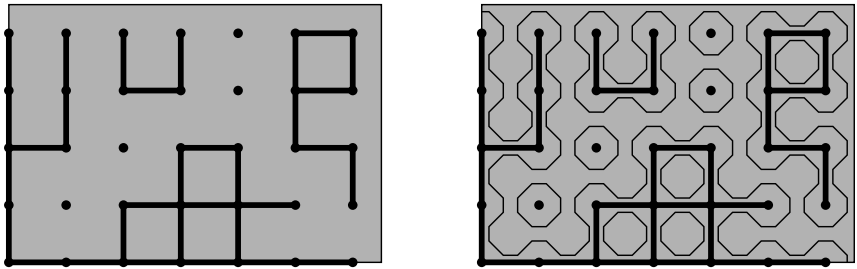}
 \caption{The Potts model in a rectangular domain. On the left we illustrate
  the graph decomposition, on the right we have drawn in the corresponding
  configuration of paths.}
 \label{fig:Potts}
\end{figure}

In order to make the connection with a path on the lattice, we express this
partition sum in a high-temperature expansion, i.e.\ in powers of a parameter
which is small when~$\beta$ is small. The first step is to write the summand
as a product:
\begin{equation} 
 Z = \sum_{\{s\}} \prod_{\langle j,k \rangle} 
	 \left[ 1 + (\e{\beta}-1)\delta_{s_j,s_k}\right].
\end{equation}
The product can be expanded in terms in which at every edge of the
lattice a choice is made between the two terms $1$ and $(\e{\beta}-1)
\delta_{s_j,s_k}$. In a graphical notation we place a bond on every edge
of the lattice where the second term is chosen, see figure~\ref{fig:Potts}.
For each term in the expansion of the product the summation over the
$s$-variables is trivial: if two sites are connected by bonds, their
respective $s$-variables take the same value, and are independent otherwise.
As a result the summation over $\{s\}$ results in a factor~$q$ for each
connected component of the graph. Hence
\begin{equation}
 Z = \sum_{{\rm graphs}} (\e{\beta}-1)^b q^c,
\end{equation}
where~$c$ is the number of connected components of the graph and~$b$ the
number of bonds. This expansion is known by the name of
Fortuin-Kasteleyn~\cite{fortuin:1972} cluster model. Note that, while~$q$ has
been introduced as the (integer) number of states, in this expansion it
can take any value.

It is convenient to rewrite the graph expansion into an expansion of paths
on a new lattice. The edges of the original lattice correspond to the vertices
of the new lattice. The graphs on the original lattice are rewritten into
polygon decompositions of the new lattice. Every vertex of the new lattice
is separated into two non-intersecting path segments. These path segments
intersect the corresponding edge of the original lattice if and only if this
edge does not carry a bond of the graph, as follows:
\begin{center}\includegraphics{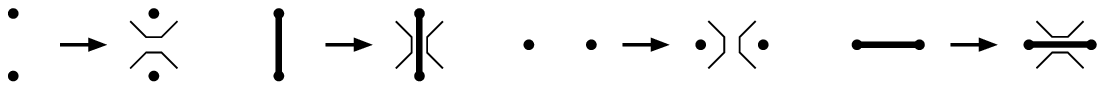}\end{center}
As a result of these transformations the new lattice is decomposed into a
collection of non-intersecting paths, as indicated in figure~\ref{fig:Potts}.
Notice that every component of the original graph is surrounded by one of these
closed paths, but also the closed circuits of the graph are inscribed by
these paths. By Euler's relation the number of components~$c$ of the original
graph can be expressed in the number of bonds~$b$, the total number of
sites~$N$ and the number of polygons~$p$: $c = (N-b+p)/2$. An alternative
expression for the partition sum is then
\begin{equation}
 Z = \sum_{{\rm graphs}} \left(\frac{\e{\beta}-1}{\sqrt{q}}\right)^b
 	 q^{(N+p)/2}.
 \label{equ:loops}
\end{equation}
At the critical point~$\beta_c$ the relation $\exp(\beta_c)=1+\sqrt{q}$ holds,
so that the partition sum simplifies. 

We will now consider this model at the critical point on a rectangular
domain. The lattice approximation of this domain is chosen such that the
lower-left corner of the rectangle coincides with a site of the lattice,
while the upper-right corner coincides with a site of the dual lattice.
The sides of the rectangle are parallel to the edges of the lattice,
as in figure~\ref{fig:Potts}. We choose as boundary condition that all
edges that are contained in the left and lower sides of the rectangle
carry bonds, and all edges that intersect the right and upper sides
perpendicularly carry no bonds. For the spin variables this means that
all the spins on the left and lower sides are in the same state, while
all other spins are unconstrained. 

In such an arrangement the diagrams in~(\ref{equ:loops}) include one path
from the lower-right to the upper-left corner. All further paths are closed
polygons, see figure~\ref{fig:Potts}. We take the scaling limit by covering
the same domain with a finer and finer mesh. It is believed~\cite{rohde:2001}
that in the scaling limit the measure on the paths approaches that of chordal
\SLE[\kappa] traces. From e.g.\ the Hausdorff
dimension~\cite{beffara:pre0211322,saleur:1987} the relation between $\kappa$
and~$q$ is
\begin{equation}
 q = 2 + 2 \cos (8\pi/\kappa)
\end{equation}
where $4\leq\kappa\leq8$. Only in a few cases this relation between
\SLE[\kappa] and the Potts partition sum has been made rigorous. For
instance, in the limit $q\to0$, the graph expansion reduces to the
uniform spanning tree, which has \SLE[8] as its scaling limit.

\subsection[The O(n) model]{The O($n$) model}
\label{ssec:OnModel}

We now turn to the O($n$) model, which is another well-known model where a
high-temperature expansion results in a sum over paths. Here the dynamic
variables are  $n$-component vectors of a fixed length, and the Hamiltonian
is invariant under rotations in the $n$-dimensional space. The simplest
high-temperature expansion is obtained when the Boltzmann weight is chosen
as
\begin{equation}
 \prod_{\langle j,k \rangle} ( 1 + x s_i \cdot s_j ),
 \label{equ:On}
\end{equation}
where the product is over nearest neighbours on a hexagonal lattice. The
partition sum is obtained by integrating this expression over the directions
of the spin vectors.  Like for the Potts model one can expand the product
and do the bookkeeping of the terms by means of graphs. In each factor
in~(\ref{equ:On}) the choice of the second term is indicated by a bond. Then
the graphs that survive the integration over the spin variables have only
even vertices, i.e.\ on the hexagonal lattice vertices with zero or two bonds.
As a result the graphs consist of paths on the lattice. In a well-chosen
normalization of the measure and the length of the spins, the partition sum
is a sum over even graphs 
\begin{equation}
 Z = \sum_{\rm graphs} x^L n^M,
 \label{equ:ZOn}
\end{equation}
where~$M$ is the number of closed loops, and~$L$ their combined length.
Note that this expression for the partition sum is well-defined also when
the number of spin components~$n$ is not integer. It is
known~\cite{baxter:1986,nienhuis:1982} that the critical point is at
$x_c = [2+(2-n)^{1/2}]^{-1/2}$ for $0\leq n\leq2$. When~$x$ is larger
that this critical value, the model also shows critical behaviour.

Consider now this model on a bounded domain, and take a correlation function
between two spins on the boundary. The diagrams that contribute to this
function contain one path between the two specified boundary points and any
number of closed polygons in the interior. We conjecture that at the critical
value of~$x$ in the scaling limit the measure on the paths between the two
boundary spins approaches that of chordal \SLE[\kappa] for $n = -2
\cos(4\pi/\kappa)$ and $8/3\leq\kappa\leq4$. For larger values of~$x$, the
scaling limit would again be \SLE[\kappa], with the same relation between
$\kappa$ and~$n$, but now with $4\leq\kappa\leq8$. 

To conclude this section, we remark that the same partition sum~(\ref{equ:ZOn})
can also be viewed as the partition sum of a dilute Potts model on the
triangular lattice, described in~\cite{nienhuis:1991}. In this variant of the
Potts model the spins take values in $\{0,1,2,\ldots,q\}$. The model is
symmetric under permutations of the~$q$ positive values. The name dilute
comes from the interpretation of the neutral value~$0$ as a vacant site.
If neighbouring sites take different values, then one of them takes the
value~$0$. The Boltzmann weight is a product over the elementary triangles
of weights that depend on the three sites at the corners of the triangle. We
take this weight to be~$1$ when all three sites are in the same state, vacant
or otherwise. Triangles with one or two vacant sites have weights $xy$
and~$x/y$, respectively. The partition sum can be expanded in terms of
domain walls between sites of different values. This expansion takes the
form of~(\ref{equ:ZOn}) for $y^{12} = q = n^2$, which is the locus of the
phase transition between an ordered phase and a disordered phase. Within this
locus, the region with $x > x_c$ is a second-order transition. In the regime
$ x < x_c$ the transition is discontinuous, and the position $x=x_c$ separates
the two regimes and is called the tricritical point. When $q=x=1$ the site
percolation problem on the triangular lattice is recovered, which is known
to converge to \SLE[6] in the scaling limit.


\section[SLE computations and results]{{\SLE} computations and results}
\label{sec:results}

In this section we discuss some of the results that have been obtained
from calculations involving {\SLE} processes. Our aim in this section is
not only to provide an overview of these results, but also to give an
impression of the typical {\SLE} computations involved, using techniques
from stochastic calculus and conformal mapping theory.

This section is organized as follows. In the first subsection we discuss
several~{\SLE} calculations independently from their connection with other
models. The results we obtain will be key ingredients for further
calculations. The second subsection gives a brief overview of how {\SLE}
can be applied to calculate the intersection exponents of Brownian motion.
Finally, we will discuss results on critical percolation that have been
obtained from its connection with~\SLE[6].

\subsection[Several SLE calculations]{Several {\SLE} calculations}
\label{ssec:SLEcalculations}

The purpose of this subsection is to explain how some typical probabilities
and corresponding exponents of events involving chordal {\SLE} processes can
be calculated. The results we find in this subsection are for whole ranges
of~$\kappa$, and might therefore have applications in various statistical
models. Some typical applications of the results for $\kappa=6$ will be shown
in the following subsections.

\subsubsection{The one-sided crossing exponent}

Consider a chordal \SLE[\kappa] process inside the rectangle $\mathcal{R}_L
:=(0,L)\times(0,\im\pi)$, which goes from $\im\pi$ to~$L$. If~$\kappa>4$ this
process will at some random time~$\tau$ hit the right edge $[L,L+\im\pi]$
of the rectangle, as in figure ~\ref{fig:Crossing}. Suppose that~$E$ denotes
the event that up to this time~$\tau$, the {\SLE} process has not hit the
lower edge of the rectangle. Then the following holds.

\begin{theorem}
 \label{the:crossingprob}
 The \SLE[\kappa] process as described above satisfies, for~$\kappa>4$,
 \begin{equation}
  \Prob[E] \asymp \exp\left[-\big(1-\frac{4}{\kappa}\big)L\right]\quad
  \mbox{as\ }L\goesto\infty,
 \end{equation}
 where $\asymp$ indicates that each side is bounded by some constant
 times the other side.
\end{theorem}

\begin{proof}
 The proof we present here is a simplification of the proof of a more general
 result which appears in~\cite{lsw:I} and which we shall discuss below.
 To prove the theorem, the problem is first translated to the upper
 half-plane. So, let $\Psi:\mathcal{R}_L\goesto\H$ be the conformal map such
 that $\Psi(0)=1$, $\Psi(L)=\infty$ and $\Psi(L+\im\pi)=0$. Then the number
 $\xi:=\Psi(\im\pi)\in(0,1)$ is determined uniquely.  This map is just the map
 of corollary~\ref{cor:rectanglemap} in appendix~\ref{ssec:rectanglemaps},
 and from this we know that $\xi\uparrow1$ as we send $L$ to infinity.

\begin{figure}
  \centering\includegraphics[scale=1.2]{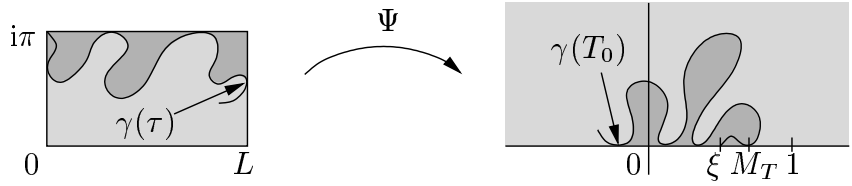}
  \caption{An {\SLE} process crossing a rectangle, and its translation to
   the upper half-plane. The darker grey areas represent the hulls of the
   processes.}
  \label{fig:Crossing}
\end{figure}

 Let~$K_t$ be the hulls of a chordal \SLE[\kappa] process in the upper
 half-plane, which is translated over the distance~$\xi$ to make it start
 in~$\xi$, and let~$\gamma(t)$ denote the trace of the process. Set
 \begin{eqnarray}
  T_0 &:=& \inf\big\{ t\geq0 : \gamma(t)\in\loival{-\infty,0} \big\}; \\
  T_1 &:=& \inf\big\{ t\geq0 : \gamma(t)\in\roival{1,\infty} \big\}; \\
  T_{\phantom{1}} &:=& \min\{ T_0, T_1 \}.
 \end{eqnarray}
 Then~$T_0$ corresponds to the time~$\tau$ when the process first crosses the
 rectangle, and~$T_1$ corresponds to the first time at which the process hits
 the bottom edge of the rectangle. Hence, the event~$E$ of the theorem
 translates to the event~$\{T_0<T_1\}$. Refer to figure~\ref{fig:Crossing}
 for an illustration.

 Now we are going to define a process which allows us to determine whether
 the event~$\{T_0<T_1\}$ or its complement occurs. A good candidate for such
 a process is the process~$Z_t$ given by
 \begin{equation}
  Z_t := \frac{W_t-g_t(0)}{g_t(1)-g_t(0)},\quad
  1-Z_t = \frac{g_t(1)-W_t}{g_t(1)-g_t(0)}
 \end{equation}
 where~$W_t$ denotes the driving process of the L\"owner evolution,
 i.e.\ $W_t$ is Brownian motion multiplied by~$\sqrt{\kappa}$, and
 $W_0=\xi$.

 Indeed, at time~$T$ either the point~$0$ or the point~$1$ becomes part of
 the hull. In the first case, $\lim_{t\upto T}Z_t=0$ because $\lim_{t\upto T}
 (g_t(0)-W_t)=0$, whereas in the second case $\lim_{t\upto T}Z_t=1$, since
 $\lim_{t\upto T}(g_t(1)-W_t)=0$. It is further clear that for all~$t<T$,
 $W_t\in(g_t(0),g_t(1))$, implying that~$Z_t\in(0,1)$ for all $t<T$. This
 means that the stopping time~$T$ conveniently translates into a stopping
 time for~$Z_t$, namely into the first time when~$Z_t$ hits $0$ or~$1$.
 The value of $Z_t$ at this stopping time tells us whether the event
 $\{T_0<T_1\}$ occurs.

 We now derive the differential equation for~$Z_t$, using stochastic
 calculus. First observe that
 \begin{eqnarray}
  && \dif{[W_t-g_t(0)]} = \dif{W_t}-\frac{2\,\dif{t}}{g_t(0)-W_t}, \\
  && \dif{[g_t(1)-g_t(0)]} = \frac{2\,\dif{t}}{g_t(1)-W_t}
     - \frac{2\,\dif{t}}{g_t(0)-W_t}.
 \end{eqnarray}
 Therefore, It\^o's formula (theorem~\ref{the:Itomulti}) tells us that~$Z_t$
 satisfies
 \begin{eqnarray}
  \dif{Z_t} &=& \frac{2\dif{t}}{(g_t(1)-g_t(0))^2}\left(
   \frac{g_t(1)-g_t(0)}{W_t-g_t(0)}-
   \left[\frac{W_t-g_t(0)}{g_t(1)-W_t}+1\right]\right) \nonumber\\
  &&\null -\frac{\dif{W_t}}{g_t(1)-g_t(0)} \nonumber\\
  &=& \frac{2\,\dif{t}}{(g_t(1)-g_t(0))^2}\left(\inv{Z_t}-\inv{1-Z_t}\right)
   -\frac{\dif{W_t}}{g_t(1)-g_t(0)}.
 \end{eqnarray}
 If we now re-parameterize time by introducing the new time parameter
 \begin{equation}
  s = s(t) := \int_0^t\frac{\dif{t}}{(g_t(1)-g_t(0))^2}\quad\mbox{for\ } t<T,
 \end{equation}
 with the inverse~$t(s)$, then it is clear that the process
 $\Ztilde_s:=Z_{t(s)}$ satisfies
 \begin{equation}
  \dif{\Ztilde_s} = \dif{X_s} +
   \left( \inv[2]{\Ztilde_s}-\inv[2]{1-\Ztilde_s} \right)\dif{s} = \dif{X_s} +
   \frac{2(1-2\Ztilde_s)}{\Ztilde_s(1-\Ztilde_s)}\,\dif{s},
 \end{equation}
 where~$X_s$ has the same distribution as the process~$W_t$, i.e. it is a
 Brownian motion multiplied by the factor~$\sqrt{\kappa}$ and starts
 in~$\xi$ (theorem~\ref{the:time-change}).

 From the above calculation we conclude that the process~$\Ztilde_s$ is a
 time-homogeneous Markov process. As we explained earlier, we are interested
 in the value of this process at the stopping time~$s(T):=\lim_{t\upto T}s(t)$,
 which is the first time when~$\Ztilde_s$ hits $0$ or~$1$. To be more precise,
 we want to calculate
 \begin{equation}
  f(\xi) := \Exp\big[1_{\displaystyle\{\Ztilde_{s(T)}=0\}}\mathrel{\big|}
   \Ztilde_0=\xi\big]
 \end{equation}
 where we take the expectation with respect to the Markov process started
 from $\Ztilde_0=\xi$. Observe that the event~$\{\Ztilde_{s(T)}=0\}$
 is equivalent to the event~$\{T_0<T_1\}$.

 Since~$\Ztilde_s$ is a time-homogeneous Markov process, the process $Y_s:=
 f(\Ztilde_s)$ (conditioned on $s<s(T)$) is a martingale with respect to the
 Brownian motion (theorem~\ref{the:Markovmartingale}). Hence, the drift term
 in its It\^o formula must vanish at~$s=0$. It follows that~$f(\xi)$ must
 satisfy the differential equation
 \begin{equation}
  \frac{\kappa}{2}\xi(1-\xi)f''(\xi) + 2(1-2\xi)f'(\xi) = 0.
 \end{equation}
 The boundary conditions are clearly given by $f(0)=1$ and~$f(1)=0$. The
 solution can be written as
 \begin{equation}
  f(\xi) =
  \frac{2^{1-\frac{8}{\kappa}}\,\Gamma\left(\frac{3}{2}-\frac{4}{\kappa}\right)}
   {\sqrt{\pi}\,\Gamma\left(2-\frac{4}{\kappa}\right)}
   (1-\xi)^{1-\frac{4}{\kappa}}\,
   \Hypergeom\left( 1-\frac{4}{\kappa},\frac{4}{\kappa};
   2-\frac{4}{\kappa}; 1-\xi \right).
 \end{equation}
 For critical percolation ($\kappa=6$) this is Cardy's
 formula~\cite{cardy:1992}.
 Note that $f(\xi)$ is exactly the probability~$\Prob[E]$, and that the
 relation between $\xi$ and~$L$ is given by the conformal mapping of
 corollary~\ref{cor:rectanglemap}. Hence, we have basically found the
 probability~$\Prob[E]$ as a function of~$L$. The asymptotic behaviour follows
 from $1-\xi=\exp[-L+O(1)]$ (corollary~\ref{cor:rectanglemap}) and the
 observation that $f(\xi)(1-\xi)^{4/\kappa-1}$ is bounded from above and
 below by some constants when~$\xi\upto1$ (consult e.g.~\cite{hypergeometric}
 for more information on the behaviour of the hypergeometric function).
\end{proof}

We can generalize the theorem in the following way. Consider again an
\SLE[\kappa] process crossing the rectangle~$\mathcal{R}_L$ from $\im\pi$
to~$L$. On the event~$E$ the trace~$\gamma$ has crossed the rectangle
without hitting the bottom edge. So conditional on this event, the
$\pi$-extremal distance between $[0,\im\pi]$ and~$[L,L+\im\pi]$
in~$\mathcal{R}_L\setminus K_\tau$ is well-defined. Let us call this
$\pi$-extremal distance~$\mathcal{L}$. Then one can prove the following
generalization of theorem~\ref{the:crossingprob}.

\begin{theorem}
 \label{the:onesidedexponent}
 For any~$\lambda\geq0$ and $\kappa>4$,
 \begin{equation}
  \Exp[1_E\,\e{-\lambda\mathcal{L}}]
   \asymp \exp\big[ -u(\kappa,\lambda)L] \qquad\mbox{as\ } L\goesto\infty,
 \end{equation}
 where
 \begin{equation}
  u(\kappa,\lambda) = \lambda +
   \frac{\kappa-4+\sqrt{(\kappa-4)^2 + 16\kappa\lambda}}{2\kappa}.
 \end{equation}
\end{theorem}

The exponent $u(\kappa,\lambda)$ is called the one-sided crossing exponent,
because it measures the extremal distance on one side of an {\SLE} process
crossing a rectangle. Observe that $u(\kappa,\lambda)$ reduces to the
exponent~$1-4/\kappa$ for~$\lambda=0$ as it should, because in this case
theorem~\ref{the:onesidedexponent} is completely analogous to
theorem~\ref{the:crossingprob}. The derivation of the one-sided crossing
exponent in~\cite{lsw:I} is rather involved, so we give only a sketch of
the proof here.

\paragraph{Sketch of the proof of theorem~\ref{the:onesidedexponent}}
 We use the same notations as in the proof of theorem~\ref{the:crossingprob}.
 Suppose that we define the conformal maps~$f_t(z)$ for~$t<T$ by
 \begin{equation}
  f_t(z) = \frac{g_t(z)-g_t(0)}{g_t(1)-g_t(0)}.
 \end{equation}
 This is a renormalized version of~$g_t$ that fixes the points $0$, $1$
 and~$\infty$. Now turn back to figure~\ref{fig:Crossing} once more, and
 let $M_T := \sup\{K_T\cap\R\}$. If we set~$N_T:=f_T(M_T)$  then it should
 be clear from conformal invariance that the $\pi$-extremal
 distance~$\mathcal{L}$ just translates into the $\pi$-extremal distance
 between the intervals $\loival{-\infty,0}$ and~$[N_T,1]$ in the upper
 half-plane.

 By corollary~\ref{cor:rectanglemap} in appendix~\ref{ssec:rectanglemaps},
 this $\pi$-extremal distance satisfies
 \begin{equation}
  \mathcal{L} = -\log[1-N_T] + O(1)
 \end{equation}
 and it follows that we have to determine the expectation value of the
 random variable~$(1-N_T)^\lambda$ on the event~$E$. Set~$x:=1-\xi$. Then we
 claim that the value of~$(1-N_T)$ is comparable to~$xf'_T(1)$. This can be
 made more precise, see~\cite{lsw:I} for the details. It follows that it is
 sufficient to calculate the expectation value of $1_E f'_T(1)^\lambda$.

 The calculation proceeds by setting~$\alpha_s:=\log f'_{t(s)}(1)$ for
 $s<s(T)$, where the time re-parameterization is the same as in the proof
 of theorem~\ref{the:crossingprob}. With It\^o's formula one can then
 calculate~$\partial_s\alpha_s$, which turns out to depend only on~$\Ztilde_s$.
 Therefore, $(\Ztilde_s,\alpha_s)$ is a two-dimensional time-homogeneous
 Markov process. So if we set
 \begin{equation}
  y(\xi,v) := \Exp\Big[
    1_{\displaystyle\{\Ztilde_{s(T)}=0\}}\,\e{\lambda\alpha_{s(T)}}
	\mathrel{\Big|} \Ztilde_0=\xi, \alpha_0=v \Big],
 \end{equation}
 then $y(\Ztilde_s,\alpha_s)$ is a martingale, and~$y(\xi,0)$ is the
 expectation value we are trying to calculate. It\^o's formula again
 yields a differential equation for~$y(\xi,v)$, and this equation can be
 solved to find the value of the one-sided crossing exponent.\hfill$\Box$

\subsubsection{The annulus crossing exponent}

\begin{figure}
 \centering\includegraphics[scale=1.2]{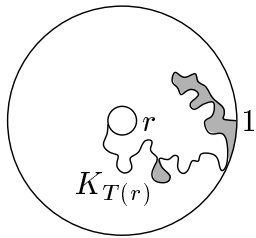}
 \caption{An {\SLE} process crossing an annulus.}
 \label{fig:Annulus}
\end{figure}

There is an analogue of the one-sided crossing exponent for radial {\SLE},
which we shall discuss only briefly here. The setup is as follows. We
consider radial \SLE[\kappa] for any $\kappa>0$, and set~$A_t:=
\partial\D\setminus K_t$. Then the set~$A_t$ is either a piece of arc of
the unit circle, or~$A_t=\emptyset$. Let~$r>0$ and let~$T(r)$ be the first
time when the {\SLE} process hits the circle~$\{z:|z|=r\}$. Denote by~$E$ the
event that~$A_{T(r)}$ is non-empty. On the event~$E$, let~$\mathcal{L}$ be
the $\pi$-extremal distance between the circles $\{z:|z|=1\}$ and
$\{z:|z|=r\}$ in $\D\setminus K_{T(r)}$, see figure~\ref{fig:Annulus}.

\begin{theorem}
 \label{the:annulusexponent}
 For all~$\lambda>0$ and~$\kappa>0$,
 \begin{equation}
  \Exp[1_E\,\e{-\lambda\mathcal{L}}] \asymp
    r^{-\nu(\kappa,\lambda)} \qquad{\rm as\ } r\downto0,
 \end{equation}
 where
 \begin{equation}
  \nu(\kappa,\lambda) =
   \frac{8\lambda+\kappa-4+\sqrt{(\kappa-4)^2 + 16\kappa\lambda}}{16}.
 \end{equation}
\end{theorem}

We call~$\nu(\kappa,\lambda)$ the annulus crossing exponent of~\SLE[\kappa].
A detailed proof of the theorem can be found in~\cite{lsw:II}. It proceeds
along the same lines as the proof of the one-sided crossing exponent.

\subsubsection[Left-passage probability of SLE]{Left-passage probability of {\SLE}}
\label{sssec:SLEformula}

So far, we have considered several crossing events of~{\SLE} processes. A
different kind of event, namely the event that the trace of {\SLE} passes
to the left of a given point~$z_0$, was studied by Schramm
in~\cite{schramm:2001}. We shall reproduce his computation of the probability
of this event below.

\begin{theorem}
 \label{the:SLEformula}
 Let $\kappa\in\roival{0,8}$ and $z_0=x_0+\im y_0\in\H$. Suppose that~$E$ is
 the event that the trace~$\gamma$ of chordal \SLE[\kappa] passes to the
 left of~$z_0$. Then
 \begin{equation}
  \Prob[E]=\ahalf +
  \frac{\Gamma\left(\frac{4}{\kappa}\right)}
       {\sqrt{\pi}\,\Gamma\left(\frac{8-\kappa}{2\kappa}\right)}\,
  \Hypergeom\left(\ahalf,\frac{4}{\kappa};\half{3};-\frac{x_0^2}{y_0^2}\right)
  \frac{x_0}{y_0}.
 \end{equation}
\end{theorem}

\begin{proof}
 Define $X_t:=\Re g_t(z_0)-\sqrt{\kappa}B_t$, $Y_t:=\Im g_t(z_0)$ and set
 $Z_t:=X_t/Y_t$. As before, we let $\tau(z_0)$ be the first time when the
 point~$z_0$ is in the hull of~\SLE[\kappa] (for $\kappa\leq4$ this never
 happens, so then $\tau(z_0)=\infty$). We consider~$\gamma$ up to the
 time~$\tau(z_0)$ only.

 Suppose that~$\omega(z_0,t)$ is the harmonic measure of the union of
 $\roival{0,\infty}$ and the right-hand side of~$\gamma\roival{0,t}$ at the
 point~$z_0$ in the domain~$H_t=\H\setminus K_t$. Then on the event~$E$,
 i.e.\ when~$\gamma$ is to the left of~$z_0$, $\omega(z_0,t)$ tends to~$1$
 when $t\upto\tau(z_0)$. To see this, note that in this limit a Brownian
 motion started from~$z_0$ is certain to first exit the domain~$H_t$ through
 the union of $\roival{0,\infty}$ and the right-hand side
 of~$\gamma\roival{0,t}$, see figure~\ref{fig:Passage}. By conformal
 invariance of harmonic measure, it follows that the harmonic measure of
 $\roival{\sqrt{\kappa}B_t,\infty}$ at the point~$g_t(z_0)$ with respect
 to~$\H$ tends to~$1$ when $t\upto\tau(z_0)$. Therefore,
 $\lim_{t\upto\tau(z_0)}Z_t=+\infty$ if and only if $\gamma$ is to the left
 of~$z_0$. In the same way we can prove that
 $\lim_{t\upto\tau(z_0)}Z_t=-\infty$ if and only if $\gamma$ is to the
 right of~$z_0$. Meanwhile, it is clear that for all $t<\tau(z_0)$, $Z_t$
 is finite.

\begin{figure}
  \centering\includegraphics[scale=1.2]{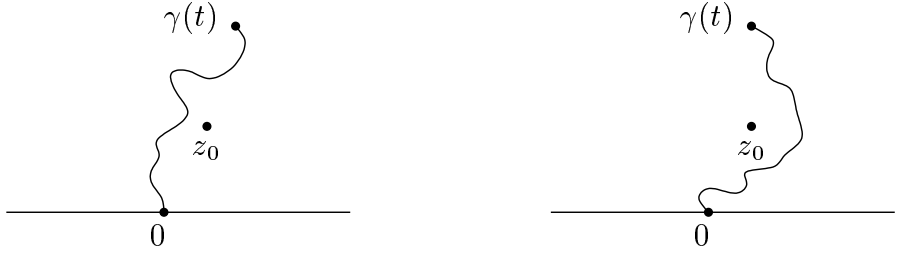}
  \caption{Two {\SLE} traces passing to the left and right, respectively, of
   a given point~$z_0$.}
  \label{fig:Passage}
\end{figure}

 Now let us look at the differential equation satisfied by~$Z_t$. To derive
 it, note first of all that $\dif{X_t}$ and~$\dif{Y_t}$ are given simply by
 taking the real and imaginary parts of L\"owner's equation. If we then apply
 It\^o's formula we find
 \begin{equation}
  \dif{Z_t} = \frac{4Z_t}{Y_t^2(1+Z_t^2)}\,\dif{t}
   -\frac{\sqrt{\kappa}}{Y_t}\,\dif{B_t}.
 \end{equation}
 If we now define $u(t):=\int_0^t Y_s^{-2}\dif{s}$ and $\Ztilde_u:=Z_{t(u)}$,
 then
 \begin{equation}
  \dif{\Ztilde_u} = \frac{4\Ztilde_u}{1+\Ztilde_u^2}\,\dif{u}
   -\sqrt{\kappa}\,\dif{\tilde{B}_u}
 \end{equation}
 where~$\tilde{B}_u$ is again standard Brownian motion
 (theorem~\ref{the:time-change}). It follows that $\Ztilde_u$ is a
 time-homogeneous Markov process. Furthermore, it is clear from the
 differential equation that~$\Ztilde_u$ does not become infinite
 in finite time. Therefore, we are interested in the probability that
 $\Ztilde_u\goesto+\infty$ when~$u\goesto\infty$.

 Now let $a<b$ be some real numbers, and define
 \begin{equation}
  h_{a,b}(x) := \Prob[\Ztilde_u \mbox{\ hits $b$ before $a$}\mid\Ztilde_0=x].
 \end{equation}
 Then the process~$h_{a,b}(\Ztilde_u)$ is a martingale, and so the drift term
 in its It\^o formula must vanish. At $u=0$ this gives us
 \begin{equation}
  \frac{\kappa}{2}h_{a,b}''(x) + \frac{4x}{1+x^2}h'_{a,b}(x) = 0,
  \quad h_{a,b}(a)=0, \quad h_{a,b}(b)=1.
 \end{equation}
 This has the unique solution
 \begin{equation}
  h_{a,b}(x) = \frac{f(x)-f(a)}{f(b)-f(a)}, \quad
  f(x) = \Hypergeom\left(\ahalf,\frac{4}{\kappa};\frac{3}{2};-x^2\right)\,x.
 \end{equation}
 The probability~$\Prob[E]$ is just $h_{a,b}(x_0/y_0)$ in the limit
 $a\goesto-\infty$, $b\goesto+\infty$. This limit exists, since the limits
 $\lim_{x\goesto\pm\infty}f(x)$ exist and are finite (see for example 15.3.4
 in~\cite{hypergeometric}). The limit values determine the constants in the
 theorem, and we are done.
\end{proof}

\subsection{Intersection exponents of planar Brownian motion}
\label{ssec:brownianmotionresults}

One of the first successes of {\SLE} was the determination of the
intersection exponents of planar Brownian motion. One way of defining these
exponents is as follows (see reference~\cite{lawler:1999b}, which also
presents alternative definitions). Let~$k\geq2$ and~$p_1,\ldots,p_k$ be
positive integers. For each~$j\in\{1,\ldots,k\}$, start~$p_j$ planar Brownian
motions from the point~$(0,j)$. Denote by~$\mathcal{B}^j_t$ the union of the
traces of these~$p_j$ Brownian motions up to time~$t$. Then we can define an
exponent~$\xi(p_1,\ldots,p_k)$ by
\begin{equation}
 \Prob\big[\forall i\neq j\in\{1,\ldots,k\},
  \mathcal{B}^i_t\cap\mathcal{B}^j_t=\emptyset\big]
  \asymp \big(\sqrt{t}\big)^{-\xi(p_1,\ldots,p_k)}
\end{equation}
when $t\goesto\infty$. The exponent~$\xi(p_1,\ldots,p_k)$ is called the
intersection exponent between~$k$ packets of~$p_1,\ldots,p_k$ Brownian
motions.

If we further require that the Brownian motions stay in the upper half-plane,
we get different exponents~$\xitilde(p_1,\ldots,p_k)$ defined by
\begin{equation}
 \Prob\big[\forall i\neq j\in\{1,\ldots,k\},
  \mathcal{B}^i_t\cap\mathcal{B}^j_t=\emptyset\mbox{\ and\ }
  \mathcal{B}^i_t\subset\H\big]
  \asymp \big(\sqrt{t}\big)^{-\xitilde(p_1,\ldots,p_k)}
\end{equation}
when $t\goesto\infty$. We could also \emph{condition} on the event that the
Brownian motions stay in the upper half-plane. The corresponding exponents
are $\hat{\xi}(p_1,\ldots,p_k)$. They are related to the previous half-plane
exponents by
\begin{equation}
 \hat{\xi}(p_1,\ldots,p_k) = \xitilde(p_1,\ldots,p_k) - (p_1+\ldots+p_k),
\end{equation}
since the probability that a Brownian motion started in the half-plane stays
in the half-plane up to time~$t$ decays like~$t^{-1/2}$.

Duplantier and Kwon~\cite{duplantier:1988} predicted the values of the
intersection exponents $\xi(p_1,\ldots,p_k)$ and $\hat{\xi}(p_1,\ldots,p_k)$
in the case where all~$p_i$ are equal to~$1$. In the series of
papers~\cite{lsw:I,lsw:II,lsw:III,lsw:2002a}, Lawler, Schramm and Werner
confirmed these predictions rigorously, and generalized them. Here, we will
only give an impression of the arguments used in the first paper~\cite{lsw:I},
and then we will summarize the main conclusions of the whole series.

\subsubsection{Half-plane exponents}

In the aforementioned article by Lawler and Werner~\cite{lawler:1999b} it
is shown how the definition of the Brownian intersection exponents can be
extended in a natural way. This leads to the definition of the exponents
$\xitilde(\lambda_1,\ldots,\lambda_k)$ for all $k\geq1$ and all non-negative
real numbers $\lambda_1,\ldots,\lambda_k$, and of the exponents
$\xi(\lambda_1,\ldots,\lambda_k)$ for all~$k\geq2$ and nonnegative real
numbers $\lambda_1,\ldots,\lambda_k$, at least two of which must be at
least~$1$.

\begin{figure}
 \centering\includegraphics[scale=1.2]{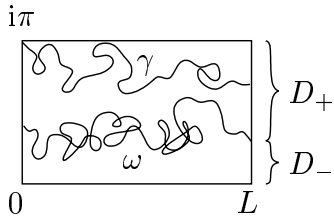}
 \caption{An \SLE[6] trace~$\gamma$ and a Brownian excursion~$\omega$ crossing
  a rectangle.}
 \label{fig:Brownian}
\end{figure}

Furthermore, the article shows how the exponents $\xitilde(\lambda_+,1,
\lambda_-)$ and $\xitilde(1,\lambda)$ can be characterized in terms of
Brownian excursions (see appendix \ref{ssec:brownianmotion} and references
\cite{lawler:1999b,lsw:I}). This characterization proceeds as follows. Let
$\mathcal{R}_L$ be the rectangle $(0,L)\times(0,\im\pi)$, and denote by
$\omega$ the path of a Brownian excursion in~$\mathcal{R}_L$. Let~$A$ be
the event that the Brownian excursion crosses the rectangle from the left
to the right. On this event, let $D_+$ and $D_-$ be the domains remaining
above and below~$\omega$ in $\mathcal{R}_L\setminus\omega$, respectively,
and let $\mathcal{L}_+$ and $\mathcal{L}_-$ be the $\pi$-extremal distances
between the left and right edges of the rectangle in these domains. We refer
to figure~\ref{fig:Brownian} for an illustration.

By symmetry, the distributions of $\mathcal{L}_+$ and $\mathcal{L}_-$ are
the same. The exponent $\xitilde(1,\lambda)$ is characterized by
\begin{equation}
 \Exp_B[1_A\,\e{-\lambda\mathcal{L}_+}] =
 \Exp_B[1_A\,\e{-\lambda\mathcal{L}_-}] \asymp
 \e{-\xitilde(1,\lambda)L} \qquad\mbox{as\ }L\goesto\infty
\end{equation}
where~$\Exp_B$ is used to indicate expectation with respect to the
Brownian excursion measure. Likewise, $\xitilde(\lambda_+,1,\lambda_-)$ is
characterized by
\begin{equation}
 \Exp_B[1_A\,\e{-\lambda_+\mathcal{L}_+}\e{-\lambda_-\mathcal{L}_-}]
 \asymp \e{-\xitilde(\lambda_+,1,\lambda_-)L}\qquad\mbox{as\ }L\goesto\infty.
\end{equation}

Another major result from~\cite{lawler:1999b} is the theorem below, which
gives the so-called cascade relations between the Brownian intersection
exponents. Together with an analysis of the asymptotic behaviour of the
exponents (theorems 11 and~12 in~\cite{lawler:1999b}), these relations
show that it is sufficient to determine the exponents $\xi(1,1,\lambda)$,
$\xitilde(1,\lambda)$ and $\xitilde(\lambda,1,\lambda)$ for~$\lambda\geq0$
to know all the intersection exponents. In this article, we shall only
explain how the exponent $\xitilde(1,\lambda)$ was determined in~\cite{lsw:I}
using {\SLE}.

\begin{theorem}
 \label{the:cascade}
 The exponents $\xitilde(\lambda_1,\ldots,\lambda_k)$ and $\xi(\lambda_1,
 \ldots,\lambda_k)$ are invariant under permutations of their arguments.
 Moreover, they satisfy the following cascade relations:
\begin{eqnarray}
 \xitilde(\lambda_1,\ldots,\lambda_k) &=&
 \xitilde\big(\lambda_1,\ldots,\lambda_{j-1},
  \xitilde(\lambda_j,\ldots,\lambda_k)\big);
 \\
 \xi(\lambda_1,\ldots,\lambda_k) &=& \xi\big(\lambda_1,\ldots,\lambda_{j-1},
  \xitilde(\lambda_j,\ldots,\lambda_k)\big).
\end{eqnarray}
\end{theorem}

We are now ready to describe how the exponent~$\xitilde(1,\lambda)$ can be
computed. To do so, suppose that we add an \SLE[6] process from $\im\pi$
to~$L$ to the same rectangle~$\mathcal{R}_L$ in which we had the Brownian
excursion~$\omega$. In what follows, it is crucial that this process has
the locality property. In our present setup, this implies that as long as
the \SLE[6] trace does not hit~$\omega$, it doesn't matter whether we regard
it as an \SLE[6] in the domain~$\mathcal{R}_L$ or in the domain~$D_+$. Since
\SLE[\kappa] has this property only for~$\kappa=6$, the following argument
works only for this special value of~$\kappa$.

Let us denote by~$\gamma$ the trace of the \SLE[6] process up to the
first time when it hits~$[L,L+\im\pi]$, and let~$E$ be the event that
$\gamma$ is disjoint from~$\omega$ and that $\omega$ crosses the rectangle
from left to right. See figure~\ref{fig:Brownian}. On the event~$E$,
the $\pi$-extremal distance between $[0,\im\pi]$ and~$[L,L+\im\pi]$ in the
domain between $\gamma$ and~$\omega$ is well-defined. We call this
$\pi$-extremal distance~$\mathcal{L}$. To obtain the value of
$\xitilde(1,\lambda)$, our strategy is to express the asymptotic behaviour
of $f(L)=\Exp[1_E\,\exp(-\lambda\mathcal{L})]$ in two different ways.

On the one hand, when~$\omega$ is given, $1_E\,\exp(-\lambda\mathcal{L})$
is comparable to $\exp[-u(6,\lambda)\mathcal{L}_+]$ by
theorem~\ref{the:onesidedexponent}. We therefore get
\begin{equation}
 f(L) \asymp \Exp_B[1_A\,\e{-u(6,\lambda)\mathcal{L}_+}]
 	  \asymp \e{-\xitilde(1,u(6,\lambda))L}.
\end{equation}
On the other hand, when~$\gamma$ is given, the distributions of $\mathcal{L}$
and~$\mathcal{L}_-$ are the same by the conformal invariance of the Brownian
excursion. But also, given~$\mathcal{L}_+$, the probability of the event~$E$
is comparable to $\exp(-\mathcal{L}_+/3)$ by theorem~\ref{the:crossingprob}.
Therefore
\begin{equation}
 f(L) \asymp \Exp_B[1_A\,\e{-\mathcal{L}_+/3}\e{-\lambda\mathcal{L}_-}]
	  \asymp \e{-\xitilde(1/3,1,\lambda)L}.
\end{equation}
By the cascade relations, $\xitilde(1/3,1,\lambda)=
\xitilde\big(1,\xitilde(1/3,\lambda)\big)$. Hence, comparing the two results
we obtain
\begin{equation}
 \xitilde(1/3,\lambda) = u(6,\lambda) = \frac{6\lambda+1+\sqrt{1+24\lambda}}{6}
\end{equation}
since $\xitilde(1,\lambda)$ is strictly increasing in~$\lambda$. Finally,
this result gives us for example $\xitilde(1,\lambda)$, because
$\xitilde(1/3,1/3)=1$, and then the cascade relations give
\begin{equation}
 \xitilde(1,\lambda) = \xitilde\big(\xitilde(1/3,1/3),\lambda\big)
  = \xitilde\big(1/3,\xitilde(1/3,\lambda)\big).
\end{equation}

\subsubsection{Summary of results}

As we mentioned before, the series of papers by Lawler, Schramm and Werner
\cite{lsw:I,lsw:II,lsw:III,lsw:2002a} led to the determination of all
Brownian intersection exponents we defined above. We state their conclusions
as a series of theorems.

\begin{theorem} 
 For all integers $k\geq2$ and all $\lambda_1,\ldots,\lambda_k\geq0$,
 \begin{equation}
  \xitilde(\lambda_1,\ldots,\lambda_k) =
  \frac{(\sqrt{1+24\lambda_1}+\cdots+\sqrt{1+24\lambda_k}-(k-1))^2-1}{24}.
 \end{equation}
\end{theorem}

\begin{theorem} 
 For all integers $k\geq2$ and all $\lambda_1,\ldots,\lambda_k\geq0$, at
 least two of which are at least~$1$,
 \begin{equation}
  \xi(\lambda_1,\ldots,\lambda_k) =
  \frac{(\sqrt{1+24\lambda_1}+\cdots+\sqrt{1+24\lambda_k}-k)^2-4}{48}.
 \end{equation}
\end{theorem}

\begin{theorem}
 For all integers $k\geq2$ and all $\lambda\geq0$,
 \begin{equation}
  \xi(k,\lambda) = \frac{(\sqrt{1+24k}+\sqrt{1+24\lambda}-2)^2-4}{48}.
 \end{equation}
\end{theorem}

By earlier work of Lawler~\cite{lawler:1996a,lawler:1996b,lawler:1999a}, it
is known that some of these exponents are related to the Hausdorff dimensions
of special subsets of the Brownian paths. Indeed, suppose that we denote by
$B[0,1]$ the trace of a planar Brownian motion up to time~$1$. Then the
Hausdorff dimension of its frontier (the boundary of the unbounded connected
component of~$\C\setminus B[0,1]$), is $2-\xi(2,0)=4/3$. The Hausdorff
dimension of the set of cut points (those points $z$ such that $B[0,1]
\setminus\{z\}$ is disconnected) is $2-\xi(1,1)=3/4$. Finally, the set of
pioneer points of~$B[0,1]$ (those points~$z$ such that for some~$t\in[0,1]$,
$z=B_t$ is in the frontier of~$B[0,t]$) has Hausdorff dimension $2-\xi(1,0)
=7/4$. This completes our overview of the {\SLE} results for Brownian motion.

\subsection{Results on critical percolation}
\label{ssec:percolationresults}

The connection between \SLE[6] and critical site percolation on the
triangular lattice can be used to verify rigorously the values of certain
percolation exponents. In this subsection we review how for example the
multi-arm exponents for percolation can be calculated from the one-sided
crossing exponent and the annulus crossing exponent of \SLE[6]. Predictions
of the values of these exponents have appeared in several places in the
physics literature, see e.g.~\cite{duplantier:1999} and references therein.
In this section we also describe Schramm's left-passage probability for
percolation. This is an example of a result that was unknown before the
introduction of~{\SLE}.

\subsubsection{Half-plane exponents}

Consider critical site percolation on the triangular lattice with fixed mesh.
Let~$A^+(r,R)$ be a discrete approximation by hexagons of the semi-annulus
$\{z:r<|z|<R,\Im z>0\}$, and denote by~$f^+_k(r,R)$ the probability that
there exist~$k$ disjoint crossings of arbitrary colours from the inner
circle to the outer circle in~$A^+(r,R)$. By a crossing we mean a sequence
of distinct connected hexagons, all in the same colour, whose first and last
hexagons are adjacent to a hexagon intersecting the inner and outer circle,
respectively. Obviously, $r$ has to be large enough if the definition
of~$f^+_k(r,R)$ is to make sense, i.e.\ $r>\mathrm{const}(k)$.

It is well-known that the probability~$f^+_k(r,R)$ does not depend on the
choice of colours of the different crossings. The reason for this is that one
can always flip the colours of crossings without changing probabilities. To
do so, one can start by considering the clockwise-most crossing. If desired,
its colour can be flipped by flipping the colours of all hexagons. Then
one proceeds by each time considering the clockwise-most crossing to the left
of the previous one. If desired, its colour can be flipped by flipping the
colours of all hexagons to the left of this previous crossing. In the end one
obtains a configuration with all crossings in the desired colours, without
changing probabilities. In particular, we can take $f^+_k(r,R)$ to be the
probability of~$k$ crossings of alternating colours.

\begin{figure}
 \centering\includegraphics[scale=1.2]{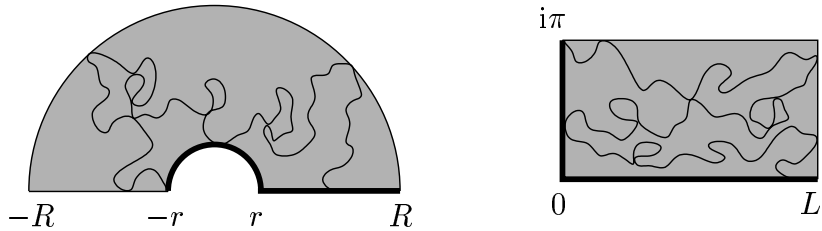}
 \caption{An \SLE[6] process which crosses a semi-annulus three times, and
  the equivalent process in a rectangle. The thick part of the boundary is
  the part coloured blue.}
 \label{fig:ThreeArm}
\end{figure}

We are now ready to make the connection with~{\SLE}. Indeed, suppose that
we colour all hexagons that intersect the boundary of the semi-annulus
blue if they are on the counter-clockwise arc from~$-r$ to~$R$, and yellow
if they are on the clockwise arc from~$-r$ to~$R$. Then the probability
$f^+_k(r,R)$ is exactly the probability that the exploration process
from~$-r$ to~$R$ makes~$k$ crossings before it hits the interval~$[r,R]$. By
Smirnov's result, this translates in the scaling limit into the probability
that a chordal \SLE[6] process from~$-r$ to~$R$ in the semi-annulus makes~$k$
crossings before it hits the interval~$[r,R]$, see figure~\ref{fig:ThreeArm}.

It is more convenient now to map the problem to a rectangle using the
logarithmic map. Suppose that~$g^+_k(L)$ denotes the probability that
an~\SLE[6] trace from~$\im\pi$ to~$L$ in the rectangle $\mathcal{R}_L:=
(0,L)\times(0,\im\pi)$ makes~$k$ horizontal crossings before it hits the
bottom. Then, by conformal invariance, we want to determine $g^+_k(L)$ for
$L=\log(R/r)$. For~$k=1$ theorem~\ref{the:crossingprob} immediately gives
$g^+_1(L)\asymp\exp(-L/3)$. Exponents for larger~$k$ can be determined
using theorem~\ref{the:onesidedexponent}.

Indeed, let~$T$ be the time at which the \SLE[6] process has crossed the
rectangle for the first time, and let~$E$ be the event that up to time~$T$
the process has not hit the bottom. Then the process still has to make~$k-1$
crossings in the domain which is left below this first crossing. Hence, if
$\mathcal{L}$ denotes the $\pi$-extremal distance between the left and
right edges in this remaining domain, we have
\begin{equation}
 g^+_k(L)=\Exp[1_E\,g^+_{k-1}(\mathcal{L})].
\end{equation}
It is now clear from $g^+_1(L)\asymp\exp(-L/3)$ and
theorem~\ref{the:onesidedexponent} that $g^+_k(L)\asymp\exp(-v^+_k L)$
for all~$k\geq1$ and some~$v_k^+$, and that the numbers~$v^+_k$ can be
determined recursively.

In terms of the one-sided crossing exponent, the recursion formula for
the~$v^+_k$ reads $v^+_k=u(6,v^+_{k-1})$. It follows that
\begin{equation}
 v^+_k=\frac{k(k+1)}{6}.
\end{equation}
Returning to the case of discrete percolation in the semi-annulus, this
result implies that
\begin{equation}
 f^+_k(r,R) \asymp R^{-k(k+1)/6}\qquad\mbox{when\ }R\goesto\infty.
\end{equation}
To make this transition to discrete percolation completely rigorous some
more work is required. We refer to~\cite{smirnov:2001a} for more details.
To complete the discussion, we finally note that for odd~$k$, $f^+_k(r,R)$
is also the probability that there exist~$j=(k+1)/2$ disjoint blue clusters
crossing the semi-annulus.

\subsubsection{Plane exponents}

We now turn to the planar case. Suppose that~$A(r,R)$ is an approximation
of the full annulus $\{z:r<|z|<R\}$ by hexagons, where~$r$ is again assumed
to be large enough. We can define an exploration process in this annulus
as follows. We colour all hexagons intersecting the inner circle blue. The
exploration process starts at~$R$ with a blue hexagon on its right, and a
yellow hexagon on its left. Each time the exploration process hits a hexagon
on the outer circle that was not visited before, we look at the argument of
the tip of the trajectory at that time (where the argument is determined
continuously, so that it makes no jumps after completing a circle). If the
argument is positive, the hexagon on the boundary is coloured blue, and
otherwise it is coloured yellow.

When the exploration process described above first hits the inner circle,
it defines unambiguously a clockwise-most blue crossing of the annulus and
a counter-clockwise-most yellow crossing, such that the point~$R$ lies between
them. Moreover, it can be seen easily that afterwards, the exploration
process continues like a chordal process in the remaining domain between
these two crossings, where the outer circle may now be assumed to be
coloured yellow. This remaining domain is equivalent to a semi-annulus.
Therefore, the probability that the process crosses this remaining domain
$k-2$ times before it disconnects the inner circle from the outer circle is
equal to the probability that there are $k-2$ crossings of arbitrary colours
of this domain, as we discussed in the previous subsection.

Let $f_k(r,R)$ be the probability that the exploration process crosses the
annulus a total number of~$k-1$ times. Then for even~$k$, $f_k(r,R)$ is
just the probability that there exist~$k$ crossings of the annulus, which
are not all of the same colour. Indeed, in this case we have the freedom
of choosing alternating colours for the crossings, and then the point~$R$
is always between a clockwise-most blue and a counter-clockwise-most yellow
crossing, which proves the point. For odd~$k$, the situation is different,
and $f_k(r,R)$ is not equal to the probability that there exist~$k$ crossings
of the annulus which are not all of the same colour. However, it can be
shown that the two probabilities differ only by a multiplicative constant,
see~\cite{smirnov:2001a}.

We now make the connection with~\SLE[6]. In the continuum limit, the
discrete exploration process converges to the following {\SLE} process.
First, we do radial \SLE[6] in the annulus from~$R$ to~$0$, up to the
first time~$T$ that the process hits the inner circle. Afterwards, the
process continues like a chordal \SLE[6] process in the remaining domain.
We further define~$E$ to be the event that up to time~$T$, the process
has not disconnected the inner circle from the outer circle. On this
event, we let~$\mathcal{L}$ denote the $\pi$-extremal distance between
the two circles in the remaining domain.

Denote by~$g_k(r,R)$ the probability that this \SLE[6] process crosses
the annulus~$k-1$ times before it disconnects the inner circle from the
outer circle. Then
\begin{equation}
 g_k(r,R) = \Exp[1_E\,g^+_{k-2}(\mathcal{L})] \asymp
  \Exp[1_E\,\e{-v^+_{k-2}\mathcal{L}}]
\end{equation}
where $g^+_k(L)$ is the probability of~$k$ crossings of the rectangle
$(0,L)\times(0,\im\pi)$, as before. Theorem~\ref{the:annulusexponent} now
tells us that~$g_k(r,R)\asymp(R/r)^{-v_k}$, where
\begin{equation}
 v_k=\nu(6,v^+_{k-2})=\frac{k^2-1}{12}.
\end{equation}
Returning to discrete percolation, it follows from this result that the
probability of~$k$ crossings of the annulus~$A(r,R)$ which are not all of
the same colour behaves like
\begin{equation}
 f_k(r,R) \asymp R^{-(k^2-1)/12}\qquad\mbox{when\ }R\goesto\infty.
\end{equation}
Again, all of this can be made rigorous~\cite{smirnov:2001a}. Observe
also that we can again interpret the result in terms of crossings of
clusters. In this case we have that for~$k$ even, $f_k(r,R)$ is comparable
to the probability that there exist~$j=k/2$ disjoint blue clusters
crossing the annulus.

So far we only considered the dichromatic exponents associated with the
probability of~$k$ percolation crossings of an annulus that are \emph{not}
all of the same colour. The corresponding monochromatic exponents for~$k$
crossings that \emph{are} of the same colour are known to have different
values. They are not so easily accessible through {\SLE} as the dichromatic
exponents. However, {\SLE} computations~\cite{lsw:2001b} have confirmed
that the one-arm exponent ($k=1$) has the value~$5/48$, and in the same
article, a description of the backbone exponent ($k=2$) as the leading
eigenvalue of a differential operator was given.

\subsubsection{Left-passage probability of critical percolation}

In section~\ref{sssec:SLEformula} we discussed the left-passage probability
of {\SLE} derived by Schramm in~\cite{schramm:2001}. From this formula he
obtained a percolation result, which was not predicted before in the physics
literature. Following Schramm, consider critical percolation on the triangular
lattice with mesh $\delta>0$ in the unit disk. Fix $\theta\in(0,2\pi)$ and let
$A_\theta$ be the arc of the unit circle between the angles $0$ and~$\theta$,
that is, $A_\theta:=\{\exp(\im s):0\leq s\leq\theta\}$. We are interested in
the probability of the event~$E_\theta$ that there is a cluster of blue
hexagons connected to~$A_\theta$, such that the union of this cluster with
the arc~$A_\theta$ surrounds the origin.

\begin{theorem}
 \label{the:percolationformula}
 \begin{equation}
  \lim_{\delta\downto0}\Prob[E_\theta] = \ahalf -
   \frac{\Gamma\left(\frac{2}{3}\right)}
    {\sqrt{\pi}\,\Gamma\left(\inv{6}\right)}\,
   \Hypergeom\left(\ahalf,\frac{2}{3};\half{3};-\cot^2\half{\theta}\right)
   \cot\half{\theta}.
 \end{equation}
\end{theorem}

\begin{proof}
 The proof of the theorem by Schramm is based on the observation that the
 event~$E_\theta$ can be written in terms of the behaviour of an exploration
 process. Indeed, consider the chordal exploration process in the unit disk
 from~$1$ to the point~$\exp(\im\theta)$. Let~$\gamma^{\mathrm{ep}}$ denote the
 trace of this process. Then~$E_\theta$ is equal to the event that the origin
 is in a connected component of~$\D\setminus\gamma^{\mathrm{ep}}$ which lies
 on the right-hand side of the exploration process. See
 figure~\ref{fig:PercFormula} for an illustration.

\begin{figure}
  \centering\includegraphics[scale=1.2]{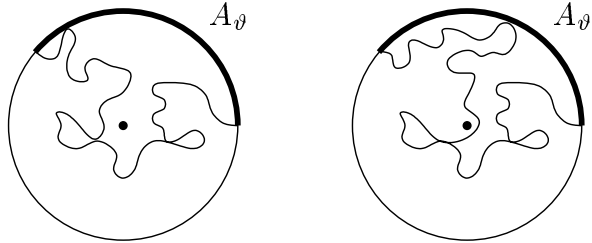}
  \caption{An exploration process passing to the left or the right of the
   origin, respectively. In the former case, there is a blue cluster connected
   to the arc~$A_\theta$ which surrounds the origin, in the latter case
   there isn't.}
  \label{fig:PercFormula}
\end{figure}

 Suppose that we now map the unit disk onto the upper half-plane in such a
 way that $1$ maps to~$0$ and $\exp(\im\theta)$ maps to~$\infty$. It is easy
 to see that an inverse map with the desired properties is
 \begin{equation}
  \phi(z) = \e{\im\theta}\frac{z+\cot\half{\theta}-\im}
  	{z+\cot\half{\theta}+\im},\quad z\in\H,
 \end{equation}
 since this is just the composition of the standard map $(z-\im)/(z+\im)$ of
 $\H$ onto~$\D$ with a translation and a rotation. Observe that the
 point~$z_0$ which maps to the origin is~$z_0=\im-\cot(\theta/2)$.

 It should be clear that in the scaling limit, the event~$E_\theta$ reduces
 to the event that the \SLE[6] trace in the half-plane passes to the left
 of the point~$z_0$. The probability of this event is given by
 theorem~\ref{the:SLEformula}. Theorem~\ref{the:percolationformula} follows
 immediately from this result.
\end{proof}


\section{Discussion}
\label{sec:discussion}

We conclude this article with a short discussion of {\SLE} and its relevance
for the study of continuous phase transitions in two dimensions. In the first
place {\SLE} appears as a serious candidate for the scaling limit of critical
models. Indeed, {\SLE} was introduced by Oded Schramm as the only possible
candidate for the scaling limit of the loop-erased random walk, and the
definition and properties of {\SLE} were sufficiently general to allow him
to conjecture that {\SLE} also describes the scaling limits of uniform
spanning trees and critical percolation. In fact, it is believed that
conformal invariance combined with the stationarity property is sufficient
for a whole range of critical models to converge to {\SLE}
(section~\ref{sec:discretemodels}).

Apart from being the candidate for the scaling limit of critical models,
{\SLE} also gives us an idea of how the convergence can be proved. One
could try to describe the discrete path of the critical model by a L\"owner
evolution, and then prove that the driving function converges to Brownian
motion. Indeed, this is the way in which the convergence of loop-erased
random walk to \SLE[2], and of the Peano curve winding around the uniform
spanning tree to \SLE[8] were proved. Recently, the harmonic explorer was
added to the list, and it seems reasonable to believe that in the future
more connections between discrete models and {\SLE} will be established.

Another important aspect of {\SLE} is that it allows us to do computations
and prove properties of critical models. Several examples have been given
in this article. We have seen that {\SLE} has not only led to rigorous
confirmations of the values of critical exponents predicted before in the
physics literature, but also to a new result in the form of Schramm's
left-passage probability. More results from {\SLE} are to be expected.

However, a limitation of {\SLE} appears to be that it is only capable
of describing a very specific aspect of the discrete models. In the
Fortuin-Kasteleyn cluster formulation of the Potts model, for example,
{\SLE} describes the boundary of one special cluster connected to the
boundary, as explained in section~\ref{ssec:PottsModel}. An interesting
question is then what {\SLE} can tell us about the full configuration of
clusters. In the case of critical percolation (\SLE[6]) a description of
the full limit appears to be possible~\cite{camia:2003}, but for other
values of~$\kappa$ it is not so clear how one should proceed. Indeed, so
far most applications of {\SLE} are restricted to the \SLE[6] case, where
the locality property allows one to ``forget'' the boundary conditions. For
other values of~$\kappa$ more work needs to be done, for example to clarify
what {\SLE} can say about correlations between spins in the Potts or O($n$)
models.

Interesting developments have taken place regarding the connection between
{\SLE} and conformal field theory (CFT), a subject not considered in this
article. Various aspects of this connection have been studied in a series
of papers by Michel Bauer and Denis Bernard~\cite{bauer:2002,bauer:2003a,
bauer:2003b,bauer:2003c}, showing for example how results from {\SLE} can
be computed in the CFT language. Another connection was proposed by John
Cardy~\cite{cardy:2003} who introduced a multiple {\SLE} process. This he
could connect with Dyson's Brownian process, and through it to the
distribution of eigenvalues of ensembles of random matrices. Using the
conformal restriction properties studied in~\cite{lsw:2003}, the work of
Roland Friedrich and Wendelin Werner~\cite{friedrich:2002,friedrich:2003,
werner:2003} further clarifies the link between the discrete systems and
conformal field theory. Thus {\SLE} may prove to be very useful in putting
the ideas of conformal field theory on a mathematically more rigorous
footing.

{\SLE} is a promising field of research, and the literature on {\SLE} is
already quite vast and still growing. In this discussion we only touched
upon some of the developments that have taken place, without the intention
of providing a complete list. In conclusion, {\SLE} seems invaluable for
adding mathematical rigour to our understanding of the scaling limits of
critical two-dimensional systems and their conformal invariance. This same
fact makes {\SLE} a mathematically and technically challenging object of
study. We hope that this article may serve as an aid to both mathematicians
and physicist for making this interesting field more accessible.

\paragraph{Acknowledgements.} We would like to thank Rob van den Berg for
organizing the {\SLE} meetings at CWI in Amsterdam, and the participants
of these meetings. Special thanks also go to Antal Jarai, Remco van der
Hofstad and Ronald Meester for their input. WK had useful conversations
about the manuscript with Debabrata Panja. This research was financially
supported by the Stichting FOM (Fundamenteel Onderzoek der Materie) in the
Netherlands.

\appendix
\section*{Appendices}
\addcontentsline{toc}{chapter}{\protect\numberline{}Appendices}


\section{Conformal mapping theory}
\label{sec:conformalmappingtheory}

This appendix gives a summary of some of the background theory we need to
study {\SLE}. We start with the general theory of conformal maps, and then
focus on specific topics regarding conformal maps of the unit disk $\D=
\{z:|z|<1\}$ and conformal maps of the complex upper half-plane $\H=
\{z:\Im z>0\}$. In the fifth subsection, we will discuss maps of rectangles
onto the upper half-plane. The material for this section is taken from the
books by Ahlfors, Gamelin and Pommerenke~\cite{ahlfors:1966,ahlfors:1973,
gamelin:2000,pommerenke:1992}, and the article~\cite{lawler:url2001} by
Lawler. Most theorems are presented without proofs, and where a proof is
provided this is done either to illustrate a technique, or because the
standard text-books do not give a proof.

\subsection{Basics of conformal mapping theory}
\label{ssec:conformalmaps}

First let us fix some terminology. A \defn{domain} is an open connected
subset of the complex plane. We call a domain \defn{simply connected} if it
contains no holes. More precisely, a domain is simply connected if its
complement in the complex plane is connected or, equivalently, if every
closed curve in the domain can be contracted continuously to a single point
of the domain.

A \defn{conformal map}~$f$ of a simply connected domain $D\neq\C$ onto another
simply connected domain $D'\neq\C$ is a one-to-one map which preserves angles.
That is, if $\gamma_0$ and~$\gamma_1$ are two curves in~$D$ which intersect at
a certain angle, then their images $f\circ\gamma_0$ and $f\circ\gamma_1$
must intersect at the same angle. In practice this means that a conformal map
$f:D\goesto D'$ is an injective and analytic function on~$D$, which has
nonzero derivative everywhere on~$D$. It has an inverse~$f^{-1}$ which is
also conformal.

The main theorem about these conformal maps is the Riemann mapping theorem,
which tells us that any simply connected domain~$D$ can be mapped conformally
onto the open unit disk~$\D$. Note that the theorem says nothing about the
behaviour of the map at the boundary~$\partial D$. However, in this article
we only consider maps whose definition can be extended to the boundary (if
there are points on the boundary that are multiple boundary points, we have
to distinguish between them, as we explain below), and in the text we may
sometimes tacitly assume this. The reason is that we only work with domains
whose boundaries are continuous curves (see chapter 2 of
Pommerenke~\cite{pommerenke:1992} for details, in particular theorems 2.1,
2.6 and 2.14).

\begin{theorem}[Riemann mapping theorem]
 \label{the:riemannmapping}
 Let~$D\neq\C$ be a simply connected domain in~$\C$. Then
 there is a conformal map of~$D$ onto the open unit disk~$\D$.
\end{theorem}

\begin{figure}
 \centering\includegraphics[scale=1.2]{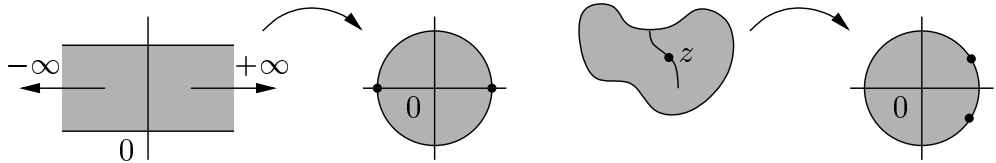}
 \caption{Some special boundary points. The infinite strip on the left has
  two distinct boundary points at~$-\infty$ and~$+\infty$, while the slit
  domain on the right has a double boundary point at~$z$.}
 \label{fig:BoundaryPoints}
\end{figure}

Note that the Riemann mapping theorem is not restricted to bounded domains.
This means that domains can have well-defined boundary points at infinity. For
example, the upper half-plane has a single boundary point at~$\infty$, and the
infinite strip $\{z:0 < \Im z < \pi\}$ has two distinct boundary points
at~$-\infty$ and at~$+\infty$. By mapping these domains onto~$\D$ it can be
made explicit that these boundary points are well-defined, see
figure~\ref{fig:BoundaryPoints}.

Likewise, the example of the slit domain depicted in the figure clarifies
that one can have a multiple boundary point at some point~$z$. In the
example, when the domain is mapped onto~$\D$ all the points along the slit
will have two images on the unit circle (except for the tip of the slit), and
are therefore double boundary points. So although the preimages happen to
coincide two-by-two, it is clear that they are distinct boundary points, and
we will treat them as such. The same holds for triple boundary points and
so on (of these there can exist only countably many).

The conformal map of a domain~$D$ onto~$\D$ is unique up to composition with
a conformal self-map of the unit disk. Therefore, the Riemann mapping theorem
together with the following theorem on the conformal self-maps of the unit
disk provide the basis for the theory of conformal maps.

\begin{theorem}
 \label{the:selfmapsofD}
 The conformal self-maps of the open unit disk~$\D$ are precisely the
 transformations of the form
 \begin{equation}
  f(z) = \e{\im\phi}\frac{z-a}{1-\abar z}, \qquad |z|<1,
 \end{equation}
 where $a$ is complex, $|a|<1$, and $0\leq\phi\leq2\pi$.
\end{theorem}

It follows from this theorem that the map~$f:D\goesto\D$ is determined
uniquely if we specify three real parameters. For example, one commonly
specifies~$f(z)=0$ and $f'(z)>0$ (that is, $f'(z)$ is real and positive) at
some specific point~$z\in D$, to make the map unique. Indeed, it should be
clear from theorems \ref{the:riemannmapping} and~\ref{the:selfmapsofD} that
such a map exists. Further, if~$g$ is another map satisfying the same
conditions, then $f\circ g^{-1}$ is a conformal self-map of~$\D$ which fixes
the origin and has positive real derivative in~$0$. But then $f\circ g^{-1}$
must be the identity, by theorem~\ref{the:selfmapsofD}, whence $f=g$. The
unique number~$1/f'(z)$ in fact defines a measure for the inner radius of
the domain~$D$, called the \defn{conformal radius}, see
section~\ref{ssec:unitdiskmaps}.

Using the Riemann mapping theorem, we can also study conformal maps between
two simply connected domains $D,D'\neq\C$ in the complex plane. A conformal
map of~$D$ onto~$D'$ is easily defined through the conformal map of $D$
onto~$\D$, and the inverse of the map of~$D'$ onto~$\D$. Again, the map is
unique if we specify three real parameters. For example, if we fix two points
$z\in D$, $w\in D'$, then there is a unique conformal map~$f$ of $D$ onto~$D'$
with $f(z)=w$ and~$f'(z)>0$.

Another way commonly used to specify a map uniquely is the following. Fix
three distinct points $z_1$, $z_2$, $z_3$ ordered counter-clockwise on the
boundary of~$D$, and three distinct points $w_1$, $w_2$, $w_3$, ordered
similarly on the boundary of~$D'$. Then there is a unique conformal map~$f$
of $D$ onto~$D'$ with $f(z_i)=w_i$, $i=1,2,3$. This may not be immediately
obvious from theorem~\ref{the:selfmapsofD}, but we shall see in
section~\ref{ssec:halfplanemaps} that this follows quite easily from the
form of the conformal self-maps of the upper half-plane.

\begin{figure}
 \centering\includegraphics[scale=1.2]{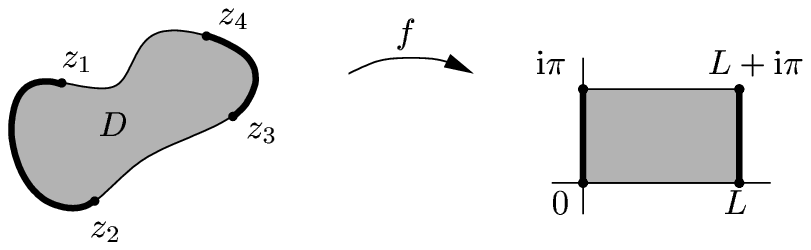}
 \caption{The $\pi$-extremal distance~$L$ between two arcs on the boundary
  of a domain~$D$ is determined by the conformal map~$f$ onto the rectangle
  of height~$\pi$.}
 \label{fig:Distance}
\end{figure}

This latter consequence of the Riemann mapping theorem suggests a way to
define a conformally invariant distance between two arcs on the boundary of a
simply connected domain~$D$. This distance is defined through the conformal
map of~$D$ onto the rectangle~$(0,L)\times(0,\im\pi)$
(see figure~\ref{fig:Distance}). Since we can choose only three real
parameters of a map, we may expect that the length~$L$ of the rectangle is
fixed uniquely. In section~\ref{ssec:rectanglemaps} we will prove that
this is indeed the case.

\begin{definition}[$\pi$-extremal distance]
 \label{def:extremaldistance}
 Let~$D$ be a simply connected domain, and let $z_1$, $z_2$, $z_3$ and~$z_4$
 be four distinct points on the boundary~$\partial D$ of~$D$, ordered
 counter-clockwise. Let~$L>0$ be the unique real number such that there is
 a conformal map~$f$ of $D$ onto the rectangle~$(0,L)\times(0,\im\pi)$ with
 $f(z_1)=\im\pi$, $f(z_2)=0$, $f(z_3)=L$ and~$f(z_4)=L+\im\pi$. Then~$L$ is
 called the \defn{$\pi$-extremal distance} between the arcs $[z_1,z_2]$
 and~$[z_3,z_4]$ on~$\partial D$.
\end{definition}

We remark that $\pi$-extremal distance is the same as $\pi$ times extremal
distance, which is itself a special case of the more general notion of
extremal length. For more information, and for some properties of extremal
distance, see Ahlfors~\cite{ahlfors:1973}. Another measure related to arcs
on the boundary of a domain~$D$ is the harmonic measure, which we define
below for a (not necessarily simply) connected domain~$D$. It is easy to
prove that the harmonic measure is invariant under conformal maps, e.g.\ by
applying the Cauchy-Riemann equations and using harmonicity of conformal maps.

\begin{definition}[Harmonic measure]
 Let $D$ be a connected domain whose boundary is continuous, and suppose
 that the boundary is divided into two parts $A$ and~$B$, each consisting
 of a finite number of arcs. Then there exists a unique bounded harmonic
 function $\omega(z)$ in~$D$ such that $\omega(z)\goesto1$ when $z$ tends
 to an interior point of~$A$ and $\omega(z)\goesto0$ when $z$ tends to an
 interior point of~$B$. The number~$\omega(z)$ is called the \defn{harmonic
 measure} of~$A$ at the point~$z$ with respect to~$D$.
\end{definition}

We complete our general introduction to conformal mapping theory with the
formulation of two basic and very useful theorems.

\begin{theorem}[Schwarz reflection principle]
 \label{the:schwarzreflection}
 Let~$D$ be a domain that is symmetric with respect to the real axis, and
 let~$D^+=D\cap\H$. Let~$f(z)$ be an analytic function on~$D^+$
 such that $\Im[f(z)]\goesto0$ as $z\in D^+$ tends to~$D\cap\R$. Then
 $f(z)$ extends to be analytic on~$D$, and the extension satisfies
 \begin{equation}
  f(\zbar)=\overline{f(z)},\qquad z\in D.
 \end{equation}
\end{theorem}

\begin{theorem}[Schwarz lemma]
 \label{the:schwarzlemma}
 Suppose that~$f(z)$ is analytic on~$\D$, that $f(0)=0$ and that
 $|f(z)|\leq1$ for~$|z|<1$. Then
 \begin{equation}
  |f(z)|\leq|z| \quad {\rm for\ } |z|<1
 \end{equation}
 and hence,
 \begin{equation}
  |f'(0)|\leq1.
 \end{equation}
 Further, if~$|f(z_0)|=|z_0|$ for some~$z_0\neq0$, then~$f(z)=\e{\im\alpha}z$
 for some real constant~$\alpha$. Moreover, $f(z)=\e{\im\alpha}z$ for some real
 constant~$\alpha$ if and only if~$|f'(0)|=1$.
\end{theorem}

\subsection{Normalized maps of the unit disk}
\label{ssec:unitdiskmaps}

In this subsection we consider two standard classes of conformal maps. The
first class is the class of one-to-one conformal maps~$f$ of~$\D$ (onto some
other domain) that are normalized by $f(0)=0$ and $f'(0)=1$. The class of
these maps is usually denoted by~$S$, and each $f\in S$ has an expansion
around $z=0$ of the form
\begin{equation}
 \label{equ:Sexpansion}
 f(z) = z + a_2 z^2 + a_3 z^3 + \ldots + a_n z^n + \ldots.
\end{equation}
The second class, denoted by~$\Sigma$, is the collection of one-to-one
maps~$F$ defined on $\{z:|z|>1\}$ that have an expansion of the form
\begin{equation}
 \label{equ:Sigmaexpansion}
 F(z) = z + \frac{b_1}{z} + \frac{b_2}{z^2} + \ldots + \frac{b_n}{z^n} + \ldots
\end{equation}
for~$z\goesto\infty$. Our purpose is to look at properties of the expansion
coefficients $a_n$ and~$b_n$, and some consequences. For more details the
reader is referred to Ahlfors~\cite{ahlfors:1973}. We start with the
class~$\Sigma$, for which the main theorem is the area theorem.

\begin{theorem}[Area theorem]
 \label{the:area}
 The coefficients in the expansion~(\ref{equ:Sigmaexpansion}) of any function
 $F\in\Sigma$ satisfy $\sum_{n=1}^\infty n|b_n|^2\leq1$.
\end{theorem}

Now we move on to the class~$S$. For this class of functions, there is a
famous conjecture of Bieberbach from 1916 on the expansion coefficients,
which was finally proved by de Branges in 1985 after many partial results.
Most notably in the present context is that L\"owner~\cite{loewner:1923}
introduced his L\"owner equation, which lies at the basis of {\SLE}, to
prove that $|a_3|\leq3$ in 1923. His method was also a key to the final
proof of the Bieberbach conjecture by de Branges.

\begin{theorem}[Bieberbach-de Branges theorem]
 \label{the:bieberbach}
 The coefficients in the expansion~(\ref{equ:Sexpansion}) of any function
 $f\in S$ satisfy $|a_n|\leq n$ for all $n\geq2$.
\end{theorem}

The following two theorems are consequences of the fact that~$|a_2|\leq2$.
The first of these theorems provides estimates for $|f(z)|$ and~$|f'(z)|$,
and is known as the Koebe distortion theorem. The second theorem is the Koebe
one-quarter theorem, which can be obtained directly from the distortion
theorem. Indeed, if we take the limit~$|z|\goesto1$ in the left-most
inequality of equation~(\ref{equ:distortion}) below, we immediately get the
desired result. The one-quarter theorem is often used in conjunction with
the Schwarz lemma to provide upper and lower bounds on some quantity.

\begin{theorem}[Koebe distortion theorem]
 \label{the:koebedistortion}
 The functions $f\in S$ satisfy
 \begin{equation}
  \label{equ:distortion}
  \frac{|z|}{(1+|z|)^2} \leq |f(z)| \leq \frac{|z|}{(1-|z|)^2},\quad
  \frac{1-|z|}{(1+|z|)^3} \leq |f'(z)| \leq \frac{1+|z|}{(1-|z|)^3}.
 \end{equation}
\end{theorem}

\begin{theorem}[Koebe one-quarter theorem]
 \label{the:koebeonequarter}
 The image of the unit disk under a mapping $f\in S$ contains the disk
 with centre~$0$ and radius~$\inv{4}$.
\end{theorem}

\begin{figure}
 \centering\includegraphics[scale=1.2]{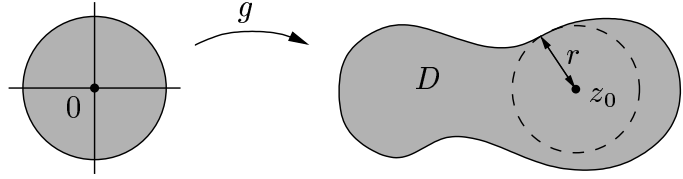}
 \caption{The in-radius~$r$ of a domain~$D$ with respect to~$z_0$, and
  the map~$g$ that defines the conformal radius with respect to~$z_0$.}
 \label{fig:In-radius}
\end{figure}
 
As an application, let us discuss the notion of conformal radius. Suppose
that~$D$ is a simply connected domain, and let~$z_0\in D$. Then the
\defn{in-radius} $r$ of~$D$ with respect to~$z_0$ is defined by $r :=
\inf\{|z-z_0|:z\not\in D\}$. It is the radius of the largest open disk
with centre~$z_0$ that fits inside~$D$, see figure~\ref{fig:In-radius}.
Now let $g$ be the conformal map of~$\D$ onto~$D$ such that $g(0)=z_0$
and $g'(0)>0$. Then the unique number $g'(0)$ is called the \defn{conformal
radius} of~$D$ with respect to~$z_0$. We now prove that this conformal
radius is determined by the in-radius up to a factor~$4$, that is,
$r\leq g'(0)\leq 4r$.

Indeed, it is clear that $g^{-1}(rz+z_0)$ is a map that satisfies the
conditions of the Schwarz lemma~\ref{the:schwarzlemma}. Therefore, it
follows that $r(g^{-1})'(z_0)\leq 1$, hence $g'(0)\geq r$. On the other
hand, the map $f(z)=(g(z)-z_0)/g'(0)$ is in~$S$, and the Koebe one-quarter
theorem says that $\inf\{|z|:z\not\in f(\D)\}\geq\inv{4}$. From this it
follows that $g'(0)\leq 4r$, and we are done.

It is clear that any conformal map $g:\D\goesto D$ can be renormalized to
yield a map $f\in S$ (onto a different domain~$D'$). We used this technique
above to prove the relation between the in-radius and the conformal radius.
Similarly, other properties of functions in~$S$ can be translated to
properties of any map~$g$ in this way.

\subsection{Conformal maps of the upper half-plane}
\label{ssec:halfplanemaps}

In this subsection we study conformal maps of a domain~$D$ onto the complex
upper half-plane~$\H$. Our first observation is that any simply connected
domain can be mapped conformally onto~$\H$. This follows from the Riemann
mapping theorem, and the fact that the map $f(w)=\im(1+w)/(1-w)$ is a
standard conformal map of~$\D$ onto~$\H$. We can also go back from the
upper half-plane to the unit disk by using the inverse map $f^{-1}(z)=
(z-\im)/(z+\im)$. In complex analysis, one often does not distinguish between
the half-plane and the unit disk, since one always has the freedom to map
conformally from the one space to the other.

Conformal maps of simply connected domains onto the upper half-plane are
unique up to composition with the conformal self-maps of the upper
half-plane. The form of these maps is given by the following theorem.

\begin{theorem}
 \label{the:selfmapsofH}
 The conformal self-maps of the upper half-plane~$\H$ are precisely the
 (fractional linear or M\"obius) transformations
 \begin{equation}
  f(z) = \frac{az+b}{cz+d}, \qquad \Im z>0,
 \end{equation}
 where $a$, $b$, $c$ and~$d$ are real numbers satisfying $ad-bc>0$.
\end{theorem}

These maps are especially effective for rearranging points on the boundary
of a domain. In particular, theorem~\ref{the:selfmapsofH} shows that the
conformal self-map of~$\H$ which takes the points~$x_1<x_2<x_3$ on the real
line to $0$, $1$ and~$\infty$, respectively, is unique. Further, the only
self-map which fixes the points $0$, $1$ and~$\infty$ is the identity. From
this one can easily deduce that any conformal map is determined uniquely
if one specifies the images of three distinct points on the boundary.

We now know how to map back and forth between the half-plane and the unit
disk, and we also know the conformal self-maps of both spaces. This
knowledge is extremely useful in deriving properties of a general conformal
map of one domain onto another. A standard procedure is to map these
domains onto $\H$ or~$\D$, and then use a conformal self-map to rearrange
the points in $\H$ or~$\D$ appropriately. As an example, let us prove
the following consequence of the Schwarz lemma.

\begin{corollary}
 \label{cor:schwarzH}
 Let~$g$ map~$\H$ into~$\H$ conformally. Then for all points $z=x+\im y\in\H$,
 \begin{equation}
  y\,|g'(z)| \leq \Im g(z).
 \end{equation}
 If~$g$ is not a conformal self-map of~$\H$, then we have strict inequality.
\end{corollary}

\begin{proof}
\begin{figure}
  \centering\includegraphics[scale=1.2]{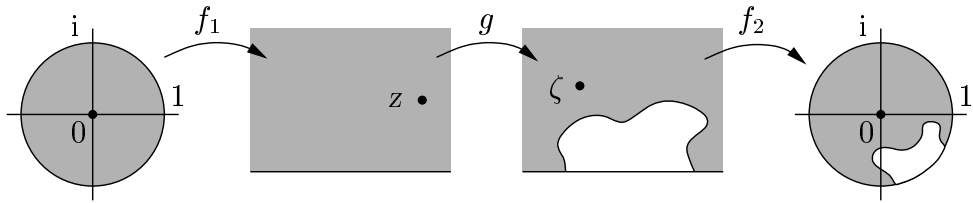}
  \caption{Illustration of how compositions can be used to derive properties
   of a conformal map~$g$.}
  \label{fig:SchwarzH}
\end{figure}
 
 The result follows by constructing a map of~$\D$ into~$\D$ satisfying
 the conditions of the Schwarz lemma~\ref{the:schwarzlemma}. First, we map
 $\D$ onto~$\H$ in such a way that~$0$ maps onto~$z$, see
 figure~\ref{fig:SchwarzH}. To find this map, we compose the standard map
 of~$\D$ onto~$\H$ with an appropriate self-map of~$\H$, which leads to
 \begin{equation}
  f_1(w) = x+\im\frac{1+w}{1-w}y, \qquad |w|<1.
 \end{equation}
 The map~$g$ then takes~$z$ to the image~$\zeta:=g(z)\in\H$. Next, we apply
 a map of~$\H$ onto~$\D$, which takes~$\zeta$ back to~$0$. To find such a
 map, we simply do a translation followed by a rescaling in the half-plane
 to move the point $\zeta$ to~$\im$, and then compose with the standard map
 of $\H$ onto~$\D$ that takes $\im$ to~$0$. This gives the map
 \begin{equation}
  f_2(w) = \frac{w-\zeta}{w-\bar{\zeta}}, \qquad w\in\H,
 \end{equation}
 Figure~\ref{fig:SchwarzH} illustrates the construction.

 Now, we note that the composite map~$f:=f_2\circ g\circ f_1$ is a map that
 satisfies the conditions of the Schwarz lemma. Indeed, $f$ is analytic on
 the unit disk, it maps~$0$ to~$0$, and it maps the unit disk into the unit
 disk (since~$g$ maps the half-plane into the half-plane). Hence, by the
 Schwarz lemma,
 \begin{equation}
  \label{equ:schwarzhproof}
  |f'(0)| = |f'_2(\zeta)|\,|g'(z)|\,|f'_1(0)| \leq 1.
 \end{equation}
 Since $f'_1(0)=2\im y$ and $f'_2(\zeta)=1/(\zeta-\bar{\zeta})$, we get
 \begin{equation}
  |f'(0)| = \frac{y}{\Im\zeta}\,|g'(z)| \leq 1.
 \end{equation}
 which is what we wanted to prove.

 Equality can only hold for a subclass of conformal self-maps of~$\H$,
 namely for those maps that correspond to rotations of the unit disk, as
 should be clear from the Schwarz lemma. Hence, if~$g$ is not a conformal
 self-map of~$\H$, we must certainly have strict inequality.
\end{proof}

\subsection{Hulls and capacity in the half-plane}
\label{ssec:hullsandcapacity}

Now let us introduce some notions and notations that are used in the
literature on \SLE. A \defn{hull} in the half-plane is a compact
set~$K\subset\close{\H}$ such that $\H\setminus K$ is simply connected
and~$K=\close{K\cap\H}$ (this latter condition ensures that~$K$ contains no
intervals of~$\R$ that are ``sticking out'' to the left or the right).
Examples of hulls in the upper half-plane are the straight line
segment~$[0,\im R]$, the closed rectangle $[0,L]\times[0,\im\pi]$ and the
closed half-disk $\{z\in\Hbar:|z|\leq R\}$.

Given a hull~$K$, according to the Riemann mapping theorem there exists
a conformal map~$g_K:\H\setminus K\goesto\H$. This conformal map can be
chosen to map infinity to infinity. Then it is clear that the map~$g_K$ has
an expansion around $z\goesto\infty$ of the form
\begin{equation}
 g_K(z) = bz + a_0 + \frac{a_1}{z} + \frac{a_2}{z^2} + \ldots.
\end{equation}
Note that the leading term must be linear in~$z$, because higher powers
of~$z$ will certainly send a part of~$\H\setminus K$ to the lower half-plane.
Further, since the map~$g_K$ maps~$\R\setminus K$ into~$\R$, the Schwarz
reflection principle applies, and the map extends to the complement
in~{\C} of
\begin{equation}
 K^\ast = \{ z : z\in K {\rm\ or\ } \zbar\in K \}.
\end{equation}
On~$\C\setminus K^\ast$, the map must satisfy~$\overline{g_K(z)}=g_K(\zbar)$,
which shows that all coefficients in the expansion of~$g_K$ have to be real.

So far, we have only specified that~$g_K$ has to map infinity to infinity.
But we would like to specify the map uniquely. Theorem~\ref{the:selfmapsofH}
tells us that this can be done by scaling and translation. A convenient choice
is to let~$g_K$ satisfy the \defn{hydrodynamic normalization}
\begin{equation}
 \lim_{z\goesto\infty} (g_K(z)-z) = 0.
\end{equation}
This fixes~$b=1$ and~$a_0=0$. The expansion of~$g_K$ around infinity is thus
of the form
\begin{equation}
 \label{equ:halfplaneexpansion}
 g_K(z) = z + \frac{a_1}{z} + \frac{a_2}{z^2} + \ldots
\end{equation}
Note that this expansion is of the same form as the expansion for functions
in the class~$\Sigma$ of section~\ref{ssec:unitdiskmaps}. This means that
we can use the area theorem to obtain bounds on the coefficients~$a_n$.

Indeed, if~$R$ denotes the radius of the hull~$K$ measured from the origin,
then the map~$g_K(Rz)/R$ is in the class~$\Sigma$. Now, as a direct
consequence of the area theorem, the coefficients~$b_n = a_n/R^{n+1}$ in the
expansion of this map around infinity satisfy~$|b_n|\leq1$ for all~$n\geq1$.
This proves the following theorem.

\begin{theorem}
 \label{the:hullarea}
 Let~$R$ be the radius of the hull~$K$ measured from the origin. Then the
 coefficients in the expansion~(\ref{equ:halfplaneexpansion}) satisfy
 $a_n\leq R^{n+1}$.
\end{theorem}

The coefficient~$a_1$, which depends only on~$K$ ($a_1=a_1(K)$), is called
the \defn{capacity} of the hull~$K$ in the half-plane~$\H$. It is clearly
invariant under translations of the hull over the real line. Thus, if~$R$
is the radius of the smallest half-disk centred on the real line that
contains~$K$, then $a_1(K)\leq R^2$ by the previous theorem. In the following
paragraphs, three more important properties of capacity will be derived.

\textbf{Positivity.}
The capacity of a nonempty hull~$K$ is a positive number, which we can prove
as follows. Observe that the map~$g_K^{-1}$ is a map of the half-plane~$\H$
into itself. Suppose that we now set~$z:=g_K(\im y)$ (where $y$ is large, and
will be sent to infinity later). Substituting this into
corollary~\ref{cor:schwarzH} gives
\begin{equation}
 \Im[g_K(\im y)]\,\Big|\big(g_K^{-1}\big)'\big(z\big)\Big| < y
 \quad\mbox{or}\quad
 y^2 - y \frac{\Im[g_K(\im y)]}{|g_K'(\im y)|} > 0.
\end{equation}
If one now uses the expansion of~$g_K$ around infinity and takes
$y\goesto\infty$, the result~$a_1(K)>0$ follows immediately.

\begin{figure}
 \centering\includegraphics[scale=1.2]{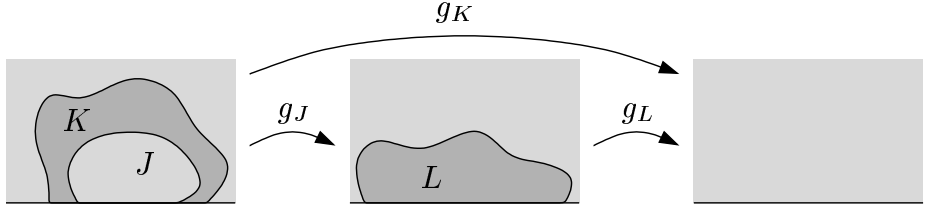}
 \caption{The capacities of two hulls~$J\subset K$ are related through a
  third hull~$L$, which is the closure of the image of~$K\setminus J$ under
  the map~$g_J$.}
 \label{fig:Composition}
\end{figure}

\textbf{Scaling rule.}
Consider the hull~$rK$ where~$r>0$, and the conformal map~$g_{rK}$ that
corresponds to this hull. It is obvious that another conformal map
of~$\H\setminus(rK)$ onto~$\H$ is given by~$g_K(z/r)$. We can easily make
this map satisfy the hydrodynamic normalization by multiplying it by a
factor~$r$, as follows from the expansion around
infinity~(\ref{equ:halfplaneexpansion}). But because the map~$g_{rK}$ of
$\H\setminus(rK)$ onto~$\H$ that satisfies the hydrodynamic normalization
is unique, the above implies that~$g_{rK}(z)=rg_K(z/r)$. Hence we obtain the
scaling relation
\begin{equation}
 \label{equ:capacityscaling}
 a_1(rK) = r^2 a_1(K)
\end{equation}
for the capacity of the hull~$K$, again using the
expansion~(\ref{equ:halfplaneexpansion}).

\textbf{Summation rule.}
The summation rule for capacities follows by considering two hulls $J$
and~$K$ in the upper half-plane such that~$J\subset K$. The corresponding
conformal maps are $g_J$ and~$g_K$. We can define a third hull~$L$
by~$L:=\close{g_J(K\setminus J)}$, which has associated with it a conformal
map~$g_L$. The conformal maps are related by~$g_K=g_L\circ g_J$, see
figure~\ref{fig:Composition}, because both $g_K$ and $g_L\circ g_J$ map
$\H\setminus K$ onto~$\H$ and satisfy the hydrodynamic normalization.
Inserting the expansions around infinity, we easily obtain
\begin{equation}
 \label{equ:capacitysum}
 a_1(K) = a_1(J) + a_1(L).
\end{equation}
Thus, if we have two hulls~$J\subset K$, the capacity of the larger hull is
the sum of the capacities of the smaller hull and a third hull~$L:=
\close{g_J(K\setminus J)}$.

We conclude this subsection with the Poisson integral representation of the
map~$g_K$ for a given hull~$K$. We should note that in text-books the Poisson
formula is often only discussed for the unit disk, while the half-plane case
is left as an exercise (see Ahlfors~\cite{ahlfors:1966} chapter~4, sections
6.3 and~6.4 and Gamelin~\cite{gamelin:2000} chapter~X, section~1 and the
exercises following these sections in both books). In the half-plane, the
Poisson integral formula tells us that
\begin{equation}
 \label{equ:invPoissonrep}
 z-g_K^{-1}(z) = \inv{\pi}\int_{-\infty}^\infty
  \frac{\Im g_K^{-1}(\xi)}{z-\xi}\,\dif{\xi}, \qquad z\in\H
\end{equation}
or, upon replacing $z$ by $g_K(z)$,
\begin{equation}
 \label{equ:Poissonrep}
 g_K(z)-z = \inv{\pi}\int_{-\infty}^\infty
  \frac{\Im g_K^{-1}(\xi)}{g_K(z)-\xi}\,\dif{\xi}, \qquad z\in\H\setminus K.
\end{equation}
Moreover, when we multiply both sides of~(\ref{equ:Poissonrep}) by~$z$
and send~$z$ to infinity we obtain the following expression for the
capacity of~$K$:
\begin{equation}
 \label{equ:capacityPoissonrep}
 a_1(K) = \inv{\pi}\int_{-\infty}^\infty \Im g_K^{-1}(\xi)\,\dif{\xi}.
\end{equation}

\subsection{Mapping rectangles onto the upper half-plane}
\label{ssec:rectanglemaps}

\begin{figure}
 \centering\includegraphics[scale=1.2]{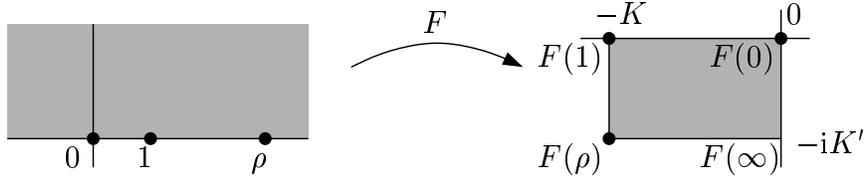}
 \caption{How~$F(w)$ of theorem~\ref{the:rectanglemap} maps the upper
  half-plane to a rectangle.}
 \label{fig:Rectangle}
\end{figure}

In this subsection we study conformal maps of rectangles onto the upper
half-plane. We are interested in these maps for two reasons. The first
reason is that the notion of $\pi$-extremal distance plays a key role in
{\SLE}, and $\pi$-extremal distance is defined through mappings to rectangles.
The second reason is that we want to study critical exponents for the
crossing of rectangles by an {\SLE} process, and we do so by mapping this
problem to the half-plane. We start our discussion from the following basic
theorem on the mapping of rectangles, which can be found in
Ahlfors~\cite{ahlfors:1966} (see figure~\ref{fig:Rectangle} for an
illustration of the map).

\begin{theorem}
 \label{the:rectanglemap}
 Let~$w\in\H$, and define the map~$F(w)$ by the elliptic integral
 \begin{equation}
  F(w) = \int_0^w \frac{\dif{z}}{\sqrt{z(z-1)(z-\rho)}},
 \end{equation}
 where $1<\rho\in\R$, and $\sqrt{z}$, $\sqrt{z-1}$ and $\sqrt{z-\rho}$ take
 on values in the first quadrant. Then $F(w)$ is the conformal map of~$\H$
 onto the rectangle $(-K,0)\times(0,-\im K')$, where
 \begin{equation}
  \label{equ:K}
  K\phantom{'} = \int_0^1 \frac{\dif{t}}{\sqrt{t(1-t)(\rho-t)}}, \quad
  \label{equ:Kprime}
  K'= \int_1^\rho \frac{\dif{t}}{\sqrt{t(t-1)(\rho-t)}},
 \end{equation}
 and $F(0)=0$, $F(1)=-K$, $F(\rho)=-K-\im K'$ and~$F(\infty)=-\im K'$.
\end{theorem}

Maps of~$\H$ onto other rectangles can be obtained from the map in the
theorem by scaling, rotation and translation. We can also rearrange the
points on the real line that map to the corners of the rectangle, by
composition with a conformal self-map of the upper half-plane. The proofs
of the following lemma and its corollary make use of these techniques.

\begin{lemma}
 \label{lem:rectanglemap}
 Let~$\Phi(z)$ be the conformal map of the rectangle $(0,L)\times(0,\im\pi)$
 onto~$\H$ such that $\Phi(\im\pi)=0$, $\Phi(0)=1$ and~$\Phi(L+\im\pi)=\infty$.
 Then $\Phi$ maps $L$ onto some point $\rho>1$ on the real line. $L$ is
 monotone increasing with~$\rho$ and, moreover, $L = \log\rho + O(1)$
 as~$L\goesto\infty$.
\end{lemma}

\begin{proof}
 From theorem~\ref{the:rectanglemap}, we know that the inverse of the
 map~$\Phi$ is of the form
 \begin{equation}
  \Phi^{-1}(w) = \im aK + \im aF(w),
 \end{equation}
 with~$a$ a scaling factor and~$K$ and~$F(w)$ as in the theorem. The aspect
 ratio of the rectangle is given by $L/\pi=K'/K$. To analyse the behaviour
 of this aspect ratio as a function of~$\rho$, observe that
 \begin{equation}
  \label{equ:Kintegral}
  \sqrt{\rho}K = \int_0^1 \frac{\dif{t}}{\sqrt{t(1-t)(1-t/\rho)}},
 \end{equation}
 while by the substitution $u = (t-1)/(\rho-1)$ we can transform the integral
 for~$K'$ into
 \begin{equation}
  \label{equ:Kprimeintegral}
  \sqrt{\rho}K'= \int_0^1 \frac{\dif{u}}{\sqrt{u(1-u)(u+(1-u)/\rho)}}.
 \end{equation}
 Clearly, $1-t/\rho$ in equation~(\ref{equ:Kintegral}) is increasing
 with~$\rho$, while $u+(1-u)/\rho$ in equation~(\ref{equ:Kprimeintegral}) is
 decreasing with~$\rho$. Hence the aspect ratio $K'/K$ is monotone
 increasing with~$\rho$.

 Moreover, equations 15.3.1 and 15.1.1 in reference~\cite{hypergeometric}
 show that
 \begin{equation}
  \sqrt{\rho}K = \Hypergeom\left(\ahalf,\ahalf;1;\frac{1}{\rho}\right)\,\pi
   = \pi + O(\rho^{-1})\quad\mbox{as\ }\rho\goesto\infty,
 \end{equation}
 while equations 15.3.1, 15.3.4 and 15.3.10 in reference~\cite{hypergeometric}
 give
 \begin{equation}
  \sqrt{\rho} K'
   = \Hypergeom\left(\ahalf,\ahalf;1;1-\frac{1}{\rho}\right)\,\pi
   = \log\rho + O(1)\quad\mbox{as\ }\rho\goesto\infty.
 \end{equation}
 Since $L=\pi K'/K$, the lemma follows.
\end{proof}

\begin{corollary}
 \label{cor:rectanglemap}
 Let~$\Psi(z)$ be the conformal map of the rectangle $(0,L)\times(0,\im\pi)$
 onto~$\H$ such that $\Psi(0)=1$, $\Psi(L)=\infty$ and~$\Psi(L+\im\pi)=0$. Then
 $\Psi$ maps~$\im\pi$ onto some point~$\xi\in(0,1)$, and
 $L = -\log(1-\xi) + O(1)$ as~$L\goesto\infty$.
\end{corollary}

\begin{proof}
 The map~$\Psi$ is obtained from the map~$\Phi$ of
 lemma~\ref{lem:rectanglemap} by composition of~$\Phi$ with the conformal
 self-map $(\rho-1)/(\rho-z)$ of the upper half-plane. This self-map sends
 $1$ to $1$, $\rho$ to infinity, and infinity to $0$, as required. It also
 sends $0$ to~$\xi=1-\rho^{-1}$. It follows that $\rho=(1-\xi)^{-1}$, and
 lemma~\ref{lem:rectanglemap} gives the result.
\end{proof}

Lemma~\ref{lem:rectanglemap} tells us that the aspect ratio~$K'/K$ of the
rectangle of theorem~\ref{the:rectanglemap} is monotone increasing
with~$\rho$. The reader may verify that this implies that the $\pi$-extremal
distance between two arcs on the boundary of a domain (recall
definition~\ref{def:extremaldistance}) is indeed determined uniquely.


\section{Theory of stochastic processes}
\label{sec:stochasticprocesses}

This appendix is devoted to the background theory of stochastic processes
that is required in the study of {\SLE}. Because not all readers may be
familiar with the measure-theoretic approach to probability theory, we will
review this background in the first two subsections. The first of these
subsections deals with probability spaces and random variables, and the
second with conditional probability and expectation. Our presentation of
this material is based on the book by Ash and Dol\'eans-Dade~\cite{ash:2000}.
Then in the third subsection we will discuss stochastic processes, and in
section~\ref{ssec:brownianmotion} we will treat Brownian motion. In
section~\ref{ssec:martingalesandOST} we go into the topic of martingales and
optional sampling. Finally, we discuss the It\^o calculus and stochastic
differential equations. References for these sections are the books by
Grimmett and Stirzaker, Lawler, and Gardiner \cite{gardiner:1983,
grimmett:2001,lawler:1995}.

\subsection{Measure-theoretic background}
\label{ssec:measuretheory}

This subsection is intended for readers who are not familiar with measure
theory or the measure-theoretic approach to probability theory. The theory
is rather technical, and we do not intend to go into all details here. Our
discussion is based on the book by Ash and Dol\'eans-Dade~\cite{ash:2000},
and we refer to this work for a completely rigorous treatment.

Suppose that we perform a random experiment. Then all possible outcomes of
this experiment together constitute a set~$\Omega$, which we call the
\defn{sample space}. It is usually treated as an abstract space, which is
often not defined explicitly but is simply assumed to exist. In probability
theory, we are typically interested in the probability that the actual
outcome of our experiment is in some given subset~$A$ of~$\Omega$, called
an \defn{event}. We denote the set of all such events by~$\fld{F}$. This
set~$\fld{F}$ has to satisfy certain conditions, imposed by our requirement
that all its elements are events whose probabilities we can talk about.

To be precise, we require~$\fld{F}$ to be a \defn{$\sigma$-field} over the
sample space~$\Omega$. This means that~$\fld{F}$ is a collection of subsets
of~$\Omega$ satisfying the conditions
\begin{enumerate}
 \item $\emptyset\in\fld{F}$;
 \item if $A\in\fld{F}$ then $\Omega\setminus A\in\fld{F}$;
 \item if $A_1,A_2,\ldots\in\fld{F}$ then $\cup_{i=1}^\infty A_i\in\fld{F}$.
\end{enumerate}
If~$\fld{G}\subset\fld{F}$ is another set satisfying the same conditions,
then $\fld{G}$ is called a \defn{sub-$\sigma$-field} of~$\fld{F}$. The
combination $(\Omega,\fld{F})$ of a set~$\Omega$ and a $\sigma$-field
$\fld{F}$ over~$\Omega$ is called a \defn{measurable space}.

So, we have now associated with our random experiment a measurable space
consisting of the sample space~$\Omega$ and a collection of events~$\fld{F}$.
To talk about probabilities, we introduce a \defn{probability measure}~$\Prob$
on the space $(\Omega,\fld{F})$. This is a function assigning a number in the
range~$[0,1]$ to every element of the $\sigma$-field~$\fld{F}$. It has to
satisfy the conditions
\begin{enumerate}
 \item $\Prob[\emptyset] = 0$, $\Prob[\Omega] = 1$;
 \item if $A_1,A_2\ldots\in\fld{F}$ are disjoint, then
  $\Prob[\cup_{i=1}^\infty A_i] = \sum_{i=1}^\infty \Prob[A_i]$.
\end{enumerate}
The triple $(\Omega,\fld{F},\Prob)$ is called a \defn{probability space}.
An event whose probability is zero is called a \defn{null event}, events
which occur with probability one are said to occur \defn{almost surely}
(abbreviated \defn{a.s.}).

At this point we would like to make the following technical remark. For a
reader with no prior knowledge of measure theory it might not be clear why
we introduced the $\sigma$-field~$\fld{F}$: why do we not just define our
probability measure on the collection of all subsets of~$\Omega\,$? The point
is that in general, not every choice of $\sigma$-field over a given
space~$\Omega$ admits the definition of an appropriate measure. For example,
there exists no translation-invariant measure (except the null-measure) on
the collection of all subsets of~$\R$, which assigns a finite number to all
bounded intervals of~$\R$ (exercise~6 in section~1.4 of~\cite{ash:2000}).
In general, we therefore have to restrict ourselves to a smaller collection
of events to keep everything consistent.

To illustrate this point further, suppose that our experiment consists in
drawing a random number taking values in the real line, so that we can
take~$\Omega=\R$. Then we will typically be interested in the probability
that the number is in some interval, or in some union of intervals, and so
on. As our $\sigma$-field we may therefore take the smallest collection of
subsets of~$\R$ which is a $\sigma$-field and which contains all intervals
$\roival{a,b}$ ($a,b\in\R$), say. This set is called the \defn{Borel
$\sigma$-field} over~$\R$, denoted by $\fld{B}=\fld{B}(\R)$, and its elements
are called \defn{Borel sets}. It contains all open, closed and compact subsets
of~$\R$, and it is the natural $\sigma$-field to work with over the real
line. In higher dimensions we can give a similar definition of the Borel
sets, and we still denote this collection of sets by~$\fld{B}$, since the
underlying space is usually clear from the context.

Now that we have captured the description of a random experiment in terms
of a probability space, we can introduce random variables. A \defn{random
variable}~$X$ on a given probability space~$(\Omega,\fld{F},\Prob)$ is
defined as a map $X:(\Omega,\fld{F})\goesto(\R,\fld{B})$. What we mean by
this notation is that~$X$ is a function assigning a real number to every
element of~$\Omega$, which has the additional property that it is
\defn{measurable} with respect to~$\fld{F}$: for every Borel set $B$, the
set $\{\omega:X(\omega)\in B\}$ must be an element of~$\fld{F}$. This
measurability ensures that the probability of all events involving~$X$ is
determined.

Indeed, it should be clear that if~$X$ is measurable, then it induces a
probability measure~$\Prob_{\!X}$ on the space~$(\R,\fld{B})$ turning it
into a probability space. The measure $\Prob_{\!X}$ is of course defined
by setting $\Prob_{\!X}(B)=\Prob[X\in B]$ for every $B\in\fld{B}$, where
$\Prob[X\in B]$ is the natural shorthand notation for $\Prob[\{\omega:
X(\omega)\in B\}]$ (we will keep on using such shorthand notations from now
on). Concepts involving the random variable~$X$ can be defined both on the
probability space $(\Omega,\fld{F},\Prob)$ and on $(\R,\fld{B},\Prob_{\!X})$.
For example, the distribution function of~$X$ is defined by
\begin{equation}
 F_X(x) := \Prob\big[X\leq x\big] = \Prob_{\!X}\big[\loival{-\infty,x}\big]
\end{equation}
and its expectation value is defined by
\begin{equation}
 \Exp[X] := \int_\Omega X(\omega)\,\dif{\Prob(\omega)}
  = \int_{\R} x\,\dif{\Prob_{\!X}(x)}.
\end{equation}
The reader is reminded that these two integrals are not ordinary Riemann
integrals, but they are Lebesgue integrals with respect to the measures
$\Prob$ and~$\Prob_{\!X}$, respectively.

The concept of a random variable can be generalized to that of a \defn{
random object} (or \defn{abstract random variable}). A random object~$X$
on a given probability space $(\Omega,\fld{F},\Prob)$ is a function
$X:(\Omega,\fld{F})\goesto(\Omega',\fld{F}')$, where $\Omega'$ is the
\defn{state space} of the object, and $\fld{F}'$ is an appropriate
$\sigma$-field over~$\Omega'$. Measurability in this case ensures that
$\{X\in B\}\in\fld{F}$ for every $B\in\fld{F}'$, and guarantees that~$X$
induces a probability measure~$\Prob_{\!X}$ on~$\fld{F}'$.

Now let us consider the collection of sets $\big\{\,\{X\in B\}:B\in\fld{F}'
\big\}$ in more detail. It is an easy exercise to show that this collection
of sets is a $\sigma$-field. We call it the $\sigma$-field \emph{generated by}
the random object~$X$, and denote it by $\sigma(X)$. Loosely speaking, it
is the smallest sub-$\sigma$-field of~$\fld{F}$ containing all information
about~$X$. An important property of this $\sigma$-field, that may help
elucidate its meaning, is the following. Suppose that $Z$ is a random
variable on $(\Omega,\fld{F},\Prob)$, which is measurable with respect
to~$\sigma(X)$. Then~$Z$ is a \emph{function} of~$X$, that is, there exists
some $f:(\Omega',\fld{F}')\goesto(\R,\fld{B})$ such that $Z(\omega)=
(f\circ X)(\omega)$. Conversely, for every such function~$f$, the random
variable $Z=f\circ X$ is measurable with respect to~$\sigma(X)$.

Suppose now that instead of a single random object, we are given a
collection $\{X_i:i\in I\}$ of random objects, where~$I$ is an arbitrary
index set and $X_i:(\Omega,\fld{F})\goesto(\Omega_i,\fld{F}_i)$. Then we
define $\sigma(X_i:i\in I)$ as the smallest sub-$\sigma$-field of~$\fld{F}$
containing all events of the form~$\{X_i\in B\}$ with $i\in I$ and
$B\in\fld{F}_i$. This set is called the $\sigma$-field \emph{generated by}
the random objects $\{X_i:i\in I\}$. Again, we have that every random
variable which is measurable with respect to $\sigma(X_i:i\in I)$ is a
function of these random objects and conversely, that every function of
these random variables is measurable with respect to $\sigma(X_i:i\in I)$.

We conclude this introductory subsection with an important example of a
random variable, namely the  \defn{indicator} (or \defn{indicator
function})~$1_A$ of an event~$A\in\fld{F}$. This is the random variable
defined by setting
\begin{equation}
 1_A(\omega) := \left\{
  \begin{array}{ll}
   1 & \mbox{if }\omega\in A; \\
   0 & \mbox{if }\omega\not\in A.
  \end{array}
 \right.
\end{equation}
In this case the state space is~$\Omega'=\{0,1\}$, and the $\sigma$-field
$\fld{F}'$ consists of all subsets of~$\Omega'$. The function of this random
object is to indicate whether the outcome of our random experiment is in~$A$.

\subsection{Conditional probability and expectation}
\label{ssec:conditional}

In this subsection we review the general notions of conditional probability
and conditional expectation on a given probability space~$(\Omega,\fld{F},
\Prob)$. The presentation of this material again follows the book by Ash
and Dol\'eans-Dade~\cite{ash:2000}. As our starting point, consider the
conditional probability of an event~$B$ given that the random object~$X$
takes the value~$x$. We would like this conditional probability to behave
like the function~$g_B(x)$ in the following theorem:

\begin{theorem}
 \label{the:conditionalprob}
 Let $X:(\Omega,\fld{F})\goesto(\Omega',\fld{F}')$ be a random object, and
 let $B\in\fld{F}$. Then there exists a function $g_B:(\Omega',\fld{F}')
 \goesto(\R,\fld{B})$ such that
 \begin{equation}
  \Prob[\{X\in A'\}\cap B] = \int_{A'} g_B(x)\,\dif{\Prob_{\!X}(x)}
   \qquad \mbox{for all }A'\in\fld{F}'.
 \end{equation}
 Furthermore, if~$h_B$ is another such function, then $g_B=h_B$ a.e.
\end{theorem}

Here, a.e.\ stands for ``almost everywhere''. This means that there is a
set~$N\in\fld{F}'$ whose measure is zero (where the relevant measure in
this case is given by~$\Prob_{\!X}$), such that $g_B=h_B$ outside~$N$. It
is clear that in probability theory we can expect many equalities to hold
only ``almost everywhere'', and this qualification is therefore usually
not mentioned explicitly. In this article we write ``a.e.'' explicitly
only for the duration of this subsection. Returning to
theorem~\ref{the:conditionalprob}, we see that the function~$g_B$ is
essentially unique (up to null events), and we define the conditional
probability of~$B$ given $\{X=x\}$, written $\Prob[B\mid X=x]$, as $g_B(x)$.

The conditional probability defined in this way reduces to the definition
we would give intuitively for simple cases. For example, suppose that~$A$
is an event having positive probability. Then we can take $X=1_A$ in the
definition, and write $\Prob[B\mid A]=\Prob[B\mid X=1]$. The reader may
verify that theorem~\ref{the:conditionalprob} then gives
\begin{equation}
 \Prob[B\mid A] = \frac{\Prob[B\cap A]}{\Prob[A]} \quad\mbox{a.e.}
\end{equation}
as we expect. But whereas the intuitive definition becomes problematic
when~$\Prob[A]=0$, theorem~\ref{the:conditionalprob} shows that it is in
fact possible to extend the definition to cover this case. Moreover, the
theorem shows that we can define $\Prob[B\mid X=x]$ for an arbitrary
random object~$X$, even when $\{X=x\}$ has probability zero for some, and
possibly all, $x$.

Having dealt with the conditional probability given $\{X=x\}$, we now
consider conditional \emph{expectation} given $\{X=x\}$. For a given random
variable~$Y$, we define $\Exp[Y\mid X=x]$ as the essentially unique
function $g_Y(x)$ in the theorem below. It can be shown that again this
definition corresponds with our intuition in simple cases. The reader may
also verify, by setting $Y=1_B$, that the theorem gives $\Exp[1_B\mid X=x]
=\Prob[B\mid X=x]$ a.e.

\begin{theorem}
 \label{the:conditionalexp1}
 Let~$Y$ be a random variable, and $X:(\Omega,\fld{F})\goesto
 (\Omega',\fld{F}')$ a random object. If $\Exp[Y]$ exists, then there is a
 function $g_Y:(\Omega',\fld{F}')\goesto(\R,\fld{B})$ such that
 \begin{equation}
  \int_{\{X\in A'\}} Y(\omega)\,\dif{\Prob(\omega)}
   = \int_{A'} g_Y(x)\,\dif{\Prob_{\!X}(x)}
   \qquad \mbox{for all }A'\in\fld{F}'.
 \end{equation}
 Furthermore, if~$h_Y$ is another such function, then $g_Y=h_Y$ a.e.
\end{theorem}

We now make the generalization to conditional expectation given a
$\sigma$-field. As a motivation for our approach, observe that the conditional
expectation given by~$g_Y$ in theorem~\ref{the:conditionalexp1} is defined
on the space~$(\Omega',\fld{F}')$. But we can turn it into a random
variable~$h_Y$ on the space~$(\Omega,\fld{F})$ by defining $h_Y(\omega)
:=g_Y(X(\omega))$:
\begin{center}
 \begin{picture}(220,45)
  \put(50,25){\makebox(0,0)[rb]{$(\Omega,\fld{F})$}}
  \put(110,25){\makebox(0,0)[b]{$(\Omega',\fld{F}')$}}
  \put(170,25){\makebox(0,0)[lb]{$(\R,\fld{B})$}}
  \put(55,30){\vector(1,0){35}}\put(70,35){\makebox(0,0)[b]{$X$}}
  \put(130,30){\vector(1,0){35}}\put(145,35){\makebox(0,0)[b]{$g_Y$}}
  \put(55,27){\line(2,-1){10}}
  \qbezier(65,22)(110,-0.5)(155,22)
  \put(155,22){\vector(2,1){10}}
  \put(110,0){\makebox(0,0)[b]{$h_Y$}}
 \end{picture}
\end{center}
Then~$h_Y(\omega)$ is the conditional expectation of~$Y$, given that~$X$
takes the value $x=X(\omega)$, and one can prove that
\begin{equation}
 \int_C Y(\omega)\,\dif{\Prob(\omega)}
  = \int_C h_Y(\omega)\,\dif{\Prob(\omega)}
  \qquad\mbox{for all }C\in\sigma(X).
\end{equation}
The random variable~$h_Y$ is a special instance of a conditional expectation
given a $\sigma$-field, namely $h_Y=\Exp[Y\mid\sigma(X)]$, which we usually
write conveniently as $h_Y=\Exp[Y\mid X]$. The general case is given by the
following theorem.

\begin{theorem}
 \label{the:conditionalexp}
 Suppose that~$\fld{G}$ is some general sub-$\sigma$-field of~$\fld{F}$.
 Let~$Y$ be a random variable such that $\Exp[Y]$ exists. Then there is a
 function $\Exp[Y\mid\fld{G}]:(\Omega,\fld{G})\goesto(\R,\fld{B})$, called
 the conditional expectation of~$Y$ given~$\fld{G}$, such that
 \begin{equation}
  \int_C Y(\omega)\,\dif{\Prob(\omega)} = \int_C \Exp[Y\mid\fld{G}](\omega)
  \,\dif{\Prob(\omega)}\qquad\textrm{for all }C\in\fld{G}.
 \end{equation}
 Moreover, any two such functions must coincide almost everywhere.
\end{theorem}

Note that we can not just take~$\Exp[Y\mid\fld{G}]=Y$ in the theorem, because
$\Exp[Y\mid\fld{G}]$ is required to be measurable with respect to~$\fld{G}$,
while~$Y$ is only required to be measurable with respect to~$\fld{F}$. In
particular, if $\fld{G}$ is the $\sigma$-field generated by a collection of
random variables, then $\Exp[Y\mid\fld{G}]$ must be a \emph{function} of these
variables. As before, we further have that $\Exp[1_B\mid\fld{G}]$ is the
conditional probability of~$B$ given the $\sigma$-field~$\fld{G}$, that is,
$\Prob[B\mid\fld{G}]=\Exp[1_B\mid\fld{G}]$. We now conclude this subsection
by stating some properties of conditional expectations given a $\sigma$-field.

\begin{theorem}
 \label{the:conditionalexpproperties}
 Let $Y$ and~$Z$ be random variables such that $\Exp[Y]$ and~$\Exp[Z]$ exist,
 and let $\fld{G}$ and $\fld{H}$ be sub-$\sigma$-fields of~$\fld{F}$. Then
 \begin{enumerate}
  \item $\Exp\big[\Exp[Y\mid\fld{G}]\big]=\Exp[Y]$;
  \item $\Exp\big[\Exp[Y\mid\fld{H}]\mathrel{\big|}\fld{G}\big]=
  	\Exp\big[\Exp[Y\mid\fld{G}]\mathrel{\big|}\fld{H}\big]=
	\Exp[Y\mid\fld{G}]$ a.e.\ if
   $\fld{G}\subseteq\fld{H}$;
  \item $\Exp[YZ\mid\fld{G}]=Z\,\Exp[Y\mid\fld{G}]$ a.e.\ if $Z$ is measurable
   with respect to~$\fld{G}$ and~$\Exp[YZ]$ exists.
 \end{enumerate}
\end{theorem}

\subsection{Stochastic processes and stopping times}
\label{ssec:stochasticbasics}

A \defn{stochastic process} is a family $X=\{X_t:t\in I\}$ of random
objects $X_t:(\Omega,\fld{F})\goesto(\Omega',\fld{F}')$ on the underlying
probability space $(\Omega,\fld{F},\Prob)$. Each of the random objects has
the same state space~$\Omega'$, so that we may refer to~$\Omega'$ or even
$(\Omega',\fld{F}')$ as the state space of the process. In this article we
only consider continuous stochastic processes, for which the index set~$I$
is $\roival{0,\infty}$. We then think of the index~$t\in I$ as time. For a
fixed~$\omega\in\Omega$, the collection $\{X_t(\omega):t\geq0\}$ is a
\defn{sample path} describing one of the possible ways in which the process
can evolve in time.

The stochastic process~$X$ is called a \defn{Markov process} if it satisfies
the Markov property. For continuous stochastic processes this is to say that
\begin{equation}
 \Prob[X_t\in B\mid X_{t_1}=x_1,\ldots,X_{t_n}=x_n]
  = \Prob[X_t\in B\mid X_{t_n}=x_n]
\end{equation}
for all $B\in\fld{F}'$, all possible states $x_1,\ldots,x_n$ and any
sequence of times $t_1<\cdots<t_n<t$. In addition, the process is called
\defn{time homogeneous} if these transition probabilities depend only on
$t-t_n$ and not on~$t_n$. The Markov condition is equivalent to the more
formal statement that for all $A\in\sigma(X_s:0\leq s\leq t)$ and
$B\in\sigma(X_s:s\geq t)$
\begin{equation}
 \Prob[B\mid A,X_t] = \Prob[B\mid X_t].
\end{equation}
In words, the Markov property states that, conditional on the present value
of the process, the future is independent of the past.

The $\sigma$-field $\sigma(X_s:0\leq s\leq t)$ we encountered above is called
the $\sigma$-field \emph{generated by} the stochastic process up to time~$t$.
This $\sigma$-field is commonly denoted by $\fld{F}_t$. The whole family
$\fil{F} = \{\fld{F}_t:t\geq0\}$ of these $\sigma$-fields constitutes a
\defn{filtration}, which is to say that $\fld{F}_s\subseteq\fld{F}_t$ if
$s\leq t$. Loosely speaking, it is a growing collection of all information
about the process~$X$ up to a given time.

Often, we will be interested in the value of some expression at a random
time~$T$, which is determined by the past and present state of the
process~$X$, but does not depend on the future. Such a random time is
called a \defn{stopping time}. More precisely, a random variable~$T$ taking
values in $[0,\infty]$ is called a stopping time with respect to the
filtration $\fil{F}$ if the event $\{T \leq t\}$ is in $\fld{F}_t$ for
all~$t\geq0$. In words, this says that we should be able to decide on
the basis of the sample path of the process up to time~$t$, whether the
stopping time has passed.

An important example of a stopping time is the first time when the
process~$X$ hits some subset~$A\in\fld{F}'$ of the state space. More
precisely, if we define $T:=\inf\{t:X_t\in A\}$, then $T$ is a stopping time
if $\{T\leq t\}$ is in $\fld{F}_t$ for all~$t\geq0$.  We sometimes call such
a stopping time a \defn{hitting time}. In practice we often consider
real-valued or complex-valued stochastic processes having continuous sample
paths, and for these processes times such as the~$T$ defined above usually
are indeed stopping times.

Suppose now that~$X$ is a Markov process, and that~$T$ is a stopping time for
this process. Then the process~$X$ is said to have the \defn{strong Markov
property} if, given the value of~$X_T$, the process after time~$T$ is again
a Markov process which is independent from the events prior to~$T$, and if
this post-$T$ process has the same transition probabilities as the process~$X$.
More formally, if we write $g_{t,B}(x):=\Prob[X_t\in B\mid X_0=x]$, then~$X$
has the strong Markov property if for all stopping times~$T$,
\begin{equation}
 \Prob[X_{T+t}\in B\mid\fld{F}_T] = g_{t,B}(X_T)
\end{equation}
for any~$t>0$ and $B\in\fld{F'}$. Here, $\fld{F}_T$ is the $\sigma$-field of
all events that are \defn{prior to}~$T$. These are the events~$A\in\fld{F}$
such that $A\cap\{T\leq t\}\in\fld{F}_t$ for all~$t\geq0$ (the reader may
check that for $T=t$ fixed, $\fld{F}_T$ is just $\fld{F}_t$).

We end this subsection with some remarks about the filtration~$\fil{F}$. It
is common practice to work with $\sigma$-fields~$\fld{F}_t$ that are somewhat
larger than the $\fld{F}_t$ we have been considering above. These larger
$\sigma$-fields arise out of the following three assumptions:
\begin{enumerate}
 \item The probability space $(\Omega,\fld{F},\Prob)$ is \defn{complete}, in
  the sense that every subset of a null event is itself an event.
 \item The~$\fld{F}_t$ contain all null events.
 \item The~$\fld{F}_t$ are right-continuous: $\fld{F}_t=\fld{F}_{t+}$
  where $\fld{F}_{t+}=\cap_{s>0}\fld{F}_{t+s}$.
\end{enumerate}
Under the last assumption, the condition for~$T$ to be a stopping time is
equivalent to requiring that~$\{T<t\}\in\fld{F}_t$ for all~$t\geq0$. In the
literature stopping times are sometimes \emph{defined} by this latter
condition.

\subsection{Brownian motion and Brownian excursions}
\label{ssec:brownianmotion}

A key role in this article is played by Brownian motion. In this subsection
we shall look at the definition of standard Brownian motion, and then
explore some basic properties. \defn{Standard Brownian motion} is defined
as a real-valued stochastic process $\{B_t : t\geq0\}$ which satisfies
the following conditions:
\begin{enumerate}
 \item for any $0 \leq s_1 < t_1 \leq s_2 < t_2 \leq \ldots \leq s_n < t_n$
  the random variables $B_{t_1}-B_{s_1},B_{t_2}-B_{s_2},\ldots,B_{t_n}-B_{s_n}$
  are independent;
 \item for any~$s<t$, the random variable $B_t-B_s$ is normally distributed
  with mean~$0$ and variance~$t$;
 \item the sample paths are almost surely continuous in time, and~$B_0=0$.
\end{enumerate}

It is not immediately obvious that the conditions of the definition are
consistent, but it can be proved that Brownian motion exists. The transition
probability for Brownian motion from the state~$x$ to a Borel set~$A$ is
given by
\begin{equation}
 \Prob[B_{s+t}\in A \mid B_s = x] =
  \inv{\sqrt{2\pi t}} \int_A \e{-(y-x)^2/2t} \dif{y}
\end{equation}
where the integrand on the right is the Gauss kernel (or heat kernel).
Knowing the transition probability density of Brownian motion, it is not
difficult to prove the following lemma.

\begin{lemma}
 \label{lem:Brownianscalings}
 Let~$\{B_t:t\geq0\}$ be standard Brownian motion. Then each of the
 following stochastic processes is also standard Brownian motion:
 \begin{center}
 \begin{tabular}{ll}
  $\{-B_t : t\geq0\}$				& reflection invariance\\
  $\{B_{s+t}-B_s : t\geq0\}\qquad$	& time homogeneity\\
  $\{aB_{t/a^2} : t\geq0\}$			& scaling property\\
  $\{tB_{1/t} : t\geq0\}$			& time inversion symmetry
 \end{tabular}
 \end{center}
\end{lemma}

It is clear from the definition that Brownian motion is a Markov process,
since given~$B_s$, the value of~$B_{s+t}$ depends only on the increment
$B_{s+t}-B_s$ which is independent from all information up to time~$s$.
Indeed, at every time~$s$ it is as if the Brownian motion starts afresh
from the position~$B_s$, as time homogeneity shows. In fact, this holds
even if~$s$ is a stopping time for the Brownian motion. More precisely,
if~$T$ is a stopping time, then $B_{T+t}-B_T$ is a standard Brownian
motion which is independent of the events prior to~$T$. In other words,
Brownian motion has the strong Markov property.

Up to now, we considered only one-dimensional Brownian motion. The extension
to more dimensions is straightforward. In~$d$ dimensions, we can consider~$d$
independent standard Brownian motions~$B^1_t,\ldots,B^d_t$ and define
$d$-dimensional standard Brownian motion by $\vec{B}_t:=(B^1_t,\ldots,B^d_t)$.
Likewise, Brownian motion in the complex plane can be defined as the random
process~$B^1_t+\im B^2_t$. It is of course equivalent to Brownian motion
in~$\R[2]$. Brownian motion in the complex plane satisfies the following
theorem.

\begin{theorem}
 \label{the:Brownianhitting}
 Let~$B_t$ be a complex Brownian motion starting in~$z$. Let $D,E$ be
 disjoint subsets of~$\C$ such that $D\cup E$ is closed. Suppose further
 that every connected component of~$\C\setminus(D\cup E)$ has a continuous
 boundary consisting of a finite number of arcs in~$D$ and a finite number
 of arcs in~$E$. Denote by~$P(z)$ the probability that the Brownian motion
 hits the subset~$D$ before it hits~$E$. Then~$P(z)$ is determined by the
 Dirichlet problem
 \begin{equation}
 \left\{\begin{array}{ll}
  \Delta P(z) = 0,\qquad	& z \in \C \setminus (D\cup E); \\
  \phantom{\Delta}P(z) = 1,	& z \in D; \\
  \phantom{\Delta}P(z) = 0,	& z \in E.
 \end{array}\right.
 \end{equation}
\end{theorem}

Note that if $z$ is in a connected component~$G$ of $\C\setminus(D\cup E)$,
then~$P(z)$ is just the harmonic measure of $\partial G\cap D$ at the
point~$z$ with respect to~$G$ (see section~\ref{ssec:conformalmaps}). We
also remark that there are analogues of theorem~\ref{the:Brownianhitting}
in higher dimensions. As an example in two dimensions, consider the
rectangle $\mathcal{R}_L:=(0,L)\times(0,\im\pi)$. Suppose that $P(z)$
denotes the probability that a Brownian motion started from $z=x+\im y$
first leaves the rectangle through the edge $(L,L+\im\pi)$. Then it can be
verified using theorem~\ref{the:Brownianhitting} that
\begin{equation}
 \label{equ:Brownianexcursion}
 P(z) = \sum_{{\rm odd\ }k>0}
 			\frac{4}{\pi k}\frac{\sinh(kx)}{\sinh(kL)}\sin(ky).
\end{equation}
Observe that for small~$x$ this probability is of order~$x$, and that for
large~$L$ it scales like~$\exp(-L)$.

Now suppose that $\Prob_z$ denotes the probability measure on Brownian paths
started from $z\in\mathcal{R}_L$ and stopped when they hit the boundary of
the rectangle. Then a measure~$\mu_L$ on Brownian paths that start from the
edge $(0,\im\pi)$ is defined by
\begin{equation}
 \mu_L(\omega) := \int_0^\pi \lim_{\epsilon\downto0} \epsilon^{-1}
  \Prob_{\epsilon+\im y}(\omega)\,\dif{y}.
\end{equation}
We call this measure the \defn{Brownian excursion measure} on paths starting
from~$(0,\im\pi)$, and we call these paths \defn{Brownian excursions} of the
rectangle starting from the edge~$(0,\im\pi)$.
Equation~(\ref{equ:Brownianexcursion}) shows that if we restrict this measure
to Brownian excursions crossing the rectangle from left to right, then it has
finite total mass, and hence can be used to define a probability measure on
these excursions.

By conformal invariance of Brownian motion, we can now easily define the
probability measure on Brownian excursions crossing an arbitrary simply
connected domain~$D$ from an arc~$A_1$ of~$\partial D$ to a disjoint
arc~$A_2$ of~$\partial D$. Mapping the domain to a rectangle, it should be
clear that the measure on such Brownian excursions is easily defined through
the measure~$\mu_{\mathcal{L}}$, where~$\mathcal{L}$ is the $\pi$-extremal
distance between the arcs $A_1$ and~$A_2$ in the domain~$D$.

\subsection{Martingales and optional sampling}
\label{ssec:martingalesandOST}

A broad class of real-valued or complex-valued stochastic processes of
interest is the class of martingales. These are ``fair'' or ``unbiased''
processes in the sense that the expected value of the process at any time in
the future, given all information about the process up to the present time, 
is equal to the present value of the process. Another way of expressing this,
is to say that martingales are processes without ``drift''. We will come back
to this point of view in the next subsection. We now give a more precise
definition of a martingale.

\begin{definition}[Martingale]
 \label{def:martingale}
 Let $Y=\{Y_t : t\geq0\}$ be a real-valued or complex-valued random process,
 and let $\fil{F}$ be a filtration such that $Y_t$ is $\fld{F}_t$-measurable
 for all~$t\geq0$. Then the pair~$(Y,\fil{F})$ is called a \defn{martingale}
 if
 \begin{enumerate}
  \item $\Exp[\,|Y_t|\,]<\infty$ for all $t\geq0$;
  \item $\Exp[Y_t\mid\fld{F}_s] = Y_s$ for all~$s<t$.
 \end{enumerate}
 If $\fil{F}$ is the natural filtration generated by the process~$Y$ itself,
 then we say simply that~$Y$ is a martingale if it satisfies the conditions
 above.
\end{definition}

The following theorem expresses a connection between a martingale and a
Markov process~$X_t$. As the theorem shows, this martingale is defined as a
function $\psi(Y_t)$ of the variable $Y_t=X_{\min\{t,T\}}$, where~$T$ is a
stopping time for~$X_t$. In practice, although this is not strictly correct,
we often say that $\psi(X_t)$ itself (conditional on~$t<T$) is a martingale
under the conditions of the theorem. A proof of the theorem is provided
since it is not so easily found in elementary text-books.

\begin{theorem}
 \label{the:Markovmartingale}
 Let $X$ be a time-homogeneous Markov process with state space~$(\Omega',
 \fld{F}')$ and~$\fil{F}$ a filtration such that $X_t$ is
 $\fld{F}_t$-measurable for all~$t\geq0$. Let~$T$ be the hitting time of
 $A\in\fld{F}'$ for this process, and let~$f:A\goesto\R$ be a bounded
 function. Define
 \begin{equation}
  \psi(y) := \Exp\big[ f(X_T) \bigm| X_0=y \big].
 \end{equation}
 Set $Y_t:=X_{\min\{t,T\}}$. Then the process $Z_t:=\psi(Y_t)$ is a
 martingale with respect to~$\fil{F}$.
\end{theorem}

\begin{proof}
 Observe that for all $y\in\Omega'$, $s\geq0$ and~$B\in\fld{F}'$ we have
 \begin{equation}
  \Prob[X_T\in B\mid Y_s=y] = \Prob[X_T\in B\mid Y_0=y]
 \end{equation}
 by time homogeneity. Using the Markov property it follows that for all
 $s\geq0$ and~$B\in\fld{F}_s$
 \begin{eqnarray}
  && \int_B f(X_T(\omega))\,\dif{\Prob(\omega)}
   = \int_B \Exp[f(X_T)\mid\fld{F}_s](\omega)\,\dif{\Prob(\omega)} \nonumber\\
  &&\qquad = \int_B \Exp[f(X_T)\mid Y_s](\omega)\,\dif{\Prob(\omega)} 
   = \int_B \psi(Y_s(\omega))\,\dif{\Prob(\omega)}.
 \end{eqnarray}
 Now suppose that $s<t$, and let $B\in\fld{F}_s$ be arbitrary. Then
 $B\in\fld{F}_t$ because $\fil{F}$ is a filtration, and it follows that
 \begin{eqnarray}
  && \int_B \Exp[\psi(Y_t)\mid\fld{F}_s](\omega)\,\dif{\Prob(\omega)}
   = \int_B \psi(Y_t(\omega))\,\dif{\Prob(\omega)} \nonumber\\
  &&\hskip4em =  \int_B f(X_T(\omega))\,\dif{\Prob(\omega)}
   = \int_B \psi(Y_s(\omega))\,\dif{\Prob(\omega)}.
 \end{eqnarray}
 This proves that $\Exp[\psi(Y_t)\mid\fld{F}_s]=\psi(Y_s)$. Finally, by the
 boundedness of~$f$ we have $\Exp[\,|\psi(Y_t)|\,]<\infty$ for all~$t$.
\end{proof}

As we said earlier, a martingale is a fair process. Thus we may expect
that if we stop the process at some stopping time~$T$, the expected
value of the process at that time is just the value at time~$0$. However,
this statement does not hold in full generality. A more precise and careful
formulation leads to the following theorem, which is called the optional
sampling theorem by some, and the optional stopping theorem by others.
We stick to the name optional sampling theorem in this article.

\begin{theorem}[Optional sampling theorem]
 \label{the:OST}
 Let~$(Y,\fil{F})$ be a martingale and let~$T$ be a stopping time for the
 process~$Y$. If $\Prob[T<\infty]=1$,
 $\Exp\big[\,|Y_T|\,\big] < \infty$ and $\lim_{s\goesto\infty}
 \Exp\big[\,|Y_s|\bigm|T>s\big]\Prob[T>s] = 0$, then
 \begin{equation}
  \Exp[Y_T] = \Exp[Y_0].
 \end{equation}
\end{theorem}

As a special application of the optional sampling theorem, consider the
following situation. Let~$D\subset\C$ be a simply connected domain with
continuous boundary, and let~$f(z)$ be a bounded harmonic function on~$D$
that extends continuously to~$\partial D$. Suppose that~$B_t$ is a complex
Brownian motion starting in~$z\in D$, and consider the process $Y_t=f(B_t)$.
It\^o's formula in two dimensions (to be discussed in the following
subsection) shows that $Y_t$ is a martingale, as long as~$B_t$ stays in~$D$.
Therefore the following theorem holds.

\begin{theorem}[Optional sampling theorem, special case]
 \label{the:OSTonC}\hfil
 Let~$D$, $f$ and~$B_t$ be as in the previous paragraph. Define the stopping
 time~$T$ by $T := \inf\{t : B_t \in\partial D \}$. Then
 \begin{equation}
  \Exp[f(B_T)] = \Exp[f(B_0)] = f(z).
 \end{equation}
\end{theorem}

\subsection{It\^o calculus and stochastic differential equations}
\label{ssec:Ito}

In this subsection we consider It\^o's definition of stochastic integration
with respect to Brownian motion. This definition lies at the basis of the
theory of stochastic differential equations. We will describe the It\^o
calculus in this context, and discuss the main results, namely It\^o's
formula and the connection with martingales.

We start with the definition of the It\^o integral. Let $\fil{F}$ be the
filtration generated by the standard Brownian motion~$B_t$, and let $Y_t$
be a real-valued process which is $\fld{F}_t$-measurable for all $t\geq0$.
In words, this is to say that $Y_t$ is completely determined by the path of
the Brownian motion up to time~$t$. Suppose further that $\Exp[Y_t^2]<\infty$
for all~$t\geq0$, and that $Y_t$ has continuous sample paths. Then the
stochastic integral
\begin{equation}
 \label{equ:stochasticintegral}
 Z_t = \int_0^t Y_s \,\dif{B_s}
\end{equation}
of~$Y_t$ with respect to~$B_t$ is defined as follows.

First, we approximate the process~$Y_s$ by a simple process, which takes on
only finitely many values in any interval~$\roival{0,t}$:
\begin{equation}
 \label{equ:Itosimple}
 Y^{(n)}_s = \left\{
  \begin{array}{ll}
   \displaystyle n\int_{(k-1)/n}^{k/n} Y_r \,\dif{r}
    &  \mbox{for }s\in\big(\frac{k}{n},\frac{k+1}{n}\big],\
   	   1\leq k\leq n^2-1;\\
	& \\
   \displaystyle 0 & \mbox{for }s\leq 1/n\mbox{ or }s>n.
  \end{array}\right.
\end{equation}
where $k$ and~$n$ are positive integers. Observe that the interval of
integration in~(\ref{equ:Itosimple}) does not match with the interval
for~$s$. This is done deliberately to make $Y_s^{(n)}$ depend only on the
history of the process, that is, to make $Y_s^{(n)}$ measurable with respect
to~$\fld{F}_s$. It can be shown that $Y^{(n)}_s$ approaches~$Y_s$ in
mean-square (that is, $\Exp[\,|Y^{(n)}_s-Y_s|^2\,]\goesto0$) if we send~$n$
to infinity. This allows us to define the stochastic
integral~(\ref{equ:stochasticintegral}) as the mean-square limit
of a simple integral.

The simple integral of~$Y^{(n)}_s$ with respect to Brownian motion is defined
by the sum
\begin{equation}
 Z^{(n)}_t = \int_0^t Y^{(n)}_s \,\dif{B_s}
  := \sum_{i=1}^{j_n}
     Y^{(n)}_{i/n}[B_{i/n} - B_{(i-1)/n}] + Y^{(n)}_t[B_t-B_{j_n/n}]
\end{equation}
where~$j_n=\floor{nt}=\max\{m\in\N:m\leq nt\}$. The stochastic
integral~(\ref{equ:stochasticintegral}) is defined as the limit of this
expression as~$n\goesto\infty$, which is a limit in mean-square.

An important property of the It\^o integral~(\ref{equ:stochasticintegral}),
defined as described above, is that it defines a martingale. More precisely,
we have that~$Z_t$ is a martingale with respect to the filtration~$\fil{F}$.
There is also a remarkable converse to this statement, which says that any
martingale $(M,\fil{F})$ satisfies an equation of the form
\begin{equation}
 M_t=M_0+\int_0^t\,Y_s\,\dif{B_s}, \qquad t\geq0
\end{equation}
for some suitable process~$Y_t$. This relation between martingales and
It\^o integrals is very valuable in the theory of stochastic processes.

The stochastic integral~(\ref{equ:stochasticintegral}) is often written
in its differential form
\begin{equation}
 \dif{Z_t} = Y_t\,\dif{B_t}.
\end{equation}
The stochastic process~$Z_t$ defined through this equation may be regarded
as a Brownian motion that at time~$t$ has a variance~$Y_t^2$. One can
also consider a stochastic process which looks like a Brownian motion with
variance~$Y_t^2$ and drift~$X_t$ at time~$t$. Such a process satisfies the
stochastic differential equation
\begin{equation}
 \dif{Z_t} = X_t\,\dif{t} + Y_t\,\dif{B_t}.
\end{equation}
The corresponding stochastic integral is
\begin{equation}
 Z_t = Z_0 + \int_0^t X_s \,\dif{s} + \int_0^t Y_s \,\dif{B_s}
\end{equation}
where the first integral is an ordinary integral, and the second an It\^o
integral. It is to be noted that the process~$Z_t$, defined in this way, is
a martingale with respect to~$\fil{F}$ if and only if the drift term~$X_t$
is zero for all~$t$. 

Suppose now that we are given a stochastic differential equation that
describes some process~$X_t$. To derive the equation satisfied by a stochastic
process~$f(X_t)$, which is a function of the process~$X_t$, one uses
the It\^o calculus. The principle is the same as in ordinary calculus:
one considers infinitesimal increments of~$X_t$ over the infinitesimal
time increment~$\dif{t}$, keeping terms up to first order in~$\dif{t}$.
However, in the It\^o calculus we must treat the stochastic increment
$\dif{B_t}=B_{t+\dif{t}}-B_t$ as an increment of order~$(\dif{t})^{1/2}$
(consult Gardiner~\cite{gardiner:1983} for a nice discussion). Keeping
this in mind, one can derive the It\^o formula.

\begin{theorem}[One-dimensional It\^o formula]
 \label{the:Ito}
 Let~$f(x)$ be a function which has (at least) two continuous derivatives
 in~$x$, and suppose that~$X_t$ satisfies the stochastic differential
 equation
 \begin{equation}
  \dif{X_t} = a(X_t,t)\,\dif{t}+b(X_t,t)\,\dif{B_t}
 \end{equation}
 where~$B_t$ is standard Brownian motion. Then the stochastic process~$f(X_t)$
 satisfies
 \begin{equation}
  \label{equ:Ito}
  \dif{f(X_t)} = \left[ a(X_t,t) f'(X_t)
   + \ahalf b(X_t,t)^2 f''(X_t) \right]\dif{t}
   + b(X_t,t) f'(X_t)\,\dif{B_t}.
 \end{equation}
\end{theorem}

This formula expresses the process~$f(X_t)$ as the sum of an ordinary
integral and an It\^o integral. The fact that any It\^o integral defines a
martingale now leads to the important conclusion that the process~$f(X_t)$
of the theorem is a martingale with respect to the Brownian motion if and
only if the drift term in its It\^o formula~(\ref{equ:Ito}) vanishes. This
relates the martingale property of~$f(X_t)$ to an ordinary differential
equation for $f$ as a function of~$x$, and is a key to many proofs in {\SLE}.

Of course, one can extend the It\^o formula to more dimensions using the
same principles as in the one-dimensional case. That is, one again considers
infinitesimal increments up to first order in~$\dif{t}$. The increments
of the Brownian motions~$B_t^i$ are to be treated as increments of order
$(\dif{t})^{1/2}$, but this time with the added constraint that products
like $\dif{B^i_t}\dif{B_t^j}$ for $i\neq j$ vanish, since the two Brownian
motions are independent. Thus one can derive the multi-dimensional It\^o
formula.

\begin{theorem}[Multi-dimensional It\^o formula]
 \label{the:Itomulti}
 Let~$f(\vec{x})$ be a function of the~$n$ variables $x_1,\ldots,x_n$
 of the vector~$\vec{x}$, which has (at least) two continuous derivatives
 in all of the~$x_i$, and suppose that the~$n$ processes~$X^i_t$
 satisfy stochastic differential equations of the form
 \begin{equation}
  \dif{X^i_t} = a_i(\vec{X}_t,t)\,\dif{t}+
   \sum_{j=1}^n b_{ij}(\vec{X}_t,t)\,\dif{B^j_t}
 \end{equation}
 where $\vec{B}_t=(B^1_t,\ldots,B^n_t)$ is standard $n$-dimensional Brownian
 motion. Then the stochastic process~$f(\vec{X}_t)$ satisfies
 \begin{eqnarray}
  \label{equ:Itomulti}
  \dif{f(\vec{X}_t)}\hskip-6pt &=& \hskip-6pt
   \left[ \sum_{i=1}^n a_i(\vec{X}_t,t)\fracpderiv{x_i}{f(\vec{X}_t)}
   + \ahalf\sum_{i,j,k=1}^n \hskip-5pt b_{ik}(\vec{X}_t,t)b_{jk}(\vec{X}_t,t)
    \frac{\partial^2{f(\vec{X}_t)}}{\partial x_i \partial x_j} \right]\dif{t}
   \nonumber\\ && \hskip-6pt\null
   + \sum_{i,j=1}^n b_{ij}(\vec{X}_t,t)
    \fracpderiv{x_i}{f(\vec{X}_t)} \,\dif{B^j_t}.
 \end{eqnarray}
\end{theorem}

To conclude this appendix, we describe the concept of a (random) time-change
of Brownian motion. As before, let~$\fil{F}$ be the filtration generated by
the standard Brownian motion~$B_t$, and let~$X_t$ be a real-valued process
with continuous sample paths which is $\fld{F}_t$-measurable for all~$t\geq0$.
Consider the stochastic process~$Y_t$ defined through
\begin{equation}
 \dif{Y_t} = X_t\,\dif{B_t},\quad Y_0=0.
\end{equation}
Then~$Y_t$ is roughly a Brownian motion that has instantaneous variance~$X_t^2$
at every time~$t$. The scaling property of Brownian motion suggests that we
can scale this variance away by a suitable re-parameterization of time, so
that the time-changed process is standard Brownian motion.

\begin{theorem}[Time-change of Brownian motion]
 \label{the:time-change}
 Let~$X_t$ and~$Y_t$ be as described above. Assume that~$X_t$ is strictly
 positive and bounded up to a given stopping time~$T$. For $t<T$ define
 $s(t):=\int_0^t X_u^2\,\dif{u}$, and let~$t(s)$ denote the inverse of this
 time-change. Then the process $\tilde{Y}_s:=Y_{t(s)}$ is a standard
 Brownian motion up to the time~$s(T):=\lim_{t\upto T}s(t)$.
\end{theorem}

\begin{proof}
 Set $W_t:=\exp\big(\im\theta Y_t
  + \frac{1}{2}\theta^2\int_0^t X_u^2\,\dif{u}\big)$, where~$\theta\in\R$
 is fixed. By standard It\^o calculus, the drift term in~$\dif{W_t}$ is zero.
 Since~$|W_t|$ is bounded, this shows that~$W_t$ is a martingale. Therefore,
 if $0\leq t_0<t_1<T$ then $\Exp[W_{t_1}\mid\fld{F}_{t_0}] = W_{t_0}$.
 Writing $s_0=s(t_0)$ and $s_1=s(t_1)$, this gives
 \begin{equation}
  \Exp\left[\exp\left(\im\theta(\tilde{Y}_{s_1}-\tilde{Y}_{s_0})\right)
  	\Bigm|\tilde{\fld{F}}_{s_0}\right]
  = \exp\left(-\frac{1}{2}\theta^2(s_1-s_0)\right)
 \end{equation}
 where we wrote $\tilde{\fld{F}}_s$ for the time-changed $\sigma$-field
 $\tilde{\fld{F}}_s:=\fld{F}_{t(s)}$. This equation is just the characteristic
 equation saying that~$\tilde{Y}_{s_1}-\tilde{Y}_{s_0}$ is a normally
 distributed random variable with mean~$0$ and variance~$s_1-s_0$
 (see~\cite{grimmett:2001}). It follows that $\tilde{Y}_s$ is standard
 Brownian motion.
\end{proof}

\end{document}